\documentclass[12pt]{article}
\usepackage{cite}
\usepackage{amsmath, amsthm, amssymb,slashed}
\usepackage{ifpdf}
\ifpdf
  \usepackage[pdftex]{graphicx}
  \usepackage{epstopdf}
\else
  \usepackage[dvips]{graphicx}
\fi
\parskip=\baselineskip
\def\Bbb{\mathbb}
\def\Tr{{\rm Tr}}
\def\16{{\bf 16}}
\def\1{{\bf 1}}
\def\2{{\bf 2}}
\def\3{{\bf 3}}
\def\4{{\bf 4}}
\DeclareMathAlphabet{\mathpzc}{OT1}{pzc}{m}{it}
\def\bar{\overline}
\def\tilde{\widetilde}
\def\R{{\Bbb{R}}}\def\Z{{\Bbb{Z}}}
\def\N{{\mathcal N}}
\def\d{{\mathrm d}}
\def\hat{\widehat}
\font\teneurm=eurm10 \font\seveneurm=eurm7 \font\fiveeurm=eurm5
\newfam\eurmfam
\textfont\eurmfam=\teneurm \scriptfont\eurmfam=\seveneurm
\scriptscriptfont\eurmfam=\fiveeurm

\font\teneusm=eusm10 \font\seveneusm=eusm7 \font\fiveeusm=eusm5
\newfam\eusmfam
\textfont\eusmfam=\teneusm \scriptfont\eusmfam=\seveneusm
\scriptscriptfont\eusmfam=\fiveeusm
\def\eusm#1{{\fam\eusmfam\relax#1}}
\font\tencmmib=cmmib10 \skewchar\tencmmib='177
\font\sevencmmib=cmmib7 \skewchar\sevencmmib='177
\font\fivecmmib=cmmib5 \skewchar\fivecmmib='177
\newfam\cmmibfam
\textfont\cmmibfam=\tencmmib \scriptfont\cmmibfam=\sevencmmib
\scriptscriptfont\cmmibfam=\fivecmmib
\def\cmmib#1{{\fam\cmmibfam\relax#1}}
\numberwithin{equation}{section}

\def\d{\mathrm d}
\def\C{{\Bbb C}}
\def\Z{{\Bbb Z}}
\def\B{{\mathcal B}}
\def\A{{\mathcal A}}
\def\bar{\overline}
\begin{document}

\begin{titlepage}
\begin{flushright}
hep-th/yymm.nnnn
\end{flushright}
\vskip 1.5in
\begin{center}
{\bf\Large{A New Look At The Path Integral}\vskip.2cm{ Of Quantum Mechanics }}
\vskip
0.5cm {Edward Witten} \vskip 0.05in {\small{ \textit{School of
Natural Sciences, Institute for Advanced Study}\vskip -.4cm
{\textit{Einstein Drive, Princeton, NJ 08540 USA}}}
\vskip.05in{and}
\vskip .05in {\textit{Department of Physics, Stanford University}}
\vskip-.15in {\textit{Palo Alto, CA 94305}}}
\end{center}
\vskip 0.5in
\baselineskip 16pt
\date{September, 2010}

\begin{abstract}
The Feynman path integral of ordinary quantum mechanics is complexified and it is
shown that possible  integration
cycles for this complexified  integral are associated with branes in a two-dimensional
$A$-model. This provides a fairly direct explanation of the relationship of the $A$-model
to quantum mechanics; such a relationship has been explored from several points
of view in the last few years.   These phenomena have an analog for Chern-Simons gauge theory in three
dimensions: integration cycles in the path integral of this theory can be derived from $\N=4$
super Yang-Mills theory in four dimensions. Hence, under certain conditions, a Chern-Simons
path integral in three dimensions is equivalent to an $\N=4$ path integral in four dimensions.
\end{abstract}
\end{titlepage}
\vfill\eject \tableofcontents
\section{Introduction}

\def\M{{\mathcal M}}
\def\H{{\mathcal H}}
\def\R{{\mathcal R}}
\def\L{{\mathfrak L}}
\def\LL{\mathcal L}
\def\frak{\mathfrak}
\def\RR{{\Bbb R}}
\def\TT{{\mathpzc T}}

The Feynman path integral in Lorentz signature is schematically of the form
\begin{equation}\label{turnox} \int\,D\Phi\,\exp\left(iI(\Phi)\right),\end{equation}
where $\Phi$ are some fields and $I(\Phi)$ is the action.  Frequently, $I(\Phi)$ is a real-valued  polynomial
function of $\Phi$ and its derivatives (the polynomial nature of $I(\Phi)$ is not really necessary
in what follows, though it simplifies things).    There is also a Euclidean version of the path integral, schematically
\begin{equation}\label{urnox}\int\,D\Phi\, \exp\left(-I(\Phi)\right),\end{equation}
where now $I(\Phi)$ is a polynomial whose real part is positive definite, and which is complex-conjugated under a reversal of the spacetime orientation.\footnote{General Relativity departs from this framework in a conspicuous way: in Euclidean
signature, the real part of its action is not positive definite.
This has indeed motivated the proposal \cite{GH} that the Euclidean path integral of General
Relativity must be carried out over an integration cycle different from the usual space of
real fields.  The four-dimensional case is difficult, but in three dimensions, the possible integration cycles can be understood rather explicitly \cite{witten}.}

The analogy between the Feynman path integral and an ordinary finite-dimensional integral
has often been exploited.  For example, as a prototype for the Euclidean version of the Feynman
integral, we might consider a one-dimensional integral
\begin{equation}\label{omorf}I=\int_{-\infty}^\infty\d x \,\exp(S(x)),\end{equation}
where $S(x)$ is a suitable polynomial, such as
\begin{equation}\label{zmorf}S(x)=-x^4/4+ax,\end{equation}
with $a$ a parameter.  One thing which we can do with such an integral is to analytically
continue the integrand to a holomorphic function of $z=x+iy$ and carry out the integral
over a possibly different integration cycle in the complex $z$-plane:
\begin{equation}\label{tmorf}I_\Gamma= \int_\Gamma \d z\,\exp(S(z)).\end{equation}
The integral over a closed contour $\Gamma$ will vanish, as the integrand is an entire function.
Instead, we take $\Gamma$ to connect two distinct regions at infinity in which $\mathrm{Re}
\,S(z)\to -\infty$.
In the case at hand,  there are four such regions (with the argument of $z$ close to $k\pi/2$, $k=0,1,2,3$) and hence
there are essentially three integration cycles $\Gamma_r$, $r=0,1,2$.  (The cycle $\Gamma_r$ interpolates between $k=r$ and $k=r+1$,
as shown in the figure.)  In general, as reviewed in \cite{witten}, the integration cycles take values in a certain relative homology group.
In the case at hand, the relative homology is of rank three, generated by $\Gamma_0,\Gamma_1,\Gamma_2$.

\begin{figure}
 \begin{center}
   \includegraphics[width=2.5in]{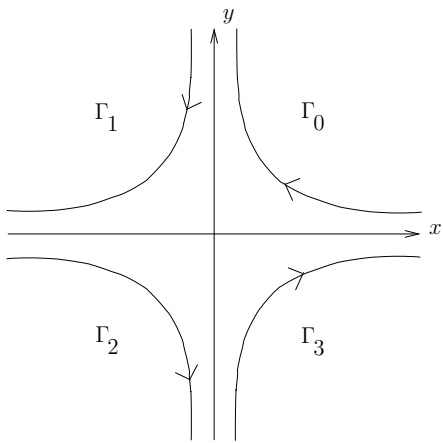}
 \end{center}
\caption{\small Integration cycles for the integral $I_\Gamma$ of eqn. (\ref{tmorf}).
The four cycles obey one relation $\Gamma_1+\Gamma_1+\Gamma_2+\Gamma_3=0$,
and any three of them give a basis for the space of possible integration cycles. }\vskip .2cm
 \label{bairy}
\end{figure}

Given a similar integral in $n$ dimensions,
\begin{equation}\label{pmorf}\int_{\RR^n}\d x_1\,\d x_2\dots\d x_n\,\exp\left(S(x_1,\dots,x_n)\right)\end{equation}
again with a suitable polynomial $S$, one can analytically continue from real variables
$x_i$ to complex variables $z_i=x_i+iy_i$ and replace (\ref{pmorf}) with an integral over
a suitable integration cycle $\Gamma\subset\C^n$:
\begin{equation}\label{rmorf}\int_\Gamma \d z_1\,\d z_2\dots\d z_n\,\exp\left(S(z_1,\dots,z_n)\right).\end{equation}
The appropriate integration cycles are $n$-cycles, simply because what we are trying to
integration is the $n$-form $ \d z_1\,\d z_2\dots\d z_n\,\exp(S(z_1,\dots,z_n))$.  Of course,
the differential form that we are trying to integrate is  middle-dimensional  simply because, in analytically
continuing from $\RR^n$ to $\C^n$, we have doubled the dimension of the space in which
we are integrating.  So what began in the real case as a differential form of top dimension
has become middle-dimensional upon analytic continuation.

In \cite{witten}, it was shown that, at least in the case of three-dimensional Chern-Simons
gauge theory, these concepts can be effectively applied in the infinite-dimensional case of a Feynman integral.
But what do we learn when we do this?
When one constructs different integration cycles for the same integral -- or the same path
integral -- how are the resulting integrals related?  For one answer, return to the original example $I_\Gamma$.  Regardless of $\Gamma$ (using only the facts that it is a cycle, without boundary, that
begins and ends in regions where the integrand is rapidly decaying, so that one can integrate by parts), $I_\Gamma$ obeys
the differential equation
\begin{equation}\label{onox} \left(\frac{\d^3}{\d a^3}-a\right)I_\Gamma=0.\end{equation}
  Indeed, $I_\Gamma$ where $\Gamma$ runs over
a choice of three independent integration cycles
gives a basis of the three-dimensional space of solutions of this third-order differential equation.

For quantum field theory, the analog of (\ref{onox}) are the Ward identities obeyed by the correlation functions.  Like (\ref{onox}), they are proved by integration by parts in function space,
and do not depend on the choice of the integration cycle.  One might think that different integration cycles would correspond to different
vacuum states in the same quantum theory, but this is not always right.   In some cases, as we will explain
in section \ref{simplint} with an explicit example, different integration cycles correspond to different quantum systems that have the same algebra of observables.   In other cases, the interpretation is more exotic.

The first goal of the present paper is to apply these ideas to a particularly basic case of the Feynman path integral.  This is the phase space path integral of nonrelativistic quantum mechanics
with coordinates and momenta $q$ and $p$:
\begin{equation}\int Dp(t)\,Dq(t)\,\exp\left(i\int\left( p\,\d q- H(p,q)\d t\right)\right).\end{equation}
($H(p,q)$ is the Hamiltonian and plays a secondary role from our point of view.)
Integration cycles for this integral are analyzed in section \ref{quantint}.   There are some
fairly standard integration cycles, such as the original one assumed by Feynman, with $p(t)$ and
$q(t)$ being real.  The main
new idea in this paper is that by restricting the integral to complex-valued paths $p(t)
,q(t)$ that are boundary values of pseudoholomorphic maps -- in a not obvious sense --
we can get a new type of integration cycle in the path integral of quantum mechanics.
Moreover, this cycle has a natural interpretation in a {\it two}-dimensional quantum field theory
-- a sigma-model in which the target is the complexification of the original classical phase space.
The sigma-model is in fact a topologically twisted $A$-model, and the integration cycle can be described
using an exotic type of $A$-brane known as a coisotropic brane \cite{KO}.

  It has
been known from various points of view \cite{BrS,Ka,KW,gualt,AZ,GW} that there is
a relationship between the $A$-model and quantization.    In the present paper, we make a new and particularly direct
proposal for what the key relation is:
 the most basic coisotropic $A$-brane gives a new integration
cycle in the Feynman integral of quantum mechanics.

The fact that boundary values of pseudoholomorphic maps give a middle-dimensional
cycle in the loop space of a symplectic manifold (or classical phase space) is one of the main
ideas in Floer cohomology \cite{floer}.  For an investigation from the standpoint  of field
theory, see \cite{FLN,FLN2,FLN3}.  The middle-dimensional cycles of Floer theory are not
usually interpreted as integration cycles, because there typically are no natural middle-dimensional forms that can be integrated over these cycles.  In the present paper, we first double the dimension by complexifying the classical phase space, whereupon the integrand of the usual Feynman integral becomes a middle-dimensional form that can be integrated over the cycles given by Floer theory of the complexified phase space.

The relationship we describe between theories in dimensions one and two
has an analog in dimensions three and four.  Here the three-dimensional theory is Chern-Simons gauge theory, with a compact gauge group $G$,
and the four-dimensional theory is $\N=4$ super Yang-Mills theory, with the same gauge group.
To some extent, the link between the two was made in \cite{witten}.  It was shown that to define
an integration cycle in three-dimensional Chern-Simons theory, it is useful to add a fourth
variable and  solve certain partial differential equations
that are related to $\N=4$ super Yang-Mills theory.  Here we go farther and show
exactly how a quantum path integral in $\N=4$ super Yang-Mills theory on a four-manifold
with boundary can reproduce the Chern-Simons path integral
on the boundary, with a certain integration cycle.
This has an application which will be described elsewhere \cite{wittenfive}.  The application involves a new way to understand the link
\cite{GSV} between BPS states of branes and Khovanov homology \cite{K} of knots.

In sections \ref{quantint} and \ref{zono} of this paper, we begin with the standard Feynman
integral of quantum mechanics and motivate its relation to a twisted supersymmetric theory in
one dimension more.
In section \ref{reverse}, we run the same story in reverse, starting in the higher dimension
and deducing the relation to a standard Feynman integral in one dimension less.  Some readers might prefer this
second explanation.  Section \ref{cs} generalizes
this approach to gauge fields and contains the application to Chern-Simons theory.

What is the physical interpretation of the Feynman integral with an exotic integration cycle? In the present paper,
we make no claims about this, except
that it links one and two (or three and four) dimensional information in an interesting way.

\section{Integration Cycles For Quantum Mechanics}\label{quantint}

\subsection{Preliminaries}\label{prelims}

A classical mechanical system is described by a $2n$-dimensional phase space $\M$,  which is endowed with a symplectic structure.
The symplectic structure is described by a two-form $f$ that is closed
\begin{equation}\label{horm}\d f= 0 ,\end{equation}
and also nondegenerate,  meaning that the matrix $f_{ab}$ defined in local coordinates $x^a, \,a=1,\dots,2n$
by $f=\sum_{a<b}f_{ab}\d x^a\wedge \d x^b$ is invertible; we write $f^{ab}$ for the inverse matrix.
Locally, if $f$ is a closed, nondegenerate two-form, one can pick canonical coordinates $p_r$, $q^s$, $r,s=1,\dots,n$ such that
\begin{equation}\label{omonk} f=\sum_{s=1}^n \d p_s \wedge \d q^s.    \end{equation}
The Poisson bracket of two functions $u$, $v$ on $\M$ is defined by
\begin{equation}\label{monk} \{u,v\}=f^{ab}\frac{\partial u}{\partial x^a}\frac{\partial v}{\partial x^b}=\sum_s\left(\frac{\partial u}{\partial q^s}
\frac{\partial v}{\partial p_s}-\frac{\partial u}{\partial p_s}
\frac{\partial v}{\partial q^s} \right). \end{equation}

In quantization, one aims to associate to that data a Hilbert space $\H$ and an algebra $\R$ of observables.  $\H$ will be finite-dimensional
if and only if $\M$ has finite volume.  In that case, to elements $U_1,\dots,U_n$ of $\R$, one can associate the trace
\begin{equation}\label{ompo}\Tr_\H U_1U_2\dots U_n,\end{equation} which describes $\R$ and its action on $\H$, up to unitary
equivalence.
In most physical applications, there is an element $H$ of $\R$ known as the Hamiltonian, and one is particularly interested
in traces
\begin{equation}\label{zompo}\Tr_\H U_1(t_1)U_2(t_2)\dots U_n(t_n)\exp(-iHt),\end{equation}
where for $U\in \R$, $U(t)$ is defined as $\exp(-iHt)U\exp(iHt)$.  However, the basic problem of quantization\footnote{Many elementary
aspects of this problem are not as well-known as they might be.  For example, the passage from classical mechanics to quantum mechanics
does not map Poisson brackets to commutators, except in leading order in $\hbar$ (which we set to 1 in the present paper) and certain
special cases such as quadratic functions on $\M$.  A review can be found in \cite{GW}.}  is to associate to $\M$
a Hilbert space $\H$ and algebra $\R$, and what we will say about this problem in the present paper mostly has nothing to do with the choice
of $H$.  We will omit $H$ (that is, set $H=0$) except in section \ref{zono}.

A more incisive approach to quantum mechanics is to consider not only a trace as in (\ref{zompo}) but
matrix elements $\langle\psi_f|U_1U_2\dots U_n|\psi_i\rangle$ between initial and final quantum states $|\psi_i\rangle$,
$|\psi_f\rangle$.  (This also avoids the technical difficulty that if $\M$ has infinite volume, $\H$ is infinite-dimensional and the
traces in (\ref{ompo}) and  (\ref{zompo}) may diverge, depending on the $U_i$ and $H$.)   However, the approach to Feynman integrals
in the present paper is most easily explained if we begin with traces rather than matrix elements.  The additional steps involved
in describing matrix elements between initial and final states are sketched in section \ref{oose}.

\subsection{The Basic Feynman Integral}\label{relims}

\def\U{{\mathcal U}}
The basic Feynman integral represents the trace (\ref{ompo}) as an integral over maps from $S^1$ to $\M$:
\begin{equation}\label{zomp}\Tr_\H \,U_1U_2\dots U_n=\int_{\mathcal U} Dp_r(t)\,Dq^r(t) \exp\left(i\oint p_s \d q^s\right)~ u_1(t_1)u_2(t_2)\dots u_n(t_n).
\end{equation}
Here  $p_r(t)$ and $q^r(t)$ are periodic with a period of, say, $2\pi$, so they define a map ${\mathcal T}:S^1\to \M$, or in other words
a point in the free loop space $\U$ of $\M$.  The $u_\alpha$ are functions on $\M$ that upon quantization will correspond
to the operators $U_\alpha$.  We have assigned a time $t_\alpha$ to each function $u_\alpha$,
but as we have taken the Hamiltonian to vanish, all that matters
about the $t_\alpha$ is their cyclic ordering on $S^1$.  As is usual, $u(t)$ is an abbreviation for
$u({\mathcal T}(t))$.

As written, the Feynman integral depends on a choice of canonical coordinates $q^r$ and momenta $p_r$.  We could, for example,
make a canonical transformation from $p,q$ to $-q,p$, replacing $\oint p_s\d q^s$ with $-\oint q^s\d p_s$.  This amounts to
adding to the exponent of the path integral a term  $\oint \d(-p_s q^s)=0$, where the vanishing holds because the functions are periodic in time.

Rather than picking a particular set of canonical coordinates, a more intrinsic approach is to  observe that, as
the two-form $f$ is closed, it can be regarded as the curvature of an abelian gauge field $b$:
\begin{equation}\label{omk}f=\d b,~~b=\sum_{a=1}^{2n} b_a\,\d x^a.\end{equation}
Then we can write (\ref{zomp}) more intrinsically as
\begin{equation}\label{uzomp}\Tr_\H\, U_1U_2\dots U_n=\int_\U Dx^a(t) \exp\left(i\oint b_a\d x^a\right)~ u_1(t_1)u_2(t_2)\dots u_n(t_n).
\end{equation}

We say that the Dirac condition is obeyed if the periods of $f$ are integer multiples of $2\pi$.  In that case,   a unitary line bundle $\L\to \M$
with a connection of curvature $f$ exists and we take $b$ to be that connection; its structure group is $U(1)$.  $\L$ is called a prequantum line bundle.  If $H_1(M,\Z)\not=0$, there are inequivalent choices of $\L$ and quantization
depends on such a choice.     When $\L$ exists, the factor $\exp\left(i\oint b_a\d x^a\right)$ is the holonomy of the connection $b$ on $\L$
(pulled back to $S^1$ via the map ${\mathcal T}:S^1\to \M$).

If the Dirac condition is not obeyed, the Feynman integral with the usual integration cycle (real phase space coordinates $x^a$) does not make
sense since the factor $\exp\left( i\oint b_a\d x^a\right)$ in the path integral is not well-defined.  We can make this factor well-defined
by replacing $\U$ by its universal cover -- or by any cover $\U^*$  on which the integrand of the path integral is single-valued.
Once we do this, the integral over the usual integration cycle of the Feynman integral  is not interesting
because all integrals (\ref{zomp}) vanish.  (They transform with a non-trivial phase under the deck transformations of the cover $ \U^*\to \U$.)
However, as analyzed in \cite{witten}, and as we will see below, after analytic continuation, there may be sensible integration cycles (related
to deformation quantization, which does not require the Dirac condition, rather than quantization, which does).  So we do
not want to assume that the Dirac condition is obeyed.

In what follows, it might be helpful to have an example in mind.  A simple example is the case that $\M=S^2$, defined by an equation
\begin{equation}\label{twos} x_1^2+x_2^2+x_3^2=j^2,\end{equation}
for some constant $j$.  We take
\begin{equation}\label{gwoos}f=\frac{\epsilon_{ijk} x_i\,\d x_j\wedge \d x_k}{3!R^2}=\frac{\d x_1\wedge  \d x_2}{x_3}.\end{equation}
The first formula for $f$ makes $SO(3)$ invariance manifest, and the second is convenient for computation.
One can verify that $\int_{S^2}f=4\pi j$, so that the Dirac condition becomes $j\in\Z/2$.  However, as already explained,
we do not necessarily want to assume this condition.   We can think of $f$ as the magnetic field due to a magnetic monopole (of magnetic
charge $2j$) located at the center of the sphere; $b$ is the gauge connection for this monopole field.

\subsection{Analytic Continuation}\label{anacon}

\def\U{{\mathcal U}}
\def\bX{{\bar X}}
\def\ba{{\bar a}}
The Feynman integral, as we have formulated it so far, is an integral over the free loop space $\U$ of $\M$, or possibly a cover of this on which the integrand
of the Feynman integral is well-defined.  Our first step, as suggested in the introduction,
is to analytically continue from $\U$ to a suitable complexification $\hat\U$.  We simply pick a  complexification $\hat\M$ of $\M$
and let $\hat \U$ be the free loop space of $\hat\M$ (or, if necessary, an appropriate cover of this).

What do we mean by a complexification of $\M$?  At a minimum, $\hat\M$ should be a complex manifold with an antiholomorphic involution\footnote{An involution is simply a symmetry whose square is the identity.}
$\tau$ such that $\M$ is a component of the fixed point set of $\tau$.
Moreover, we would like $\hat\M$ to be a complex symplectic
manifold endowed with a holomorphic two-form $\Omega$ that is closed and nondegenerate and has the property that
its imaginary part, when restricted to $\M$, coincides with $f$. We introduce the real and imaginary parts of $\Omega$ by
\begin{equation}\label{inthem}\Omega=\omega+if.\end{equation}  The fact that $\Omega$ is closed and nondegenerate implies that
 $\omega$ and $f$ are each closed and nondegenerate.  And we assume that
   under $\tau$, $\Omega$ is mapped
 to $-\bar\Omega$, so in other words
 \begin{equation}\label{them}\tau^*(\omega)=-\omega,~~\tau^*(f)=f.\end{equation}
It follows that on the fixed point set $\M$, $\omega$ must vanish:
\begin{equation}\label{hem}\omega|_\M=0.\end{equation}

We denote local complex coordinates on $\hat\M$ as
$X^a$, and their complex conjugates as $\bX{}^a$, and we denote a complete set of local real coordinates (for example, the real and imaginary parts of the $X^a$) as $Y^A$.  Since $\Omega$ is a closed form,
we can write it as the curvature of a complex-valued abelian gauge field:
\begin{equation}\label{curvox}\Omega=\d\Lambda,~~\Lambda=\sum_A \Lambda_A\d Y^A.\end{equation}
If the original model obeyed the Dirac condition and the complexification introduces no new topology, we can regard $\Lambda$ as a
gauge field with structure group $\C^*$ (the complexification of $U(1)$).  In general, however, as stated in section \ref{relims}, we do not assume this to be the case, and instead we replace $\hat \U$ by a suitable cover on which the integrand of the integral (\ref{orox}) introduced
below is well-defined.  It is convenient to introduce the real and imaginary parts of $\Lambda$
just as we have done for $\Omega$.  So we write
\begin{equation}\label{polx}\Lambda=c+ib,\end{equation}
where $b$ and $c$ are real-valued connections.  Thus
\begin{equation}\label{polix}\d c = \omega,~~\d b=f.\end{equation}
We will slightly sharpen (\ref{hem}) and assume that there is a gauge with
\begin{equation}\label{pem}c|_\M=0.\end{equation}

 Let us describe what these definitions  mean for our example with $\M=S^2$.  We define $\hat\M$ by the same
equation (\ref{twos}) that we used to define $\M$, except\footnote{For a critical discussion of the sense in which this analytic continuation
is or is not natural, as well as a discussion of the class of observables considered below, see \cite{GW}.}
we regard it as an equation for complex variables $X_i$ rather than real variables $x_i$:
\begin{equation}\label{omig}X_1^2+X_2^2+X_3^2=j^2.\end{equation}
And we define $\Omega$ by the same formula as before except for a factor of $i$:
\begin{equation}\label{gwos}\Omega=i\frac{\epsilon_{ijk} X_i\,\d X_j\wedge \d X_k}{3!R^2}=i\frac{\d X_1\wedge  \d X_2}{X_3}.\end{equation}
(The factor of $i$ in the definition of $\Omega$ is a minor convenience; the formulas that follow are slightly more elegant if we take
$f$ to be the imaginary part of $\Omega$ -- restricted to $\M$ -- rather than the real part.)

In addition to the conditions that we have already stated, $\hat\M$ must have one additional property.  Some condition of completeness
of $\hat\M$ must be desireable, since certainly we do not expect to get a nice theory if we omit from $\hat\M$ a randomly chosen $\tau$-invariant closed set that is disjoint from $\M$.    It is not obvious {\it a priori} what the right condition should be, but as we will find  (and as found in \cite{GW} in another
way) the appropriate condition is that $\hat\M$, regarded as a real symplectic manifold with symplectic structure $\omega$, must have a well-defined
 $A$-model.   For noncompact symplectic manifolds, this is a non-trivial though in general not well understood condition.  Our example
of the complexification of $S^2$ certainly has a good $A$-model, since in fact this manifold admits a complete hyper-Kahler metric
-- the Eguchi-Hansen metric.

As for the functions $u_\alpha$ that appear in the path integral (\ref{zomp}), to make sense of them in the context of an analytic
continuation of the Feynman integral, they must have an analytic continuation to holomorphic functions on $\hat \M$ (which we still denote
as $u_\alpha$).  Moreover, so as not to affect questions involving convergence of the path
integral,
the analytically continued functions should not grow too fast at infinity.  For example, in the case of $S^2$, a good class of functions are the polynomial functions $u(x_1,x_2,x_3)$.  The analytic continuation of
such a polynomial is simply the corresponding polynomial $u(X_1,X_2,X_3)$.  These are the best observables to consider, because they
are the (nonconstant)  holomorphic functions with the slowest growth at infinity.

Having analytically continued the loop space $ \U$ of $\M$ to the corresponding complexified loop space $\hat\U$ of $\hat\M$,
and having similarly analytically continued the symplectic structure and the observables, we can formally write down the Feynman integral over an arbitrary integration cycle
$\Gamma\subset\hat\U$:
\begin{equation}\label{orox}\int_\Gamma D Y^A(t) \exp\left(\oint \Lambda_A\d Y^A\right)~u_1(t_1)\dots u_n(t_n).\end{equation}
$\Gamma$ is any middle-dimensional cycle in $\hat\U$ on which the integral converges.
Eqn. (\ref{pem}) ensures that if we take $\Gamma$ to be the original integration cycle
$\U$ of the Feynman integral, then (\ref{orox}) does coincide with the original Feynman integral.

\subsection{The Simplest Integration Cycles}\label{simplint}

The reason that there is some delicacy in choosing $\Gamma$ is that the real part of
the exponent in (\ref{orox}) is not bounded above.

The troublesome factor in (\ref{orox}) comes from the real part of $\Lambda$.
(We assume that the observables $u_\alpha$ do not grow so rapidly as to affect the following
discussion.)  The integration cycle $\Gamma$ must be chosen so that the dangerous factor
\begin{equation}\label{prox}\exp\left(\oint c_A\d Y^A\right)=\exp\left(\mathrm{Re}\oint\Lambda_A\,\d Y^A\right)\end{equation}
does not make the integral diverge.  The reason that this is troublesome is that
the exponent
\begin{equation}\label{rox} h=\oint c_A\d Y^A \end{equation}
is unbounded above and below.  For example, the map
\begin{equation}\label{hefalo}Y^A(t)\to \tilde Y^A(t)=Y^A(nt)\end{equation} multiplies
$h$ by an arbitrary integer $n$, which can be positive or negative.

Leaving aside the question of convergence, how can we find a middle-dimensional
cycle $\Gamma\subset \hat\U$?  The most elementary approach is to define $\Gamma$
by a local-in-time condition.  We pick a middle-dimensional submanifold $\M_0\subset\hat\M$
and take $\Gamma\subset\hat\U$ to be the free loop space of $\M_0$.  In other words,
$\Gamma$ parametrizes maps ${\mathcal T}:S^1\to\hat\M$ whose image lies in $\M_0$ for all time.

This type of choice cannot be wrong, since if we take $\M_0=\M$, we get back to the original
Feynman integral.  Now let us ask for what other choices of $\M_0$ we get a suitable integration
cycle.

When restricted to the loop space of $\M_0$, the function $h$ must be identically zero, or the
argument around eqn. (\ref{hefalo}) will again show that it is not bounded above or below.  The variation of
$h$ under a small change in a map ${\mathcal T}:S^1\to \hat\M$ is
\begin{equation}\label{orpz}\delta h=\oint \omega_{AB}\delta Y^A \d Y^B.\end{equation}
For this to vanish identically when evaluated at any loop in $\M_0$, we require that
$\omega$ restricted to $\M_0$ must vanish.\footnote{This is enough to ensure that $h$ is
constant in each connected component of the free loop space of $\M_0$.  For it to vanish
identically, we need in addition that $c$ is a pure gauge when restricted to $\M_0$.}

The requirement, in other words, is that $\M_0$ must be a Lagrangian submanifold with
respect to $\omega$.  This is a familiar condition in the context of the two-dimensional
$A$-model: it is a classical approximation to the condition for $\M_0$ (endowed with a trivial
Chan-Paton bundle) to be the support of an $A$-brane.   This is no
coincidence, but a first hint that the possible integration cycles for the path
integral are related to the $A$-model.

Upon picking $\M_0$ so that the real part of the exponent of the path integral is bounded
above, we are not home free: to make sense of the infinite-dimensional path integral,
the phase factor that comes from the imaginary part of the exponent must be nondegenerate.
For this, we want $f$ to be nondegenerate when restricted to $\M_0$.  In other words, $\M_0$
should have some of the basic properties of $\M$: when restricted to $\M_0$, $\omega$ vanishes
and $f$ is nondegenerate.\footnote{\label{borx} We imposed one more condition on $\M$: it is
a component of the fixed point set of an antiholomorphic involution $\tau$.  As explained in
\cite{GW}, this is needed so that quantization of $\M$ in the sense of the $A$-model
admits a hermitian structure.  So if $\M_0$ does not obey this condition, its ``quantization''
via the $A$-model -- this operation is reviewed in section \ref{oose} -- is not unitary.}   Under these conditions, the Feynman integral (\ref{orox}) for
$\Gamma$ equal to the loop space of $\M_0$ is simply the original Feynman integral (\ref{uzomp})
with $\M_0$ replacing $\M$.  In other words, it is the usual Feynman integral
associated with quantization of $\M_0$.

In this situation, different quantum mechanics problems with different choices of $\M_0$
correspond to different integration cycles in the same complexified Feynman integral.
For a concrete example of how this can happen, let us return
to the familiar example, with $\hat\M$ defined by
the equation \begin{equation}\label{bonkers}X_1^2+X_2^2+X_3^2=j^2.\end{equation}   The condition on $\M_0$ is that
${\mathrm{Re}}\,\Omega|_{\M_0}=0$, or in other words
\begin{equation}\label{congo}\left.\mathrm{Im}\left(\frac{\d X_1\wedge \d X_2}{X_3}\right)\right|_{\M_0}=0.
\end{equation}
A sufficient condition for this is  that $X_1,X_2,X_3$ are all real.  This brings us back to the original
phase space $\M$.    To get another example, we define $\M'$ by requiring that $X_1$ is
real and positive while $X_2$ and $X_3$ are imaginary.  It is not hard to see that while $\M$
is a two-sphere $S^2$, $\M'$ is a copy of the upper half plane $H^2$.  In quantization of either
$\M$ or $\M'$, the observables are the same, namely polynomials $u(X_1,X_2,X_3)$ (modulo
the relation (\ref{bonkers})), restricted
to $\M$ or $\M'$.  Moreover, any identities in correlation functions of these observables (apart
from (\ref{bonkers})) arise as Ward identities in the path integral and are proved by integration
by parts in field space.  So just like the differential equation (\ref{onox}), the same identities
hold regardless of which integration cycle we pick.  So we have arrived at a pair of quantum
mechanical systems -- associated to quantization of $\M$ and $\M'$ -- that have the same
algebra of quantum observables, though with inequivalent representations.

Finally, we make  a few remarks about this example that are more fully explained in \cite{GW} and
are not really needed for the present paper.
Concretely, at the classical level, the only relation that the $X_i$ obey, apart from (\ref{bonkers}),
is that they commute.  Quantum mechanically, commutativity of the $X_i$ is deformed
to the $\frak{sl}(2)$ relations $[X_1,X_2]=X_3$ and cyclic permutations thereof.  (The constant $j^2$ in
(\ref{bonkers}) is also modified quantum mechanically.)   This means that the algebra of
observables is the universal enveloping algebra of $\frak{sl}(2)$ --  whether we quantize
$\M$ or $\M'$.  This gives a concrete explanation of why the two systems have the same algebra
of observables.  An important detail  is that the equivalence between the algebra of observables in quantizing
$\M$ with that in quantizing $\M'$ does not map hermitian operators to hermitian operators.
Most simply, this is because the polynomials $u(X_1,X_2,X_3)$ that are real when restricted
to $\M$ are not the same as the ones that are real when restricted to $\M'$.

In footnote \ref{borx}, we noted that in general $\M_0$ may fail to obey one condition
that we imposed on the original $\M$: there may not be an antiholomorphic involution with
$\M_0$ as a component of its fixed point set.  However, in the case of $\M'$, there is
such an involution $\tau'$, acting by $X_1,X_2,X_3\to \bar X_1,-\bar X_2,-\bar X_3$.
Hence $\M'$ actually obeys all of the conditions satisfied by $\M$.  The relation between
the two is completely symmetrical.   We may consider $\hat\M$ to arise by analytic
continuation from either $\M$ or $\M'$.

\subsection{Review Of Morse Theory}\label{morse}

To construct more interesting integration cycles, we will use Morse theory and steepest descent.  A detailed
review of the relevant ideas is given in \cite{witten}.  Here we give a brief synopsis to keep this paper self-contained.

\def\O{{\mathcal O}}
\def\CC{{\mathcal C}}
Let $Z$ be an $m$-dimensional manifold with local coordinates $w^i,\,i=1,\dots,m$ and a Morse function $h$.  A Morse function is simply a real-valued function
whose critical points are nondegenerate.  A critical point of $h$ is a point $p$ at which its derivatives all vanish.   $p$ is called a nondegenerate
critical point of $h$ if the matrix of second derivatives $\partial^2 h/\partial w^i\partial w^j$ is invertible at $p$.  If so, the number of negative
eigenvalues of this matrix is called the Morse index of $h$ at $p$; we denote it as $i_p$.

Pick a Riemannian metric $g_{ij}\d w^i\d w^j$ on $Z$.  Introducing a ``time'' coordinate $s$ (we reserve the name $t$ for the time in the original quantum
mechanics problem, to which we return later), we define the Morse theory flow equation:
\begin{equation}\label{flowq}\frac{\d w^i}{\d s}=-g^{ij}\frac{\partial h}{\partial w^j}.\end{equation}
The first property of the flow equation is that the Morse function is always decreasing along
any nonconstant flow, since
\begin{equation}\label{lowq}\frac{\d h}{\d s}=-g^{ij}\frac{\partial h}{\partial w^i}\frac{\partial h}{\partial
w^j}.\end{equation}
The right hand side is negative unless $\partial h/\partial w^i=0$, in which  case the  flow  sits at a critical point for all
$s$.  In this statement, of course, we rely on positivity of the metric $g_{ij}\d w^i\d w^j$.
The fact that $h$ decreases along the flow is the reason that the flow equation will be useful.

Let us look at the flows in the neighborhood of a critical point $p$.  After diagonalizing the matrix of second derivatives, we can find a
 system of Riemann normal
coordinates $w^i$ centered at $p$   in which $h=h_0+\sum_{i=1}^m e_i w_i^2+\O(w^3)$, $g_{ij}=\delta_{ij}+\O(w^2)$, with constants
$h_0$, $e_i$.  The flow equations
become
\begin{equation}\label{flowb}\frac{\d w^i}{\d s}=- e_i w^i,\end{equation}
with the solution
\begin{equation}\label{lowb} w^i=r^i\exp(-e_is), \end{equation}
with constants $r^i$.  A solution of the flow equation, if it does not sit identically at $p$ (that is, at $w^i=0$) for all $s$,
can only reach the point $p$ at $s=\pm\infty$.  Let us focus on solutions that flow from the critical point $p$ at $s=-\infty$.
From (\ref{lowb}), clearly, the condition for this is that $r^i$ must vanish whenever $e_i>0$.  This leaves $i_p$ undetermined parameters,
so  the solutions that
start at $p$ at $s=-\infty$ form a family of dimension $i_p$.    We define an $i_p$-dimensional subspace $\CC_p$ of $Z$ that
consists of the values at $s=0$ of solutions of the flow equations that originate at $p$ at $s=-\infty$.
The point $p$ itself lies in $\CC_p$, since it is the value at $s=0$ of the trivial flow
that lies at $p$ for all $s$.
Since $h$ is strictly decreasing along any non-constant flow, the maximum value of $h$
in $\CC_p$ is its value at $p$.

In favorable situations, the closures of the $\CC_p$ are homology cycles that generate the homology of $Z$.
As reviewed in \cite{witten}, a very favorable case is that $Z$ is a complex manifold, say of complex dimension $n$ and real dimension $m=2n$, of a type that admits many holomorphic
functions, and $h$ is the real part of a generic holomorphic function $S$.  The local form of $h$ near a nondegenerate critical
point $p$ is  $h=h_0+\mathrm{Re}\,\left(\sum_{i=1}^n z_i^2\right)$, with local complex coordinates $z_i$.  Setting $z_i=x_i+iy_i$, with $x_i$, $y_i$
real, and noting that $\mathrm{Re}\,z_i^2=x_i^2-y_i^2$, we see that stable and unstable directions for $h$ are paired.  As a result, the
Morse index $i_p$ of any such $p$ always equals $n=m/2$, and the corresponding  $\CC_p$  is middle-dimensional.  In this
situation, the $\CC_p$ are closed (but not compact) for generic\footnote{A sufficient criterion, as explained in \cite{witten}, is that there are no flows
between distinct critical points. Since $\mathrm{Im}\,S$ is conserved along the flow lines, this is the case if distinct critical points (or distinct components of the critical set, if the critical points
are not isolated) have different values of $\mathrm{Im}\,S$.
 The exceptional case with flows between critical points leads to Stokes phenomena, which were
 important in \cite{witten}, but will not be important in the present paper.} $S$ and furnish a basis of the appropriate relative
homology group, which classifies cycles on which $h$ is bounded above and goes to $-\infty$ at infinity.

Now let us take $Z=\C^n$ and consider an exponential integral of the sort described in the introduction:
\begin{equation}\label{romorf}\int_\Gamma \d z_1\,\d z_2\dots\d z_n\,\exp\left(S(z_1,\dots,z_n)\right).\end{equation}
Setting $h=\mathrm{Re}\,S$, the main problem with convergence of the integral comes from the fact that the integrand has modulus
$\exp(h)$.
Convergence is assured if $h\to-\infty$ at infinity along $\Gamma$, so the cycles $\CC_p$ just described give a basis
for the space of reasonable integration cycles.   For instance, let us consider the one-dimensional
integral that was discussed in the introduction:
\begin{equation}\label{coef}\int_\Gamma\d z\,\exp(S(z)),~~S(z)=-z^4/4+az.\end{equation}
The equation for a critical point of $h=\mathrm{Re}\,S$ is the cubic equation $\d S/\d z=0$.  This equation has three roots in the complex
$z$-plane, in accord with the fact that the space of possible integration cycles has rank three, as is evident in fig. \ref{bairy} of the introduction.

\def\N{{\mathcal N}}
In our application,  our Morse function $h$ will be the real part of a holomorphic function $S$, but its critical points will not be isolated.
The above discussion then needs some small changes.  Let $\N$  be a component of the critical point set.  $\N$ will be a complex submanifold
of $Z$, say of complex dimension $r$; we still take $Z$ to have complex dimension $n$.  In this case, there are $n-r$ complex dimensions or $2(n-r)$ real dimensions normal to $\N$.
We assume that $h$ is nondegenerate in the directions normal to $\N$, meaning that
the matrix of second derivatives of $h$ evaluated at a point on $\N$ has $2n-2r$ nonzero eigenvalues.  In that case, the local
form of $h$ is $h=h_0+\mathrm{Re}\sum_{i=1}^{n-r}z_i^2$, and the matrix of second derivatives of $h$ has precisely $n-r$
negative eigenvalues.   The space $\CC_\N$ of solutions of the flow equation that begin on $\N$ at $s=-\infty$
will have real dimension $2r+(n-r)=n+r$, where $2r$ parameters determine a point on $\N$ at which the flow begins and $n-r$ parameters
arise because the flow has $n-r$ unstable directions.   Thus, $\CC_\N$ is a cycle that is above the middle dimension.  To get a middle-dimensional cycle, we have to impose $r$ conditions, by requiring the flow to begin on  a middle-dimensional cycle $V\subset \N$.  The values at $s=0$
 of solutions
of the flow equation on the half-line $(-\infty,0]$ that begin on $V$ form a cycle  $\CC_V\subset\C^n$ that is of middle dimension.

Let us consider a simple example.  With $Z=\C^3$, we take
\begin{equation}\label{pooko}S=(z_1^2+z_2^2+z_3^2-j^2)^2.\end{equation}
Then $S$ has an isolated nondegenerate critical point at the origin.  In addition it has a family $\N$ of critical points given by $z_1^2+z_2^2+z_3^2=j^2$.
$\N$ has complex dimension two, and $h=\mathrm{Re}(S)$ is nondegenerate in the directions normal to $\N$.  $\N$ happens to be equivalent
to  the complex manifold $\hat\M$ that was introduced in eqn. (\ref{omig}), so for examples of middle-dimensional cycles in $\N$, we can take
our friends $\M$ and $\M'$, defined respectively by setting the $z_i$ to be real, or by taking $z_1$ to be real and positive while
$z_2$ and $z_3$ are imaginary.

\subsection{A New Integration Cycle For The Feynman Integral}\label{tono}

Hopefully it is clear that in attempting to describe integration cycles for the Feynman integral,
we are in the situation just described.  The exponent of the Feynman integral is a holomorphic
function $\oint \Lambda_A\d Y^A$ on the complexified free loop space $\hat \U$.  We want to take
its real part, namely
\begin{equation}\label{realp}h=\mathrm{Re}\oint\Lambda_A\d Y^A=\oint c_A\d Y^A\end{equation}
as a Morse function and
use the flow equations to generate an integration cycle on which $h$ is bounded above.

The first step is to find the critical points of $h$.  This is easily done.  We have
\begin{equation}\label{crit}\delta h =\oint \delta Y^A\d  Y^B\omega_{AB}.\end{equation}
Since $\omega$ is nondegenerate, the condition for $\delta h$ to vanish for any $\delta Y^A$ is
that $\d Y^B=0$.  In other words, a critical point is a constant map ${\mathcal T}:S^1\to \hat\M$.
This should be no surprise.  Since we have taken the Hamiltonian to vanish, Hamilton's
equations say that the coordinates and momenta are independent of time.  The space of critical
points is thus a copy of $\hat\M$, embedded in its free loop space $\hat\U$ as the space of constant maps.  Let us write $\hat\M_*$ for this copy of $\hat \M$.
As explained in section \ref{morse}, to get an integration cycle, we pick a middle-dimensional
cycle $V\subset\hat\M_*$ and consider all solutions of the flow equation on a half-line
that start at $V$.

The difference from the practice  case discussed in section \ref{morse} is that the flow equations will be two-dimensional.  The objects
that are flowing are functions $Y^A(t)$, describing a map ${\mathcal T}:S^1\to \hat\M$. To describe a flow
we have to introduce a second coordinate $s$, the flow variable, which will take values in $(-\infty,0]$, and consider functions
$Y^A(s,t)$ that describe a map ${\mathcal T}:C\to \hat\M$.  Here $C$ is the cylinder $C=S^1\times \RR_+$, where $\RR_+$ is the half-line $s\leq 0$.

The flow equations will not automatically be differential equations on $C$, but this will happen
with a convenient choice of metric  on $\hat\U$.
 We pick an ordinary metric $g_{AB}\d Y^A\d Y^B$ on $\hat\M$.  In principle, any metric
will do, but we will soon find that choosing a certain type of metric leads to a nice simplification.
We also pick a specific angular variable $t$ on $S^1$ with $\oint \d t=2\pi$.  (Until this
point, since we have taken the Hamiltonian to vanish, all formulas have been invariant under
reparametrization of the time.)  Then we define a metric on $\hat\U$ by
\begin{equation}\label{metdef}|\delta Y|^2=\oint\d t \,g_{AB}(Y(t))\delta Y^A\delta Y^B.\end{equation}

 With the metric that we have picked, the flow equation becomes
\begin{equation}\label{hormox}\frac{\partial Y^A(s,t)}{\partial s}=-g^{AB}\omega_{BC}\frac{\partial Y^A(s,t)}{\partial t}.\end{equation}
The boundary condition at $s\to-\infty$ is that $Y^A(s,t)$ approaches a limit, independent of
$t$, that lies in the subspace $V$ of the critical point set $\hat\M_*$.  There is no restriction on
what the solution does at $s=0$ (except that it must be regular, that is well-defined).
For any choice of metric $g_{AB}$, the space of  solutions of the flow equation on the cylinder $C$
with these conditions gives an integration cycle for the path integral.  If we change $g_{AB}$ a little, we get a slightly different but homologically equivalent integration cycle.

However, something nice happens if we pick $g_{AB}$ judiciously.  Let
\begin{equation}\label{orlo} I^A{}_C=g^{AB}\omega_{BC}.\end{equation}
We can think of $I$ as an endomorphism (linear transformation) of the tangent bundle of $\hat \M$.
The flow equation is
\begin{equation}\label{circo}\frac{\partial Y^A}{\partial s}=-I^A{}_B\frac{\partial Y^B}{\partial t}.\end{equation}
Now suppose we pick $g$ so that $I$ obeys
\begin{equation}\label{turlo}I^2=-1,\end{equation}
or more explicitly $\sum_B I^A{}_B I^B{}_C=-\delta^A{}_C$.  (The space of $g$'s that has this property is always nonempty and
contractible; the last statement means that there is no information of topological significance in the choice of $g$.)
 This condition means that  $I$ defines an almost
complex structure on $\hat\M$.  When that is the case, the flow equation is invariant under conformal
transformations of $w=s+it$.  Indeed, since $I$ is real-valued and obeys $I^2=-1$, it is
a direct sum of $2\times 2$ blocks of the form
\begin{equation}\label{pm}\begin{pmatrix} 0 & -1\\ 1 & 0 \end{pmatrix}.\end{equation}
In each such $2\times 2$ block, the flow equations look like
\begin{equation}\label{trock}\frac{\partial u}{\partial s}=\frac{\partial v}{\partial t},~~\frac{\partial v}{\partial s}=-\frac{\partial u}{\partial t}.\end{equation}
These are Cauchy-Riemann equations saying that $(\partial_s+i\partial_t)(u+iv)=0$ or in
other words that $u+iv$ is a holomorphic function of $w=s+it$.  Their invariance under
conformal mappings is familiar.

Actually, to literally interpret the flow equation as saying that the map ${\mathcal T}:C\to \hat\M$ is
holomorphic, we need $I$ to be an integrable complex structure.  If $I$ is a nonintegrable
almost complex structure, then (\ref{circo}) is known as the equation for a pseudoholomorphic
map (or an $I$-pseudoholomorphic map if one wishes to be more precise).  This is a well-behaved, elliptic, and conformally invariant equation whether $I$ is integrable or not.  Thus, as soon as $I^2=-1$, the flow
equations are invariant under conformal mappings of $w$.

There is in fact a very convenient
conformal mapping: we set $z=\exp(w)$, mapping the cylinder $C$ to the punctured
unit disc $0<|z|\leq 1$.  Generically, a map from the punctured unit disc to $\hat\M$ would
not extend continuously over the point $z=0$.  In this case, however, the boundary
condition that $Y^A(s,t)$ is a constant independent of $t$ for $s\to-\infty$ precisely means
that $Y^A$, when regarded as a function of $z$, does have a continuous extension across
$z=0$.  Moreover, this extended map is still pseudoholomorphic, by the removeable singularities theorem
for pseudoholomorphic maps.

Thus, we arrive at a convenient description of an integration cycle $\CC_V$ for the Feynman integral of quantum mechanics. $\CC_V$ consists of the boundary values of $I$-pseudoholomorphic  maps
${\mathcal T}:D\to\hat\M$, where $D$ is the unit disc $|z|\leq 1$, and ${\mathcal T}$ maps the point $z=0$ to the prescribed subspace $V\subset\hat\M$.

\subsection{$I$ And The $A$-Model}\label{whati}

What sort of complex or almost complex structure is $I$?  $\hat \M$ is
by definition a complex manifold; it was introduced as a complexification of $\M$. The defining
conditions on $\hat\M$ were that it is a complex  symplectic manifold, with a complex structure that
was previously unnamed and which we will now call $J$, and with a holomorphic two-form
$\Omega$ whose imaginary part, when restricted to the original classical phase space
$\M\subset\hat\M$, coincides with the
original symplectic form $f$ of $\M$.  (There were some additional conditions that we do not
need right now.)  For example, in the familiar case $\M=S^2$, $J $ is the complex structure in
which the coordinates $X_1,X_2,X_3$ of eqn. (\ref{omig}) are holomorphic.

Since we already know about one complex structure on $\hat\M$, namely $J$, one might wonder
if we can pick the metric $g$ so that $I=J$.   This is actually not possible.  Since $\omega=
{\mathrm {Re}}\,\Omega$, where $\Omega$ is of type $(2,0)$ with respect to $J$, it follows
that $\omega$ is of type $(2,0)\oplus (0,2)$ with respect to $J$.  Therefore, for $g^{AB}\omega_{BC}$ to coincide with $J^A{}_C$, $g_{AB}$ would also have to be of type
$(2,0)\oplus (0,2)$ with respect to $J$.  But this would contradict the fact that $g$ is supposed to be a positive-definite Riemannian metric.  (See the discussion of
eqn. (\ref{lowq}), where this positivity was a key ingredient in the Morse theory construction
of appropriate integration cycles.)

So $I$ will have to be something new, that is not something that was introduced along
with $\hat\M$.  On the other hand, the conditions obeyed by $g_{AB}$ and $I^A{}_B$
are famlliar in one branch of two-dimensional quantum field theory.  This is the $A$-model,
in fact in the present case the $A$-model obtained by twisting a two-dimensional sigma-model in which the
target space is $\hat\M$ and the symplectic structure is $\omega$.  We have already encountered
this $A$-model in a naive way in section \ref{simplint}, and it will now enter our story in a more interesting fashion.

 In defining the
$A$-model of a symplectic manifold  $X$ -- such as  $X=\hat\M$ -- with a given symplectic
structure $\omega$, one introduces an almost complex structure $I$ on $X$ with respect to which
$\omega$ is of type $(1,1)$ and positive.  Positivity means that the metric $g$ defined by
$I^A{}_B=g^{AB}\omega_{BC}$ is in fact a positive-definite Riemannian metric on $X$.  The nicest
case is that one can choose $I$ to be an integrable complex structure.  In that case, the metric
$g$ is Kahler.   In general, one cannot pick $I$ to be integrable and one can define the $A$-model
for any almost complex structure $I$ such that $\omega$ is of type $(1,1)$ and positive.

Indeed, to make sense of the $A$-model, one only needs the equation
for an $I$-pseudoholomorphic map.  (The non-integrable case was important in the early
mathematical applications of the $A$-model \cite{gromov,floer} as well as many more recent
ones and was described from a quantum
field theory perspective in \cite{topsigma}.)
In general, $A$-model computations are localized on such maps.
Usually, one encounters finite-dimensional families of $I$-pseudoholomorphic maps.  In the $A$-model with target $X$ on a Riemann surface $\Sigma$ without boundary, one encounters the
moduli
spaces of $I$-pseudoholomorphic maps ${\mathcal T}:\Sigma\to X$; these  are finite-dimensional.
If $\Sigma$ has a boundary, we usually consider boundary conditions associated with Lagrangian $A$-branes, and in this case the moduli spaces of $I$-pseudoholomorphic maps are again
finite-dimensional.  What may be unfamiliar about the present problem from the point of view
of the $A$-model is that our integration cycle $\CC_V$ is actually an infinite-dimensional
space of $I$-pseudoholomorphic maps.  The relation of this cycle to the $A$-model is
explained more fully in sections \ref{zondo} and \ref{loose}.

In the meanwhile, let us give a  concrete example of what $I$ can be, given that it cannot
coincide with the complex structure $J$ by which $\hat\M$ was defined.  We return first
to the familiar example in which $\hat\M$ is defined, in complex structure $J$, by the
equation $X_1^2+X_2^2+X_3^2=j^2$.  In fact, this complex manifold admits
a complete hyper-Kahler metric, the Eguchi-Hansen metric.  The original complex
structure $J$ and the holomorphic two-form $\Omega$ are part of the hyper-Kahler structure of
$\hat\M$.  Indeed, a hyper-Kahler manifold has a triple of complex structure $I,J,K$
obeying the quaternion relations $I^2=J^2=K^2=IJK=-1$.  It also has a triple of real symplectic
structures $\omega_I,\omega_J,\omega_K$, where $\omega_I$ is of type (1,1) and positive
with respect
to $I$, and similarly for $\omega_J$ and $\omega_K$.  Finally, $\Omega_I=\omega_J+i\omega_K$
is a holomorphic two-form with respect to $I$, and similarly with cyclic permutations of indices $I,J,K$.  Now, in our example, define $\Omega$ by eqn. (\ref{gwos}) and normalize the hyper-Kahler
structure of $\hat\M$ so that $\Omega=\omega_I-i\omega_K$ (this is $-i(\omega_K+i\omega_I)$,
so it is holomorphic with respect to $J$).  So $\omega=\mathrm{Re}\,\Omega$ is equal
to $\omega_I$, and is of type $(1,1)$ and positive with respect to $I$.  In other words, in this
example, we can take the metric $g$ on $\hat\M$ to be the Eguchi-Hansen hyper-Kahler metric, and $I$
to be one of the complex structures for which that metric is Kahler.  The other real
symplectic structure of $\hat \M$ in  our original description is $f=\mathrm{Im}\,\Omega=-\omega_K$.

We have used no special property of $\hat\M$ except that its complex symplectic
structure $J,\Omega$ extends to a hyper-Kahler structure.  Whenever this is so,
the corresponding hyper-Kahler metric on $\hat\M$ is a very convenient choice.  (It can happen
that the extension of $J,\Omega$ to a hyper-Kahler structure is not unique; varying it in a continuous
fashion will give a family of equivalent and convenient integration cycles.)   Since $\hat\M$ is complex symplectic, its real dimension is
always divisible by four, but it may not admit a hyper-Kahler structure that extends its complex symplectic structure.
 While we cannot necessarily pick $I$ to be integrable, we can always
pick it so that $IJ=-JI$.  In this case, defining $K=IJ$ and $\omega_J=Jg$, we arrive at what one might call an almost hyper-Kahler structure.
The three almost complex structures $I,J,K$ and the three  two-forms $\omega_I,\omega_J,\omega_K$ obey all the usual algebraic relations.
$J$ is integrable and $\omega_I$ and $\omega_K$ are closed; $I$ and $K$ may not be integrable and $\omega_J$ may not be closed.

\subsection{Interpretation In Sigma-Model Language}\label{zondo}

At this point, we are supposed to do a Feynman integral
\begin{equation}\label{omitop}\int_{\CC_V }DY^A(t) \exp\left(\oint \Lambda_A\d Y^A\right)\prod_\alpha u_\alpha(t_\alpha) \end{equation}
where $Y^A(t)$ is a one-dimensional field, but the integration cycle $\CC_V$ is described in two-dimensional
terms, in terms of boundary values of $I$-pseudoholomorphic maps.

This is a hybrid-sounding recipe.  A natural idea is to try to reformulate (\ref{omitop}) as a two-dimensional path integral,
with a field $Y^A(s,t)$ that describes a map from $D$ to $\hat\M$, and certain additional fields that we will need to introduce along the
way.    The first step is obvious -- after extending $Y$ to a function defined on $D$, we introduce another field $T$ that
will be a Lagrange multiplier enforcing the desired equation (\ref{circo}) of an $I$-pseudoholomorphic map.

\def\sign{{\mathrm{sign}}}
We can write (\ref{circo}) in the
form  $U^A=0$ where
\begin{equation}\label{polygo} U^A=\d Y^A+\star I^A{}_B\d Y^B;\end{equation} here $\star$ is the Hodge star operator on $D$ (defined so that $\star\d s=\d t,~
\star\d t=-\d s$).
We view $U$ as a one-form on $D$ with values in the pullback of the tangent bundle of $\hat\M$.  $U$ obeys the identity
\begin{equation}\label{starx} U^A=\star  I^A{}_B U^B.\end{equation}
We introduce a Lagrange multiplier field $T_A$ that is a one-form on $D$ with values in the pullback of the cotangent bundle of $\hat\M$,
and further obeys the dual relation
\begin{equation}\label{lygo} T_B=\star T_A I^A{}_B.\end{equation}
Then if we include a term in the two-dimensional action of the form $\int_D T_A\wedge U^A$, the integral over $T_A$ will
give a delta function setting $U^A=0$.

So a rough approximation to the two-dimensional path integral that we want would be
\begin{equation}\label{zolygo} \int D T_A(s,t)\,DY^B(s,t) \,\exp\left(i\int_D T_A\wedge U^A\right)\,\,\left(\cdots\right),\end{equation}
where the ellipses refer to the integrand in (\ref{omitop}).

\def\cF{{\cmmib F}}
After integrating out $T$, this expression will lead to an integral over $Y$ that is supported on the cycle $\CC_V$, but it is not   the integral we want.
In fact, the integral (\ref{zolygo}) will depend on the details of the almost complex structure $I$ that is used in defining $U^A$.
The problem is that the integral over $T_A(s,t)$ indeed generates a delta function setting $U^A(s,t)$ equal to zero,
but this delta function multiplies an unwanted determinant $1/|\det (\delta U/\delta Y)|$.  (An analog of the appearance of this determinant in the case of an ordinary
integral is that, if $f(x)$ is a function that vanishes precisely at $x=a$, then the integral $\int \d\lambda/2\pi\, \exp(i\lambda f(x))$
equals not $\delta(x-a)$ but $\delta(x-a)/|f'(a)|$.)  To cancel this determinant, we add fermions with a kinetic energy that is precisely
the linearization of the equation $U=0$.  As a result, the fermion determinant will cancel the boson determinant up to sign.
In the present problem, the sign will not do anything essential; this is because $\CC_V$ is connected, and we can pick the sign of the fermion
measure so that the sign is $+1$.  In a more general $A$-model problem, fermion and boson determinants cancel only up to sign
and contributions of $I$-pseudoholomorphic curves are weighted by the sign of the fermion determinant (the boson determinant is
always positive).

\def\D{{\mathcal D}}
The fermions will also carry a new $U(1)$ quantum number (``fermion number'') that we call $\cF$.  We need fermions $\psi^A$ of
$\cF=1$ that take values in the pullback to $D$ of the tangent bundle of $\hat\M$.  And we need fermions $\chi_A$ of $\cF=-1$ that have
the same quantum numbers as the bosons $T_A$: they are a one-form on $D$ with values in the pullback of the cotangent bundle
of $\hat\M$, and they obey a constraint $\chi_B=\star\chi_A I^A{}_B$.  We take the fermion action to be  $i\int_D  \chi_A \D\psi^A$,
where the operator $\D$ is defined as the linearization of the equation $U^A=0$.  This means that if we vary $Y^A$ around a solution
of $U^A=0$, we have to first order $\delta U^A=\D\delta Y^A$.  Concretely, if $\hat \M$ is Kahler (so that $I$ is covariantly constant, which
makes the formulas look more familiar),
then
\begin{equation}\label{orxo} \D \psi^A=\frac{D\psi^A}{Ds}+I^A{}_B\frac{D\psi^B}{Dt},\end{equation}
where $D/Ds$ and $D/Dt$ are defined using the pullback to $D$ of the Riemannian connection on the tangent bundle to
$\hat\M$.    We now consider a two-dimensional path integral
\begin{equation}\label{zing}\int DT\,DY\,D\chi\,D\psi \exp\left(i\int_D(T_A\wedge U^A-\chi_A\wedge \D\psi^A)\right)\,\,
\exp\left(\oint \Lambda_A\d Y^A\right)\prod_\alpha u_\alpha(t_\alpha)\,\,\O_V(0).\end{equation}
Here we have included explicitly all the factors from the original path integral (\ref{omitop}).  We have also included an operator
$\O_V(0)$ -- the details of which we will describe presently -- that is supposed to incorporate the constraint that $Y^A(z)$ lies in $V$
at the point $z=0$.

The two-dimensional action that we have arrived at has a fermionic symmetry that squares to zero.  It is invariant under
\begin{equation}\label{ponzo}\delta Y^A=\psi^A,~~\delta\psi^A=0,\end{equation}
together with, roughly speaking,\footnote{\label{zork} We are engaging here in a small sleight of hand and omitting terms of higher
order in fermions.  As $\chi$ is a section of the pullback of the cotangent
bundle of $\hat\M$, to keep it ``constant'' when $Y$ is varied, we must transport it using some connection; we will use the Riemannian
connection.  So we measure $\delta\chi$ relative to parallel transport by the Riemannian connection.  With this understanding, we should
write  the first part of  eqn. (\ref{onzo}) in the form $\delta\chi_A-\Gamma^C_{AB}\delta Y^B\chi_C =T_A$, where $\Gamma^C_{AB}$ are the Christoffel symbols.
Since $\delta Y^B=\psi^B$, we may write this as $\delta\chi_A-\Gamma^C_{AB}\psi^B\chi_C=T_A$.  To ensure that $\delta^2=0$, we
must take $\delta T_A=-\delta(\Gamma^C_{AB}\psi^B\chi_C)$.  This is equivalent to $\delta T_A-\Gamma^C_{AB}\delta Y^B T_C=-(1/2) R^C_{ADB}\psi^D\psi^B\chi_C$, where $R^C_{ADB}$ is the Riemann tensor.  In deriving eqn. (\ref{torque}) below, the terms proportional
to $\Gamma$ cancel because the connection is metric-compatible, leaving a four-fermi term proportional to $R$.}
\begin{equation}\label{onzo}\delta \chi_A=T_A,~~\delta T_A=0.\end{equation}
Clearly $\delta^2=0$.
The exponent in (\ref{zing}) is
\begin{equation}\label{king}i\int_D\left(T_A\wedge U^A-\chi_A\wedge \D\psi^A\right)=\delta\int_Di\chi_A\wedge U^A,\end{equation}
which  makes clear its invariance under $\delta$.  This also makes it clear that the path integral (\ref{zing}) is invariant under deformations
of the  almost complex structure $I$ (or equivalently  of the metric $g$).  Indeed, $I$ appears only in terms in the action of the form
$\delta(\cdots)$; varying such a term does not change the value of the path integral.

Without changing anything essential, we can add to the action a further exact term
\begin{equation}\label{torque}\delta\left(\frac{\epsilon}{2}\int_D g^{AB}\chi_A\star T_B\right)=\frac{\epsilon}{2}\int_D g^{AB}T_A\wedge\star T_B
+\frac{\epsilon}{4}\int_D R^{CA}_{DB}\chi_C\wedge\chi_A\psi^D\psi^B,\end{equation}
with an arbitrary parameter $\epsilon$.  (The origin of the four-fermi term is explained in footnote \ref{zork}.) After performing the Gaussian integral over $T$, we get an equivalent path integral for the other
fields
\begin{align}\label{pling}\int DY\,D\chi\,D\psi~ & \exp\left(-\frac{1}{2\epsilon}\int_D g_{AB}U^A\wedge\star U^B-i\int_D\chi_A \D\psi^A-\frac{\epsilon}{4}\int_D R^{CA}_{DB}\chi_C\wedge\chi_A\psi^D\psi^B\right)\cr &
\exp\left(\oint \Lambda_A\d Y^A\right)\prod_\alpha u_\alpha(t_\alpha)\,\,\O_V(0).\end{align}

\def\V{{\mathcal V}}
\def\FS{{\mathcal S}}
In fact, the fermionic symmetry  that was introduced in eqns. (\ref{ponzo}), (\ref{onzo}) is the usual topological supersymmetry of the $A$-model with target $\hat\M$. The symmetry
generated by $\delta$ is usually denoted as $Q$ (and called the BRST operator or topological supercharge).  The action is a standard $A$-model action, as we discuss in section
\ref{loose}. When the model is interpreted as an $A$-model,
the operator $\O_V(0)$ that imposes the constraint that the point $z=0$ is mapped to $V$ is a conventional closed string observable
of the $A$-model.  This comes about in a standard way, which we sketch for completeness.  Operators
$\FS(Y(z_0),\psi(z_0))$ that depend on $Y$ and $\psi$ only (evaluated at some point $z=z_0$) are naturally associated to differential
forms on $\hat\M$. Here one simply thinks of $\psi^A$ as the one-form $\d Y^A$.  An arbitrary such $\FS$ is a finite linear combination of expressions of the form $\FS_{A_1A_2\dots A_k}(Y)
\psi^{A_1}\psi^{A_2}\dots\psi^{A_k}$; we associate such an expression to the differential form
$\FS_{A_1A_2\dots A_k}(Y)
\d Y^{A_1}\d Y^{A_2}\dots\d Y^{A_k}$.  The relations
 $[Q,Y^A]=\psi^A$, $\{Q,\psi^A\}=0$ imply that $Q$ acts on differential forms as the exterior derivative $\d$.  To be more
precise, if $\gamma$ is a differential form on $Y$ and $\FS_\gamma(Y,\psi)$ is the corresponding quantum field operator, then $[Q,\FS_\gamma]
=\FS_{\d\gamma}$.  So $Q$-invariant local operators of the $A$-model of this type correspond to closed differential forms on
$\hat\M$. (Similarly, cohomology classes of $Q$ acting on local operators of this type correspond to the de Rham cohomology of $\hat\M$.)   Now, given any submanifold $V\subset \hat\M$, let $\V$ be a differential form (of degree equal to the codimension of $V$)
that is Poincar\'e dual to $V$.  (This notion is explained in eqn. (\ref{palooka}).) Then $\V$ is a closed differential form with delta function support on $V$, and the operator
$\O_V$ that we need in (\ref{pling}) is simply $\FS_\V$.  So it is indeed a standard $Q$-invariant local operator of the $A$-model.

\subsection{The Physical Model}\label{loose}

Let us explicitly evaluate the bosonic kinetic energy term $K=\frac{1}{2\epsilon}\int_D g_{AB}\,U^A\star U^B$  in (\ref{pling}), using the definition (\ref{polygo}) of $U$.  We find
\begin{equation}\label{forz}K= \frac{1}{\epsilon}\int_D\d s\,\d t \,g_{AB}\left(\frac{\d Y^A}{\d s}\frac
{\d Y^B}{\d s}
+\frac{\d Y^A}{\d t}\frac{\d Y^B}{\d t}\right)+\frac{2}{\epsilon}\int \d s\,\d t\,\omega_{AB}\frac{\d Y^A}{\d s}\frac{\d Y^B}{\d t}.\end{equation}
With $\omega_{AB}=\partial_A c_B-\partial_B c_A$, the last term in (\ref{forz}) is
\begin{equation}\label{orz} \frac{2}{\epsilon}\int \d s\,\d t\,\omega_{AB}\frac{\d Y^A}{\d s}\frac{\d Y^B}{\d t}=\frac{2}{\epsilon}\int_D\d s \,\d t \left(\frac{\partial}{\partial s}\left(c_B\frac{\partial Y^B}{\partial t}\right)-\frac{\partial}{\partial t}\left(c_B\frac{\partial Y^B}{\partial s}\right)\right)
=\frac{2}{\epsilon}\oint c_B\,\d Y^B=\frac{2h}{\epsilon }.\end{equation}

Now let us examine the purely bosonic factors in the integrand of the path integral (\ref{pling}), ignoring the insertions of operators $u_i(t_i)$ and
$\O_V(0)$ (which will not affect the convergence of the path integral). Those factors are
\begin{equation}\label{malfond}  \exp(-K)\exp\left(\oint \Lambda_A\d Y^A\right)=\exp(-K)\exp\left(\oint (c_A+ib_A)\d Y^A\right)=
\exp\left(-K+h+i\oint b_A\d Y^A\right).\end{equation}
 The terms proportional to $h$ in the exponent cancel out if we eliminate $K$ using eqns. (\ref{forz}), (\ref{orz}) and also set $\epsilon=2$.
Then the  bosonic factors become
\begin{equation}\label{zorny} \exp\left( -\frac{1}{2}\int_D\d s\,\d t \,g_{AB}\left(\frac{\d Y^A}{\d s}\frac
{\d Y^B}{\d s}
+\frac{\d Y^A}{\d t}\frac{\d Y^B}{\d t}\right)+i\oint_{\partial D} b_A\d Y^A  \right).\end{equation}

At this particular value of $\epsilon$, our path integral is essentially that of an ordinary supersymmetric nonlinear sigma-model with a particular
boundary condition that is compatible with unitarity.  The bulk integral in the exponent in (\ref{zorny}) is the ordinary bulk bosonic
action of a sigma-model:
\begin{equation}\label{irz} \frac{1}{2}\int_D\d s\,\d t \,g_{AB}\left(\frac{\d Y^A}{\d s}\frac
{\d Y^B}{\d s}+\frac{\d Y^A}{\d t}\frac
{\d Y^B}{\d t}\right).\end{equation}
The boundary contribution to the exponent is imaginary precisely at this value of $\epsilon$, where it reduces to
\begin{equation}\label{arz} i\oint_{\partial D} b_A\d Y^A \end{equation}
Being imaginary means that this contribution can be interpreted as the coupling of the sigma-model to a rank one {\it unitary}
Chan-Paton bundle $\L\to \hat\M$.  The connection on this bundle is $b$ and its curvature is $f=\d b$.
As for the fermion kinetic energy $-i\int \chi_A\wedge \D\psi^A$, it is not the standard fermion kinetic energy of the usual
supersymmetric sigma-model,
but it is the standard fermion kinetic energy in the $A$-twisted version of this model.  Finally, the four-fermion coupling in (\ref{pling}) is a standard part of the supersymmetric nonlinear sigma-model, again written in $A$-twisted notation.

In short, at this value of $\epsilon$, it must be possible to interpret the boundary interaction that we have at $s=0$ as a supersymmetric
brane in the usual supersymmetric sigma-model with target $\hat\M$. (Twisting does not affect the classification of branes, because on
a flat worldsheet, the twisting is only a matter of notation; near the boundary of a Riemann surface $\Sigma$, we can always consider $\Sigma$
 to be flat.)  More specifically, what we have at $s=0$  is a brane in the usual sigma-model that preserves $A$-type supersymmetry.

  Which brane is it?  The support of the brane is all of $\hat\M$
(since the bosonic fields are locally allowed to take any values at $s=0$) and the curvature of its Chan-Paton bundle is $f$.
These properties uniquely identify this brane: it is the most simple coisotropic $A$-brane constructed in the original paper on that
subject by Kapustin and Orlov \cite{KO}.
Their presentation  contains more or less the same ingredients as in our derivation, but arranged quite differently.
Their starting point   was the $A$-model of a  symplectic manifold $X$ with symplectic structure $\omega$.   To define the
$A$-model, an almost complex structure $I$ is chosen such that $\omega$ is of type $(1,1)$ and positive.   The question asked was then what branes of rank 1 are possible in this $A$-model.  Such a brane is characterized by its support and by the curvature
of the Chan-Paton line bundle that it carries.  For brevity, let us state the answer in \cite{KO} only for the case of a rank one brane $\B$
whose support is all of $X$.  The answer turned out to be that the Chan-Paton curvature $f$ of such a brane must have the property that
$J=\omega^{-1}f$ is an integrable complex structure on $X$ (which will necessarily be different from $I$, which may or may not be integrable).
This was  quite a surprising answer at the time; previously the only known branes of the $A$-model were Lagrangian $A$-branes, supported
on a middle-dimensional submanifold of $X$.

\def\B{{\mathcal B}}
Saying that $J=\omega^{-1}f$ is equivalent to saying that $\Omega=\omega+if$ is a holomorphic two-form with respect to $J$.
So whenever we study the $A$-model of a symplectic manifold $X$ -- such as $X=\hat\M$ -- endowed with a symplectic form $\omega$
that is the real part of a holomorphic two-form $\Omega$, in some complex structure $J$, there is a canonical way to obey
the Kapustin-Orlov conditions.
We can define a rank one $A$-brane whose support is all of $X$ by taking the Chan-Paton curvature to be $f=\mathrm{Im}\,\Omega$.
Since this is the simplest way to satisfy the relevant  conditions, and also the brane constructed this way seems to arise in many applications,
this brane has been called the\footnote{Calling this brane ``the'' canonical coisotropic $A$-brane is perhaps a little too
cavalier, since it depends on the choice of $J$.  Depending on the context, a distinguished $J$ may or may not present itself.}
 canonical coisotropic $A$-brane and denoted $\B_{cc}$.
In our formulation in the present paper, we have arrived at the same structure from a different end.  We started with a classical symplectic
manifold $\M$ with symplectic form $f$.  Seeking to analytically continue the Feynman integral that arises in quantizing $\M$,
we replaced $\M$ by a complexification $\hat\M$ and analytically continued from $f$ to $\Omega=\omega+if$.   Then we found
an integration cycle in the complexified Feynman integral that has a natural interpretation via a path integral
on the unit disc $D$ in the complex $z$-plane.  From the standpoint of \cite{KO}, what we have constructed is the path integral of the $A$-model on $D$
with a boundary condition set by the $A$-brane $\B_{cc}$, with boundary insertions of open string vertex operators $u_i(t_i)$, and with
a closed string $A$-model vertex operator $\O_V$ inserted at $z=0$.  The closed string insertion is needed since otherwise,
because of an anomaly in the fermionic quantum number $\cF$, the path integral on the disc would vanish.  As for the open string
vertex operators $u_\alpha(t_\alpha)$ that are inserted on the boundary of the disc, it is a result of \cite{KO} that the $(\B_{cc},\B_{cc})$ strings correspond to holomorphic
functions in complex structure $J$.

The reader may want to compare this discussion to the analysis in \cite{FLN,FLN2,FLN3} of localization of sigma-model path integrals on a middle-dimensional subspace of the loop space of a target manifold.  (The target space was not assumed to be a complex
symplectic manifold, so the middle-dimensional cycles were not interpreted as integration cycles.)   The approach was the reverse
of what we have explained here -- and more like what we will explain in section \ref{reverse} -- in the sense that the starting point
was taken to be a two-dimensional supersymmetric
sigma-model, rather than the problem of finding an integration cycle for a path integral in dimension one.

\subsubsection{The Boundary Condition And Localization}\label{zumu}

Here we will add a few remarks on the boundary condition obeyed by the field $Y^A$ at the boundary of the punctured disc $D$.  (We reconsider the boundary conditions
in a related problem and include the fermions  in section \ref{sigmaone}.)  The purpose
is to clarify the meaning of $A$-model localization in the presence of coisotropic branes.

We read off from eqn. (\ref{malfond}) that the bosonic part of the integrand of the path integral is $\exp(-\hat K)$, where the
``action'' $\hat K$, with boundary contributions included, is
\begin{equation}\label{ubonkers}\hat K= K - h -i \oint b_A\d Y^A.\end{equation}
When we vary $\hat K$ with respect to $Y^A$, we get \begin{align}\label{onkers}\delta\hat K=&-\frac{2}{\epsilon}\int_D\d s\,\d t\,\delta Y^A g_{AB}\left(\frac{D^2Y^B}{Ds^2 } +\frac{D^2Y^B}{Dt^2}\right)
 \\ \notag &+\oint_{\partial D}\d t\,\delta Y^A\left(\frac{2}{\epsilon}\left(g_{AB}\frac{\d Y^B}{\d s}+\omega_{AB}\frac{\d Y^B}{\d t}\right)
-(\omega_{AB}+if_{AB})\frac{\d Y^B}{\d t}\right).\end{align}
Setting this to zero, we find that the bulk equation of motion is the equation $D^2Y^B/Ds^2+D^2Y^B/Dt^2=0$ for a harmonic map
(this equation is satisfied for $I$-pseudoholomorphic maps).  Moreover, as we wish to  place no local restriction on the boundary values of
$\delta Y^A$, to set the boundary term to zero, the boundary condition on $Y^A$ must be
\begin{equation}\label{omonkey}\frac{2}{\epsilon}\left(g_{AB}\frac{\d Y^B}{\d s}+\omega_{AB}\frac{\d Y^B}{\d t}\right)
-(\omega_{AB}+if_{AB})\frac{\d Y^B}{\d t}=0.\end{equation}

If we set $\epsilon=2$, the boundary condition becomes
\begin{equation}\label{pmonkey}  g_{AB}\frac{\d Y^B}{\d s}-if_{AB}\frac{\d Y^B}{\d t}=0,\end{equation}
which is the boundary condition\footnote{The meaning of this boundary condition in quantum theory is a little subtle, because of the
factor of $i$ multiplying the second term, while classically the field $Y^B$ is real.  This subtlety is a standard phenomenon in Euclidean field theory
and has nothing to do with issues considered in the present paper.  If we replace $D$ by a two-manifold of Lorentz signature,
the factor of $i$ disappears and the interpretation of the boundary condition becomes straightforward.}
 of the physical sigma-model with a coisotropic brane as presented in \cite{KO}.  On the other hand,
 to get the simplest topological
 field theory description,
we take $\epsilon\to 0$.  Precisely in the limit $\epsilon=0$, the boundary condition merely says that the equation
for a pseudoholomorphic map should be obeyed on $\partial D$.  This is not really a boundary condition at all,
since at $\epsilon=0$, the equation for a pseudoholomorphic map is obeyed everywhere.

The best way to understand what happens precisely at $\epsilon=0$ is  to go back to the form (\ref{zing}) of the path integral with
an auxiliary field $T$.  When we repeat the above exercise, assuming no constraint on the boundary values of $\delta Y^A$ or $\delta T_A$,
we find a boundary condition for $T$ but none for $Y$.  (This  happens because the boundary contribution to the
variation of the action has a $T\delta Y$ term
but no $Y\delta T$ term.)  So in this form of the theory, $Y$ obeys no boundary condition at all.  The only condition on $Y$ is the equation for an $I$-pseudoholomorphic map, which is enforced when we do the $T$ integral.

The question ``on what class of $I$-pseudoholomorphic maps does the $A$-model localize in the presence of a coisotropic brane?'' does not
seem to have been  answered in the literature.  The answer for the case that the
coisotropic brane is the canonical one considered here is that localization occurs in and only in the limit $\epsilon\to 0$ and the localization is on the infinite-dimensional space of all $I$-pseudoholomorphic maps.
 For any $\epsilon\not=0$, we can do a path integral that gives the same results for all $A$-model observables, but it
arrives at these results in a more complex way, not by a simple localization on a space of
$I$-pseudoholomorphic maps.

Another way to describe the localization that occurs at $\epsilon\to 0$ is that the localization
is on quantum mechanics -- the two-dimensional $A$-model path integral localizes on
a new integration cycle for a path integral of ordinary quantum mechanics.

\subsection{Recovering The Hilbert Space}\label{oose}

We have defined an integration cycle by flowing in $s$ over a semi-infinite interval $(-\infty,0]$.  An obvious question is to ask
what we would get if we flow in $s$ for a finite interval only, say the interval $[-s_0,0]$.  In other words, what happens
if we replace the semi-infinite cylinder $C=S^1\times \RR_+$ with a compact cylinder $C_{s_0}=S^1\times [-s_0,0]$?

We now need a different type of boundary condition at $s=s_0$.  We cannot start the flow from a critical point, since as
we observed in section \ref{morse}, a nonconstant flow  can only start from or leave a critical point at $s=\pm\infty$.

Let us approach the problem from the standpoint of the $A$-model.
Since $C_{s_0}$ has a second boundary component at $s=-s_0$, we need a boundary condition there.  This boundary condition
will have to correspond to a second $A$-brane.  Though we could consider the case that the second brane is a coisotropic one
(like the canonical coisotropic brane $\B_{cc}$ that we will continue to use at  $s=0$), let us consider instead the case that
the second brane is a Lagrangian $A$-brane -- supported on a submanifold $\LL\subset \hat\M$ that is Lagrangian with respect to $\omega$.
We denote the second brane as $\B_\LL$.

The boundary condition associated to $\B_\LL$ requires the boundary values of the map ${\mathcal T}:C_{s_0}\to \hat\M$ at $s=-s_0$ to lie in $\LL$.
Those boundary values define an arbitrary\footnote{This is a slight simplification.  The boundary values at $s=-s_0$ must be initial
values of a solution of the flow equation that is regular at least up to $s=0$.} point in the free loop space of $\LL$.
Then we solve the flow equations for a ``time'' $s=s_0$.  The resulting integration cycle $\Gamma_{s_0}$ in the free loop space of $X$ consists of
all possible boundary values at $s=0$ of the solution of the flow equation.  As $s_0$ is varied, we get a  one-parameter family of integration
cycles.  The value of $s_0$ does not matter, since in general the integral of a holomorphic differential form of top dimension
-- in this case the integration form of the Feynman integral -- over a middle-dimensional cycle is invariant under continuous deformations
of that cycle.  (Alternatively, the value of $s_0$ does not matter, since the metric of $C_{s_0}$ is immaterial in the $A$-model.)

We can try to evaluate the integral in the limit $s_0\to 0$.  The limit is particularly simple if $\LL$, while Lagrangian for $\omega=\mathrm{Re}\,
\Omega$, is symplectic from the standpoint of $f=\mathrm{Im}\,\Omega$.  (If $\hat\M$ is the familiar example given by $X_1^2+X_2^2+X_3^2=j^2$,
then $\M$ and $\M'$ as described in section \ref{simplint} are possible choices of $\LL$.)  In the limit that $s_0\to 0$, the flow equation does not
do anything (since no time is available for the flow) and the limit of $\Gamma_{s_0}$ is simply the free loop space of $\LL$.  The integral over this free loop space is simply
the original Feynman integral (\ref{uzomp}), with $\LL$ replacing $\M$.  This path integral is associated with quantization of $\LL$.
So the result is that the $A$-model path integral for strings stretched between the brane $\B_{cc}$
and a Lagrangian brane $\B_\LL$ (with $f$ nondegenerate on $\LL$)  is associated with quantization of $\LL$ (with respect to symplectic
structure $f$).

The corresponding statement from a Hamiltonian
point of view is that the space of $(\B_{cc},\B_\LL)$ strings is the  Hilbert space that arises in quantization of $\LL$.
As is usual, the path integral on the cylinder can be interpreted in terms of a trace in this Hilbert space.
If rather than a trace, we wish to consider specific initial and final quantum states,
we can consider a path integral on a  disc $D$ with its boundary $\partial D$ divided into an interval $\partial D_1$
 labeled by $\B_{cc}$ and a second interval $\partial D_2$ labeled
by $\B_\LL$. (The associated $A$-model path integral will localize on $I$-pseudoholomorphic maps that send $\partial D_2$ to $\LL$
and whose boundary values on  $\partial D_1$ are unconstrained. The space of such maps is infinite-dimensional.)
 At the two points where $\partial D_1$ and $\partial D_2$ meet, we must insert $(\B_{cc},\B_\LL)$
  and $(\B_{\LL},\B_{cc})$ vertex operators, which correspond to
initial and final quantum states in the quantization of $\LL$.  Insertions of $(\B_{cc},\B_{cc})$ strings on $\partial D_1$ will give matrix elements of quantum observables between initial and final quantum states.

None of these assertions really require the analysis in the present paper, and indeed fuller and more direct explanations have been given in \cite{GW}, following a variety of earlier clues and examples \cite{BrS,Ka,KW,gualt,AZ}.
In our brief and somewhat cavalier explanation here, we have omitted some key details (involving the conditions for the space
of $(\B_{cc},\B_\LL)$ strings to have a hermitian structure, the role of the flat
Chan-Paton line bundle of $\B_\LL$, etc.) that can be found in \cite{GW}.

If $f$ when restricted to $\LL$ is degenerate, then the integral (\ref{uzomp}) (with $\LL$ replacing $\M$) is not
well-defined.  The limit $s_0\to 0$ needs to be taken more carefully; higher order bosonic terms in the Lagrangian cannot be omitted,
and fermion fields cannot be dropped as they have zero modes.    The extreme case that $f$ restricted to $\LL$ is zero was treated in \cite{KW};
it leads to ${\mathcal D}$-modules rather than to quantization.

We can also consider the opposite limit of $s_0\to\infty$.  In this limit, the finite cylinder $C_{s_0}$ approaches the semi-infinite
cylinder $C$, whose compactification is a disc $D$.
It is natural to compare the path integral on the cylinder $C_{s_0}$ with boundary conditions set by $\B_\LL$ at $s=-s_0$
to a path integral on $D$
with an insertion of the closed string vertex operator $\O_\LL$ at the center.  For $s_0\to\infty$, the path integral on $C_{s_0}$ converges to the corresponding
path integral on $D$, but some information is lost.    This is related to the fact that the path
integral on $D$ makes sense for any middle-dimensional cycle $\LL$, while the path integral on $C_{s_0}$ can only be defined if $\LL$ is Lagrangian.
The path integral on either  $C_{s_0}$ or $D$ with boundary insertions at $s=0$ gives a cyclically symmetric trace-like function on the noncommutative algebra $\mathcal R$
that arises by deformation quantization of the ring of holomorphic functions on $\hat \M$.  (A trace-like function is a family of cyclically
symmetric functions $f_n(u_1,u_2,\dots,u_n)$, defined for a cyclically ordered $n$-plet of elements $u_1,\dots,u_n\in\mathcal R$, for any $n\in \Bbb{N}$,
and obeying $f_{n-1}(u_1u_2,u_3,\dots,u_n)=f_n(u_1,u_2,\dots,u_n)$.)  The path integral on $C_{s_0}$ gives a trace-like function
that actually can be interpreted as a trace in an $\mathcal R$-module (namely the $\mathcal R$-module obtained by quantizing the
$(\B_{cc},\B_\LL)$ strings), while the path integral on $D$, in the general case that $\LL$ is not Lagrangian, gives a trace-like function that is not
necessarily the trace in any $\mathcal R$-module.

\section{Hamiltonians}\label{zono}

Once we associate a Hilbert space $\mathcal H$ and an algebra of observables $\mathcal R$ to a classical phase space $\M$,
we can take an element of $\mathcal R$ and call it the Hamiltonian.   However, one may ask if there is some useful way to include
a Hamiltonian in the analysis from the beginning.

Here we will approach this question in two ways.  In section \ref{desc}, we rerun the analysis of section \ref{quantint} in a straightforward
fashion with a Hamiltonian
in place from the beginning.  The analysis is not difficult, and makes sense for any Hamiltonian, but is probably not very enlightening.
In section \ref{superham}, we do something that is probably more useful.  We ask whether, in the construction of section
\ref{quantint}, it is possible to relate the Hamiltonian of a one-dimensional description to the superpotential of a two-dimensional
description.  This is interesting when it is possible, though it is not usually possible.

\def\mH{{\mathfrak H}}
\subsection{Rerunning The Story With A Hamiltonian}\label{desc}

In the presence of a Hamiltonian $H(p,q)$, the integrand in the quantum path integral (\ref{zomp}) is modified in the familiar
fashion: the integral $\oint p_i\,\d q^i$ is replaced by $\oint\left(p_i\,\d q^i-H(p,q)\,\d t\right)$.
(We now take the $t$ coordinate to have period $\tau$; this parameter is meaningful
when the Hamiltonian is nonzero.)  The path
integral with the Hamiltonian included is then
\begin{equation}\label{nuromp}   \int_\U D p_i(t)\,D q^i(t)\exp\left(i\oint\left(p_i\,\d q^i-H(p,q)\,\d t\right)\right)\, u_1(t_1)u_2(t_2)\dots u_n(t_n).
\end{equation}
Actually, this is the real time version of the path integral, related to the evaluation of $\exp(-i \tau H)$.  The imaginary time
version of the path integral, related instead to $\exp(-\tau H)$, differs only by dropping the factor of $i$ in front of $-H(p,q)\,\d t$:
\begin{equation}\label{luromp}   \int_\U D p_i(t)\, D q^i(t)\exp\left(\oint\left(i\,p_i\,\d q^i-H(p,q)\,\d t\right)\right)\, u_1(t_1)u_2(t_2)\dots u_n(t_n).
\end{equation}
Most of our considerations in this section apply equally to the real or imaginary time
version of the path integral.  For definiteness, we consider the real time version until section \ref{anex}, where it is useful to consider both cases.

As in section \ref{anacon}, we begin by analytically continuing the classical phase space $\M$ to a complex symplectic phase space
$\hat\M$.  We will certainly not be able to incorporate a Hamiltonian in the discussion unless it too can be analytically continued,
so we assume that $H$ can be analytically continued to a holomorphic function $\mH$ on $\hat\M$.  It is convenient to introduce
the real and imaginary parts of $\mH$.  We write
\begin{equation}\label{mfh}\mH=H+iG,\end{equation}
with $H,G$ real.

The analog of the analytically continued path integral (\ref{orox}) is
\begin{equation}\label{norox}\int_\Gamma D Y^A(t)\,\exp\left(\oint\left(\Lambda_A\d Y^A-i \mH \d t\right)\right)\,u_1(t_1)u_2(t_2)\dots
u_n(t_n).\end{equation}
The dangerous real exponential factor in the path integral is $\exp(h)$ where now
\begin{equation}\label{gelx}h=\oint\left( c_A\d Y^A + G\,\d t\right).\end{equation}
Just as in section \ref{quantint}, we need to pick the integration cycle $\Gamma$ such that this exponential factor does not cause
the path integral to diverge.

The most obvious type of integration cycle is given by the sort of local-in-time condition considered in section \ref{simplint}.
We take $\Gamma$ to be the loop space of a middle-dimensional submanifold $\M'\subset\hat\M$.  We can practically
borrow the analysis of section \ref{simplint}. Under the transformation \ref{hefalo}, the term $\oint c_A\d Y^A$ in $h$ is multiplied by an arbitrary integer $n$, while the term $\oint G\d t$ is invariant.  To ensure that $h$ is bounded above, a necessary condition is  that $\oint c_A\d Y^A$ should vanish identically (for paths in $\M'$).  Just
as in section (\ref{simplint}), this means that $\M'$ should be Lagrangian with respect to $\omega$.  The exponent in (\ref{gelx})
then reduces to $\oint G\,\d t$, which is bounded above precisely if $G$ is bounded above when restricted to $\M'$.
So that is our answer: $\M'$ must be a Lagrangian submanifold (with respect to $\omega$) on which $G$ is bounded above.

We can also imitate the construction of integration cycles associated to flow equations.  The only real differences are that we have
to include the Hamiltonian in the condition for a critical point and in the flow equations.  Let us first write the conditions
for a critical point.  Prior to analytic continuation, requiring the functional $\oint\left(p_i\,\d q^i-H\,\d t\right)$
 to be stationary gives Hamilton's equations:
\begin{align}\label{mormo}\frac{\d q^i}{\d t}& =\frac{\partial H}{\partial p_i}\\
                  \notag                          \frac{\d p_i}{\d t}& =-\frac{\partial H}{\partial q^i}.\end{align}
So critical points correspond to arbitrary periodic solutions of Hamilton's equations.

Of course, it does not really suffice to consider only the real Hamilton equations.  Integration cycles in the analytically continued
path integral can be derived from all critical points of the function $h$ on the free loop space of $\hat\M$, real or not.  The general critical point is a  periodic solution of the complexified Hamilton equations, which we can write
\begin{equation}\label{zormo}-i\Omega_{AB}\frac{\d Y^B}{\d t}=\frac{\partial \mH}{\partial Y^A}.\end{equation}
To every critical point, one associates an integration cycle in the Feynman integral.
It is defined by solving the flow equations
\begin{equation}\label{pormo}\frac{\partial Y^A}{\partial s}=-g^{AB}\frac{\delta h}{\delta Y^B(s,t)}
=-I^A{}_B\frac{\partial Y^B}{\partial t}-g^{AB}\frac{\partial G }{\partial Y^B}\end{equation}
(for some choice of metric $g$ on $\hat \M$) for flows that start at $s=-\infty$ at
 the given critical point.

Moreover, by using some more information about solutions of the flow equations \cite{witten},
the original integration cycle of the Feynman integral -- with real $p$'s and $q$'s -- can be expressed as a linear combination
of these critical point cycles.  So if one asks,  ``Can the path integral
of a quantum system be expressed in terms of properties of the classical orbits?'' then this procedure gives an answer of sorts.

There are two problems which will tend to make this answer unuseful.
First, it requires an unrealistic degree of knowledge about the classical system.  Except for an integrable system, we cannot
even describe the generic periodic solutions of Hamilton's equations, even when we restrict to real $p$'s and $q$'s.    And it will also
be hard to say a lot about the solutions of the flow equations.   What is more, it is not clear what one can say about the Feynman integral
evaluated on a cycle associated to a critical point (except that perturbation theory for
this integral is likely to be Borel-summable).

Second, for a generic Hamiltonian, the flow equations lack the two-dimensional symmetry which was the main reason for the power
of the analysis in section \ref{quantint}.

But is there a class of special Hamiltonians for which the flow equations will once again have
 two-dimensional symmetry?  We consider this question next.

\subsection{Hamiltonians And Superpotentials}\label{superham}

To incorporate a Hamiltonian while preserving two-dimensional symmetry, we will
need to consider the $A$-model with a superpotential.

As usual, we consider the $A$-model with target $X$ and symplectic structure $\omega$.
(In our application, $X$ is a complex symplectic manifold $\hat\M$ and $\omega$ is the real part of a holomorphic
symplectic form $\Omega$ on $\hat\M$.)
To include a superpotential in the $A$-model is most natural when $X$ is actually
Kahler, so we henceforth assume that the metric $g$ of $X$ is Kahler and in particular
the almost
complex structure $I=g^{-1}\omega$ is integrable.   We let $W:X\to \C$ be a holomorphic
function\footnote{One can define the equation for a holomorphic function on any
almost complex manifold.  However, this equation is overdetermined and has no nonconstant
solutions on a generic almost complex manifold.  This is why we assume that $I$ is integrable.}
 that we call the superpotential.

 Consider the $A$-model on a Riemann surface $\Sigma$ with local complex coordinate $w$.
 In the absence of a superpotential, the $A$-model localizes on holomorphic maps ${\mathcal T}:
 \Sigma\to X$.  In the presence of a superpotential, the condition for a holomorphic map is
 perturbed and becomes
 \begin{equation}\label{corno}\frac{\partial x^i}{\partial \bar w}+g^{i\bar j}\frac{\partial\bar W}{\partial
 x^{\bar j}}=0,\end{equation}
 where $x^i$ are local $I$-holomorphic functions on $X$.  For the case of a single chiral
 superfield, this equation was studied in \cite{wm}.  The more general case was studied
 from a physical point of view in \cite{gs}.  For mathematical studies of  this
 equation, with an elucidation of some important points, see \cite{fjr,fjrtwo}.  The derivation of
this equation will be sketched in section \ref{construction}.

 To make sense of eqn. (\ref{corno}) globally along $\Sigma$, we cannot simply interpret
 it as an equation for a map from $\Sigma$ to $X$.  To explore this point, consider the case
 of a single chiral superfield $x$ with Kahler metric $|\d x|^2$ and superpotential $W(x)=x^n/n$.  The equation is
 \begin{equation}\label{orno}\frac{\partial x}{\partial \bar w}+\bar x^{n-1}=0.\end{equation}
 If $x$ is regarded as a scalar function, then the first term in this equation, namely
 $\partial x/\partial\bar w$, is a $(0,1)$-form on $\Sigma$, while the second term, namely
 $\bar x^{n-1}$,  is a scalar function. So with that interpretation the equation does not make
 sense globally. To make sense of (\ref{orno}) globally, while preserving two-dimensional
 symmetry, we
 need  a line bundle $\L\to\Sigma$ with an isomorphism $\L^n\cong K$, where $K$ is the
 canonical bundle of $\Sigma$.  We also need a Kahler metric on $\Sigma$, compatible
 with its complex structure; this gives an isomorphism between $\bar K$ (the space of $(0,1)$-forms
 on $\Sigma$) and $K^{-1}$.   We interpret $x$ as a section of $\L$.  Given this structure,  $\bar x^{n-1}$ and $\partial x/\partial \bar w$
 are both sections of $\L^{-n+1}$, so eqn. (\ref{orno}) makes sense.  The choice of Kahler metric
 on $\Sigma$ is inessential in the $A$-model, in the same sense that the Kahler metric
 on $X$ is inessential: a choice is needed to define the $A$-model, but the results are
 independent of the choice.  By contrast, the choice of $n^{th}$ root $\L$ of the canonical
 bundle is an important part of the structure.  The isomorphism of $\L^n$ with $K$ is allowed
 to have certain singularities at insertion points of vertex operators, as described in \cite{wm};
 there is a constraint on the genus of $\Sigma$ and the choices of vertex operators such
 that $\L$ exists globally.

 In the general case with several chiral superfields, one proceeds similarly.  One needs an
 action on $X$ of $U(1)$ (or in general of a covering group of $U(1)$) that preserves
 its Kahler structure and under which $W$
 transforms with charge 1;  that is, under the element $e^{i\theta}$ of $U(1)$, $W$
 maps to $e^{i\theta}W$.  Given  an action of $U(1)$ on $X$, let $K_1$ be the subbundle of $K$
 consisting of unit vectors (with respect to the metric on $\Sigma$), and define
 a fiber bundle  $\mathcal Y\to \Sigma$, whose fiber is isomorphic to $X$, by
 \begin{equation}\label{zorx}\mathcal Y=K_1\times_{U(1)}X.\end{equation}
(Thus an element of the fiber is a pair $(k,x)\in K_1\times X$ with
an equivalence relation $(k,x)\cong (ka^{-1},ax)$ for $a\in U(1)$.)  If the group that acts
on $X$ is actually an $n$-fold cover $U(1)_n$ of $U(1)$, for some integer $n$,  then as above we pick
a line bundle $\L
\to\Sigma$ with an isomorphism  $\L^n\cong K$, and denote its subbundle of unit
vectors as $\L_1$.  The action of $U(1)$ on $K_1$ lifts to an action of $U(1)_n$ on $\L_1$, and $\mathcal Y$ is defined as $\L_1\times_{U(1)_n}X$. The $A$-model
is defined for sections of the bundle $\mathcal Y$.

In our application, $\Sigma$ is actually the cylinder $C=S^1\times\RR_+$.
Its canonical bundle is naturally trivial, so the details of the last two paragraphs will
not play an important role.  We have explained them to make clear in what sense
 the eqn. (\ref{corno}) that we will be using does have two-dimensional
symmetry.

With $w=s+it$, so that $\partial/\partial\bar w=\frac{1}{2}(\partial_s+i\partial_t)$,  eqn. (\ref{corno}) can be written
\begin{equation}\label{orf}\frac{\partial x^i}{\partial s}=-i\frac{\partial x^i}{\partial t}
- 2g^{i\bar i}\frac{\partial}{\partial\bar{ x^{i}}}\left(W+\bar W\right).\end{equation}
After combining $x^i $ and $\bar {x^i}$ to real coordinates $Y^A$ and
replacing the complex number $i$ with the complex structure $I$, we can write
\begin{equation}\label{norf}\frac{\partial Y^A}{\partial s}=-I^A{}_B\frac{\partial Y^B}{\partial t}
   - g^{AB}\partial_B\left(4\,{\mathrm {Re}}\,W\right).\end{equation}
 (The part of this equation that is of type $(1,0)$ with respect to $I$ coincides with
 (\ref{orf}), and the $(0,1)$ part is the complex conjugate.)

Comparing this to (\ref{pormo}), we see that the supersymmetry condition of the $A$-model
with a superpotential coincides with the flow equation for quantum mechanics with a
Hamiltonian precisely if
\begin{equation}\label{ugh}G=4\,\mathrm{Re}\,W.\end{equation}
In other words, $G$, which was defined as the imaginary part of the $J$-holomorphic
function $\mH$, must also be the real part of the $I$-holomorphic function $4W$.

\subsubsection{Hyper-Kahler Symmetries}\label{hypsym}

This condition might sound impossibly restrictive, but it actually leads to something
relatively nice.  We can analyze the condition fully in case the metric $g$ on $X=\hat \M$
is actually hyper-Kahler; as described in section \ref{whati}, this is the most elegant
case of our construction.  We begin by considering a special situation, and show in section
\ref{moregen} that this special situation  actually is typical.

Let us suppose that $X$ admits a Killing vector field $V$
that preserves its hyper-Kahler structure.  This means in particular that the corresponding Lie
derivative $\mathpzc L_V$ annihilates the three symplectic structures $\omega_I,\omega_J,\omega_K$.
This is equivalent to saying  that, with $\iota_V$ the operation of contraction with $V$,  the one-forms
$\iota_V\omega_I$, $\iota_V\omega_J$, $\iota_V\omega_K$ are closed.  Integrating
those closed one-forms,  we find functions\footnote{
If necessary, we replace $\hat\M$ by a cover on which these functions are single-valued.
The logic is the same as it was in section \ref{relims}
where we chose not to impose the Dirac condition.}  $\mu_I,\mu_J,\mu_K$ that  up to
inessential additive constants are defined by
\begin{align}\label{zalign}\d \mu_I& = \iota_V\omega_I   \notag \\
                                              \d \mu_J& = \iota_V\omega_J \\   \notag
                                               \d\mu_K & =\iota_V\omega_K.\end{align}

The triple of functions $\vec\mu=(\mu_I,\mu_J,\mu_K)$ define the hyper-Kahler
moment map; their main properties are described in \cite{HKLR}.
Of particular importance to us, $\nu_I=\mu_J+i\mu_K$ is $I$-holomorphic
and $\nu_J=\mu_K+i\mu_I$ is $J$-holomorphic.  (These statements have an obvious
analog in complex structure $K$ and indeed in any complex structure that is a linear
combination of $I,J,$ and $K$.)
So if we set
\begin{equation}\label{heq} \mH=i\nu_J,\end{equation}
or in other words
\begin{equation}\label{teq} H =-\mu_I,~~ G=\mu_K,\end{equation}
then $G$ is indeed the real part of an $I$-holomorphic function.
In fact, $G=4\,{\mathrm{Re}}\,W$ with
\begin{equation}\label{peq}W=-\frac{i\nu_I}{4}.  \end{equation}

Finally, when can we find a $U(1)$ symmetry of $X$ that preserves the $A$-model
complex structure $I$ but rotates $W=(\mu_K-i\mu_J)/4$?  Such a $U(1)$ symmetry
does {\it not} preserve complex structures $J$ or $K$; rather, in $IJK$ space, it acts
by a rotation around the $I$ axis.  Many hyper-Kahler manifolds admit such a symmetry.

\subsubsection{An Example}\label{anex}

 An example is the familiar
case of the Eguchi-Hansen manifold $\hat \M$, defined in complex structure $J$ by
$X_1^2+X_2^2+X_3^2=j^2$.  This manifold, for any value of its hyper-Kahler
moduli, has an $SU(2)$
group of isometries that preserves its hyper-Kahler structure.
We take $V$ to be the vector field that generates a one-parameter
subgroup of this $SU(2)$.  In addition, for generic moduli, $\hat \M$ has a $U(1)$
symmetry that perserves one complex structure and rotates the other two.  Which complex structure is preserved
depends on the values of the hyper-Kahler moduli of $\hat\M$.   For our
purposes, we want to pick the moduli so that $\hat\M$ admits a $U(1)$ symmetry that acts by
 rotation about the $I$ axis.  (Concretely, $\hat\M$ has three real moduli, namely the real and imaginary
parts of $j$ and a third real modulus that is a Kahler parameter from the point of view of $J$;
 we set to zero $\mathrm{Im}\,j$ and the
third modulus. In this case, in complex structure $I$, $\hat\M$ is a blowup rather than deformation
of the $A_1$ singularity and has the desired symmetry.)   The $U(1)$ symmetry that rotates about the $I$ axis does not
preserve complex structure $J$, so it is not easily visible in that complex structure.

A typical choice of $V$ is
\begin{equation}\label{torox}V=X^1\frac{\partial}{\partial X^2}-X^2\frac{\partial}{\partial X^1}+\mathrm{complex~conjugate}
\end{equation}
(we add the complex conjugate because $V$ is supposed to generate an isometry).
According to eqn. (\ref{gwos}) and section \ref{whati},
the holomorphic two-form $\Omega=\omega_I-i\omega_K$
of $\hat\M$  is  $i\d X_1\wedge \d X_2/X_3$.  So we compute
that $\iota_V\Omega=-i(X_1\d X_1+X_2\d X_2)/X_3=i\d X_3$.  Thus $\mu_I-i\mu_K
=iX_3$, so $\mu_K=-\mathrm{Re}\,X_3$, $\mu_I=-\mathrm{Im}\,X_3$.
In view of (\ref{teq}), we then have
\begin{equation}\label{guffo} H=\mathrm{Im}\,X_3,~~G=-\mathrm{Re}\,X_3.\end{equation}

Now let us interpret the Hamiltonian $H$ in terms of real quantum mechanics.
In doing this, we want to interpret $\hat\M$ as the complexification of an ``underlying
real classical phase space'' $\M_0$, and $\mH$ as the analytic continuation to $\hat\M$
of a real Hamiltonian $H$ on $\M_0$.  The candidates for $\M_0$ that we will consider
are the two that have been discussed throughout this paper: $\M_0$ may be $\M$, defined
by $X_1,X_2,X_3$ real, or $\M'$, defined by $X_1$ real and positive, and $X_2,X_3$ imaginary.
For $\mH$ to be the analytic continuation to $\hat\M$
of a real function $H$ on $\M_0$, we at least need $\mH=H$ on $\M_0$.  Thus, a necessary
condition is that $G=\mathrm{Im}\,\mH$ vanishes on $\M_0$.  Otherwise, the path integral
(\ref{norox}) that we have investigated is not an analytic continuation of the ordinary Feynman
integral (\ref{nuromp}).

Looking at (\ref{guffo}), we see that this condition is obeyed by $\M'$ and not by $\M$.
(In fact, $H$ vanishes identically on $\M$ so $H$ is not very interesting as a Hamiltonian
on $\M$.)  So for the case at hand, we should take the mechanical system of interest
to be the one with phase space $\M'$.    $\M'$ has $SL(2,\RR)$ symmetry, and $H$ is one
of the generators of $SL(2,\RR)$.

If we want an example well-adapted to $\M$, we must do the Feynman integral  in imaginary
time rather than real time.  Changing from real time to imaginary time in (\ref{norox}) means
that we replace $i\,\d t$ by $\d t$.  This is equivalent to replacing $\mH$ by $-i\mH$,
so we can repeat the whole analysis for imaginary time by making that simple replacement.
Then (\ref{heq}) becomes $\mH=\nu_J$ and (\ref{teq}) becomes $H=\mu_K$, $G=\mu_I$.
In our example, we now have $H=\mathrm{Re}\,X_3$, $G=\mathrm{Im}\,X_3$.
Now the vanishing of $G$ on $\M_0$ allows $\M_0$ to be $\M$ but not $\M'$.
$\M$ is a two-sphere, and $H$ generates its rotations around the $X_3$ axis.

We motivated the condition that $G=0$ when restricted to $\M_0$ by asking that the space of
solutions of the flow equations can be intepreted as an integration cycle for
the Feynman integral of $\M_0$ with Hamiltonian $H$.  However, this condition has
another interpretation  \cite{hvi}: it is the condition for $\M_0$ to be the support of a Lagrangian
$A$-brane in the presence of the superpotential $W$.  Consider the path integral on a finite
cylinder $S^1\times [-s_0,0]$, with boundary conditions set at $s=-s_0$ by a Lagrangian
brane of support $\M_0$ and at  $s=0$ by a coisotropic brane (the coisotropic brane is defined by allowing
any solution of the flow equations near $s=0$, as in sections \ref{zondo}, \ref{loose}).
This path integral
computes $\Tr\,\exp(-it H)$, with the trace taken in the quantum Hilbert
space of $\M_0$.  The argument is the same as it was in section \ref{oose}.
The path integral on $S^1\times [-s_0,0]$ with the indicated boundary conditions is independent of
$s_0$, and for $s_0\to 0$ it goes over to the standard Feynman integral representation of  the trace.

\subsubsection{General Analysis}\label{moregen}

Now we would like to show that, for $\hat\M$ hyper-Kahler, the construction that we have described using a
vector field $V$ that preserves the hyper-Kahler structure is the most general possibility.

First of all, the assertion that $H+iG$ is $J$-holomorphic is equivalent to\footnote{We regard a complex
structure such as $J$ as a linear transformation that acts on the tangent bundle; its transpose $J^t$
acts on the cotangent bundle.} $(1+iJ^t)\d(H+iG)=0$, since $1+iJ^t$ projects onto the $(0,1)$ part of
a one-form.  Equivalently, we can write
\begin{equation}\label{omp}\d H = J^t \d G.\end{equation}
Similarly, the fact that $G$ is the real part of an $I$-holomorphic function means that there is a function $U$ such that
\begin{equation}\label{tzomp}\d G = I^t \d U.\end{equation}

The fact that $\omega_K+i\omega_I$ is $J$-holomorphic means that $J^t(\omega_K+i\omega_I)=i(\omega_K+i\omega_I)$,
or
\begin{equation}\label{romp} J^t\omega_K=-\omega_I,~~J^t\omega_I=\omega_K.\end{equation}
Similarly,
\begin{equation}\label{tomp} I^t\omega_J=-\omega_K,~~I^t\omega_K=\omega_J.\end{equation}

The assertion that a vector field $V$ preserves the hyper-Kahler structure of $\hat\M$ is equivalent to saying that $V$ preserves the three symplectic
structures $\omega_I$, $\omega_J$, $\omega_K$.  Indeed, using (\ref{romp}) and (\ref{tomp}), we can compute the complex structures
from the symplectic structures:
\begin{equation}\label{womp} J^t=-\omega_I\omega_K{}^{-1},~~ I^t=\omega_J\omega_K{} ^{-1},~~K=IJ.\end{equation}
And the metric is $g=I^t\omega_I=J^t\omega_J=K^t\omega_K.$  So $I,J,K$ and $g$ are all $V$-invariant if the three symplectic forms are.

In general, a vector field $V$ preserves a symplectic structure $\omega$ if there is a function $\eusm Q$ (called the moment map)
with $V=\omega^{-1}\d \eusm Q$.
Given functions $G, H, U$ obeying (\ref{omp}), (\ref{tzomp}), we will find a vector field $V$ that preserves the three symplectic
structures and has  $G, H, U$ as moment maps.

We simply define $V=\omega_K^{-1}\d G$, so $V$ certainly preserves $\omega_K$.  Then we  compute using the
above formulas that $\omega_I^{-1}\d H
=\omega_I^{-1}J^t\d G=-\omega_I^{-1}\omega_I\omega_K^{-1}\d G=-V$.  So $V$ preserves $\omega_I$ also.  Finally,
$\omega_J^{-1}\d U=-\omega_J^{-1}I^t\d G=-\omega_J^{-1}\omega_J\omega_K^{-1}\d G = -V$.  So again $V$ preserves $\omega_I$.
Finally, the relations $V=\omega_K^{-1}\d G=-\omega_I^{-1}\d H=-\omega_J^{-1}\d U$ that we have just found
imply that $H, G, U$ are the moment maps for the action of $V$: $H=-\mu_I$, $U=-\mu_J$, $G=\mu_K$.

\section{Running The Story In Reverse}\label{reverse}

So far, our point of view has been to start with a hopefully natural question -- find a new integration cycle for the Feynman
integral of quantum mechanics -- and express the answer in terms of a sigma-model with an extra spacetime dimension.

In the present section, we will run the story in reverse.  We start with a sigma-model and pick boundary conditions in the sigma-model
so that the sigma-model path integral has an interesting interpretation as an integral over boundary data.  In section \ref{sigmaone},
we consider a one-dimensional sigma-model.  So the boundary is a point and the integration over boundary data is an ordinary finite-dimensional integral.  In section \ref{sigmatwo}, we consider sigma-models in two dimensions, so that the boundary integral is a quantum mechanical
path integral, such as we considered so far.

We will hopefully emerge from this analysis not just with a better understanding of why
the constructions of section \ref{quantint} and \ref{zono} work, but a better understanding of why the machinery in those sections is
necessary -- that is, why some simpler tries do not work.

\subsection{Sigma-Models In One Dimension}\label{sigmaone}

To begin with, we consider supersymmetric quantum mechanics -- a sigma-model in one dimension with a target space $Z$,
endowed with a Riemannian metric $g$.  The model is a supersymmetric theory of maps ${\mathcal T}:L\to Z$, where $L$ is a one-manifold,
parametrized by a ``time'' coordinate $s$.  (We take $L$ to have Euclidean signature.)  We describe ${\mathcal T}$ by bosonic fields $x^I(s)$ that correspond to local coordinates $x^I$ on $Z$.
The fermi fields  are two sections $\psi^I(s)$ and $\chi^I(s)$ of ${\mathcal T}^*(TZ)$, the pullback to $L$ of the tangent bundle of $Z$.
The model has a ``fermion number'' symmetry $\cF$; $\psi$ and $\chi$ respectively have $\cF=1$ and $\cF=-1$.  There is also
a ``superpotential'' $h$, which is a real-valued function on $Z$.

\def\D{{\mathcal D}}
\def\DD{{\mathpzc D}}
In superspace, $x^I,\psi^I,\chi^I$ and an  auxiliary field can be combined to a superspace field $ X^I(s,\theta,\bar\theta)$ and the action
can be written concisely
\begin{equation}\label{porto}I=\frac{1}{2}\int \d s\,\d\theta \,\d\bar\theta\left(g_{IJ} \DD X^I \bar \DD X^J +h\right),\end{equation}
with superspace derivatives $\DD$, $\bar\DD$.
For our purposes, we will simply describe the theory in terms of the component fields, emphasizing its relation to Morse theory
\cite{wittenmorse}.

There are two supersymmetry operators $Q$ and $\bar Q$, of respectively $\cF=1$ and $\cF=-1$:
\begin{align}\label{twox}\notag        Q & =\psi^I\left(\frac{\partial}{\partial x^I}-\frac{\partial h}{\partial x^I}\right) \\
                                                           \bar Q & = \chi^I\left(-\frac{\partial}{\partial x^I}-\frac{\partial h}{\partial x^I}\right).\end{align}
They obey $Q^2=\bar Q^2=0$, $\{Q,\bar Q\}=2H$, where $H$ is the Hamiltonian.   For our purposes, we focus on $Q$ and consider
path integrals with $Q$-invariant boundary conditions.

The commutation relations generated by $Q$ are
\begin{align}\label{ork}\notag  [Q,x^I] & = \psi^I \\  \{Q,\psi^I\}& =0\\
                        \notag               \{Q,\chi^I\}&=-\left(\frac{\d x^I}{\d s} +g^{IJ}\frac{\partial h}{\partial x^J}\right). \end{align}
To get these relations, we use $\{\chi^I,\psi^J\}=g^{IJ}$ and
$g_{IJ}\partial x^J/\partial s=-\partial/\partial x^I$ (the last formula depends on the fact that in Euclidean signature, the usual relation
$p=-i\partial/\partial x$ becomes $p=-\partial/\partial x$).
One of the main observations in \cite{wittenmorse} is that, if we identify $\psi^I$ with the one-form $\d x^I$, we can identify
$Q$ with a conjugated version of the exterior derivative $\d=\sum_I \d x^I\partial/\partial x^I$:
\begin{equation}\label{ofox} Q=\d_h=\exp(h)\d \exp(-h).\end{equation}
To derive this result, one simply observes that it reproduces the commutation relations (\ref{ork}).
In particular, the form of $Q$ ensures that the condition for a map ${\mathcal T}:L\to Z$ to be $Q$-invariant --  which is
that  $\{Q,\chi^I\}$ must vanish --  is the flow equation of
Morse theory:
\begin{equation}\label{ropox} \frac{\d x^I}{\d s}=-g^{IJ}\frac{\partial h}{\partial x^J}.\end{equation}

We take $L$ to be the half-line $s\leq 0$.  We assume that $h$ is a Morse function (its critical points are isolated and nondegenerate) and
we let $p$ be one of the critical points.  We consider a path integral on $L$ with the boundary condition that $x^i(s)\to p$ for $s\to-\infty$.
We are going to assume that all flow lines starting at $p$ flow to infinity (rather than to another critical point), in which case, as reviewed
in section \ref{morse},
the possible boundary values $x^i(0)$ of the flow
form a cycle $\CC_p\subset Z$.  The dimension of this cycle is $i_p$, the Morse index of $h$ at $p$.  A typical example is that $Z$ is a noncompact
complex
manifold that admits lots of holomorphic functions (for instance, $Z=\C^n$ for some $n$) and $h$ is the real part of a generic holomorphic function on $Z$.

One may consider two points of view about the path integral.  Regarded as a function of the boundary values of the fields
at $s=0$, the path integral on $L$ computes a $Q$-invariant physical state $\Upsilon$.  Alternatively, if we integrate over the boundary values (with
some boundary condition), the path integral computes a number.

\def\phys{\mathrm{phys}}
\def\top{\mathrm{top}}
We  first explore the first point of view.  Exactly what the state $\Upsilon$ will turn out to be depends on exactly what action we take.
In the most physically natural version of the theory, the bosonic part of the action is
\begin{equation}\label{tomxo}I_{\phys}=\frac{1}{2\epsilon}\int \d s\left(g_{IJ}\dot x^I\dot x^J+g^{IJ}\frac{\partial h}{\partial x^I}\frac{\partial h}{\partial x^J}
\right),\end{equation}
with a parameter $\epsilon$.  From a topological field theory perspective,
it is natural to use a slightly different action that differs from $I_\phys$ by a boundary term:
\begin{equation}\label{omxo} I_{\top}=\frac{1}{2\epsilon}\int \d s\,\left(\frac{\d x^I}{\d s}+g^{IJ}\frac{\partial h}{\partial x^J}\right)^2
=I_\phys+\left.\left(h/\epsilon\right)\right\vert_{s=-\infty}^0.\end{equation}
Finally, there is an equivalent action with an auxiliary field $T$:
\begin{equation}\label{womox}I_{\mathrm {aux}}=-i\int_L\d s \,T_I\left(\frac{\d x^I}{\d s}+g^{IJ}\frac{\partial h}{\partial x^J}\right)
+\frac{\epsilon}{2}\int_L \d  s \, g^{IJ}T_IT_J.\end{equation}
Integrating out $T$ brings us back to (\ref{omxo}).  The parameter $\epsilon$ is inessential (and has been set to 1 in formulas such
as (\ref{twox}) above), because it can be absorbed in rescaling $g\to g/\epsilon$, $h\to h/\epsilon$.  But it is convenient
to make this parameter explicit, since localization of the path
integral on the solutions of the flow equations occurs for $\epsilon\to 0$. Henceforth, we write $h_\epsilon$ for $h/\epsilon$.

The path integral on the half-line $(-\infty,0]$ gives a $Q$-invariant state\footnote{This assertion can fail if there are flow
lines that interpolate from $p$ to some other critical point $q$.  We assume that there are none, for instance because $h$ is the
real part of a generic holomorphic function on a complex manifold.} in the Hilbert space associated to the boundary.
In the physical formalism with the action $I_\phys$, the  path integral gives a state $\Upsilon_\phys$ that is annihilated
by $Q=e^{h_\epsilon}\d e^{-h_\epsilon}$.
\begin{equation}\label{orny}\d \left(e^{-h_\epsilon}\Upsilon_\phys\right)=0.\end{equation} Since the relation between path integrals in the physical and topological actions comes from (\ref{omxo}) or
\begin{equation}\label{proton}\exp(-I_\top)=\exp(-I_\phys)\exp(-h_\epsilon(x(0))+h_\epsilon(x(-\infty))),\end{equation}
the result of replacing $I_\phys$ by $I_\top$ is simply to multiply $\Upsilon_\phys$ by the function $\exp(-h_\epsilon+h_\epsilon(p))$.
Thus the output of the topological path integral is $\Upsilon_\top=\exp(-h_\epsilon+h_\epsilon(p))\Upsilon_\phys$, and it obeys
simply
\begin{equation}\label{omork}\d \Upsilon_\top=0.\end{equation}
The state $\Upsilon_\top$ depends on $\epsilon$, but only by $Q$-exact terms.  It is easiest to compute $\Upsilon_\top$ if we
use the version (\ref{womox}) of the path integral with an auxiliary field $T$.  In the limit $\epsilon\to 0$, the path integral over $T$
\begin{equation}\label{pomork}\int DT\exp\left( i\int_L T_I\left(\frac{\d x^I}{\d s}+g^{IJ}\frac{\partial h}{\partial x^J}\right)\right)\end{equation}
gives a delta function supported on the solutions of the flow equations.  This means that the wavefunction $\Upsilon_\top$ has (for $\epsilon=0$)
delta function support on $\CC_p$, the locus of boundary values of solutions of the flow equations.  The closed differential form supported
on $\CC_p$ with the smallest possible degree (and the only one that can be defined without more information) is known as the Poincar\'e dual
to $\CC_p$.  This concept is defined more systematically in \cite{botttu}, but in brief, if $\CC_p$ is defined locally by
equations $x^1=x^2=\dots=x^m=0$ (where $m$ is the codimension of $\CC_p$), then
the Poincar\'e dual is locally
\begin{equation}\label{palooka}\delta(x^1)\delta(x^2)\cdots \delta(x^m)\psi^1\psi^2\dots \psi^m.\end{equation}
Since, for a fermionic variable $\psi$, a delta function is simply a linear function $\delta(\psi)=\psi$, we can equally well write the
Poincar\'e dual as
\begin{equation}\label{alooka}\delta(x^1)\delta(x^2)\cdots\delta(x^m)\delta(\psi^1)\delta(\psi^2)\cdots\delta(\psi^m).\end{equation}
To get the fermionic delta functions, simply do the integral over $\chi$, which takes the form
\begin{equation}\label{looka}\int D\chi\,\exp\left(i\int_L\chi_I \D\psi^I\right),\end{equation}
where $\D$ is the linearization of the flow equations.  This gives a delta function setting $\D\psi=0$.  When restricted to $s=0$,
this delta function forces $\psi^I(0)$ to be tangent to $\CC_p$, so it sets to zero precisely the modes $\psi^1,\psi^2,\dots,\psi^m$ of $\psi$ that
are valued in the normal bundle to  $\CC_p$.
This explains why $\Upsilon_\top$ is the Poincar\'e dual to $\CC_p$.  (For more on this type of integral, see the discussion of
eqn. (5.2) in \cite{FLN}.)

So far, we have identified the state $\Upsilon_\top$ that emerges when the path integral (with the action $I_\top$) is computed
as a function of the boundary values.  Now we would like to integrate over the boundary values, perhaps with the help of some additional
boundary condition or boundary couplings, to get a number.

If $\CC_p$ has positive dimension $i_p$, its Poincar\'e dual has the wrong degree to be integrated.  We will choose
boundary couplings that involve a choice of closed form $\Lambda$ of degree $i_p$, and we will aim to give a path integral recipe for computing
\begin{equation}\label{pool} \int_Z \,\Upsilon_\top\wedge\Lambda.\end{equation}
Actually, we will take $Z$ to be a noncompact Calabi-Yau manifold of complex dimension $n$, endowed with a holomorphic volume
form $\Omega$.  (We take the metric of $Z$ to be hermitian, possibly Kahler.)
For example, we might take $Z=\C^n$, with complex coordinates $z^1,\dots,z^n$, and $\Omega=\d z^1\d z^2\cdots
\d z ^n$.   We take
\begin{equation}\label{zonkers}\Lambda=\Omega\,\exp(S), \end{equation}
where $S$ is a holomorphic function on $Z$.   Then the finite-dimensional integral that we will represent by a path integral on $L=(-\infty,0]$ is
\begin{equation}\label{ondo}\int_Z\Upsilon_\top \wedge \Lambda =\int_{\CC_p}\Lambda=\int_{\CC_p}\Omega\,\exp(S).  \end{equation}  (We have reduced
an integral over $Z$ to one over $\CC_p$ using the fact that $\Upsilon_p$ has delta function support on $\CC_p$.)
Since $\Omega$ is an $n$-form, the integral vanishes unless $\CC_p$ is $n$-dimensional.  In addition, of course, the Morse function $h$
must be related to $S$ in such as way that the integral converges.   A simple way to ensure this is to set
\begin{equation}\label{rime} h =\mathrm{Re}\,S,\end{equation}
as discussed in \cite{witten} and in section \ref{morse}.  But actually, any $h$ that is close enough to $\mathrm{Re}\,S$ will work
just as well.  In the example discussed in the introduction, with $n=1$ and $S(z)=-z^4+az$, with a constant $a$, we could take
\begin{equation}\label{polz}h=\mathrm{Re}\,(-z^4+\tilde a z),\end{equation}
with some other constant $\tilde a$; changing $h$ by subleading terms does not affect the convergence of the integral.

To represent (\ref{ondo}) as a path integral, we simply write the usual Feynman integral for this supersymmetric system, but with
a boundary insertion of $\exp(S)\Omega$.  Of course, $\Omega$ is represented in the path integral as $\Omega_{i_1i_2\dots i_n}
\psi^{i_1}\psi^{i_2}\dots\psi^{i_n}$.  Thus, we consider the path integral
\begin{equation}\label{omorox}\int DX\,D\psi\,D\chi\,\,\exp(-\hat I_\top)\, \exp(S(x(0))\,\Omega_{i_1i_2\dots i_n}\psi^{i_1}(0)\psi^{i_2}(0)
\dots\psi^{i_n}(0).\end{equation}
Here $\hat I_\top$ is the supersymmetric completion of $I_\top$ defined in (\ref{omxo}), including fermionic terms.

We would like to verify directly that the path integral in (\ref{omorox}) is supersymmetric, that is, $Q$-invariant.  It is reasonable to expect this, since this path
integral equals the period $\int_{\CC_p}\Omega\,\exp(S)$, which has a topological meaning, that is, it is invariant under small
deformations of the integration cycle $\CC_p$.

First, let us just directly verify $Q$-invariance.  Since $[Q,I_\top]=0$, we only need to worry about $Q$-invariance of the boundary
insertion $\exp(S(x(0))\,\Omega_{i_1i_2\dots i_n}\psi^{i_1}(0)\psi^{i_2}(0)
\dots\psi^{i_n}(0)$.  Since $\{Q,\psi\}=0$, we only need to worry about $[Q,x^I]=\psi^I$, or in terms of local complex coordinates $x^i$,
$[Q,x^i]=\psi^i$, $[Q,\bar {x^i}]=\psi^{\bar i}$.  Actually, $[Q,\bar{x^i}]$ will not enter, since $S$ and $\Omega$ are holomorphic. And
the contributions that come from $[Q,x^i]$ all vanish by fermi statistics, since the boundary insertion is already proportional to the
product of all $n$ fermions of type $(1,0)$, namely $\psi^{i_1}\psi^{i_2}\cdots \psi^{i_n}$.  So the boundary condition is $Q$-invariant.

Now we will obtain the same result in a longer but illuminating way.   Let us work out the boundary conditions on bosons and fermions that arise
naturally from the above path integral.  Because the boundary insertion is proportional to the product of all fermion fields $\psi^i$ of type $(1,0)$,
and a fermion field $\psi^i$ is equivalent to a delta function $\delta(\psi^i)$, the boundary condition sets the $(1,0)$ part of $\psi$ to zero
at $s=0$:
\begin{equation}\label{recent}\psi^{(1,0)}\vert_{s=0}=0.\end{equation}
What about $\chi$?  The proper boundary condition on $\chi$ is determined by setting to zero the boundary contribution
in the equations of motion.  The fermion kinetic energy is $I_f=i\int_{s=-\infty}^0\d s\, \chi_I \D\psi^I$ , and when we vary it, we get bulk equations
of motion $\D\psi=\D\chi=0$.  After imposing these equations, we still find a nonzero boundary contribution to the variation of $I_f$,
and we must pick the boundary conditions to set this to zero.  The boundary contribution is $\chi_I\delta\psi^I|_{s=0}$.  Since the only
restriction on $\delta\psi$ is that its $(1,0)$ part $\delta\psi^i$ is equal to zero, the boundary condition on $\chi$ must be
vanishing of the $(0,1)$ part $\chi_{\bar i}$.

We can analyze the bosonic boundary conditions in the same way.  The purely bosonic part of the exponent of the path integral is
\begin{equation}\label{hurley} -\frac{1}{2\epsilon}\int_{-\infty}^0\d s\,g_{IK}\left(\frac{\d x^I}{\d s}+g^{IJ}\frac{\partial h}
{\partial x^J}\right)\left(\frac{\d x^K}{\d s}+g^{KL}\frac{\partial h}{\partial x^L}\right) +S(x)|_{s=0}.\end{equation}
Upon varying this and imposing the Euler-Lagrange equations to set the bulk part of the variation to zero, we are left with a boundary
variation at $s=0$, which takes the form
\begin{equation}\label{peroxide}\left.\left(\left( -\frac{1}{\epsilon}\left(g_{IJ}\frac{\d x^J}{\d x}+\frac{\partial h}{\partial x^J}\right) +\frac{\partial S}
{\partial x^I}\right)\delta x^I\right)\right|_{s=0}.\end{equation}
Assuming that we want free boundary conditions, so that $\delta x^I$ is unconstrained, the boundary condition must be
\begin{equation}\label{eroxide}   -\frac{1}{\epsilon}\left(g_{IJ}\frac{\d x^J}{\d s}+\frac{\partial h}{\partial x^I}\right) +
\frac{\partial S}{\partial x^I
}=0.\end{equation}
Let us verify that this condition, combined with the fermionic boundary conditions, is in fact supersymmetric.

In general, supersymmetry of a boundary condition means that the component of the supercurrent normal to the boundary vanishes.
But here, as we are in spacetime dimension one, the normal component of the supercurrrent is simply the supercharge $Q=\left(
g_{IJ}\d x^J/ds+\partial_I h\right)\psi^I$.  Using the bosonic boundary condition (\ref{eroxide}), this is equivalent at $s=0$ to
$\epsilon \partial_I S\psi^I$.  Because $S$ is holomorphic, this is equivalent to $\epsilon\partial_i S\psi^i$, and vanishes at $s=0$
 since the $(1,0)$ part of $\psi$ vanishes at $s=0$.  So again we have established the $Q$-invariance of the boundary condition.  This experience
will stand us in good stead in more
complicated examples.

\subsection{Sigma-Models In Two Dimensions}\label{sigmatwo}

The sigma-model studied in section \ref{sigmaone} can be regarded as a reduction to one dimension of a model defined in two
(or even more) dimensions.  Let us consider the two-dimensional case.  For brevity, rather than considering unintegrable structures,
we will assume that the target space $Z$ is a Kahler manifold, with complex structure $I$ and Kahler metric $g$.
Then a sigma-model with target $Z$ has four supercharges, and admits
a topologically twisted $A$-model.  Just as in one dimension, the bosonic fields $x^I$ describe a map ${\mathcal T}:\Sigma\to Z$,
for some Riemann surface $\Sigma$, and the fermi fields $\psi^I$ of $\cF=1$ are sections of the pullback to $\Sigma$ of $TZ$,
the tangent bundle of $Z$.   The fermi fields of $\cF=-1$ are a one-form $\chi$ on $\Sigma$ with values in the pullback of $T^*Z$
(the cotangent bundle of $Z$); as in section  \ref{zondo}, $\chi$ obeys an algebraic constraint $\chi_J=\star\chi_KI^K{}_J$.

Let us ask whether we can naively imitate the construction of section \ref{sigmaone}
and add to the bosonic part of the action a boundary coupling
\begin{equation}\label{wondo} \int_{\partial \Sigma}\d t \,S(x(t)),\end{equation}
where $S$ is an $I$-holomorphic function on $Z$, and $\d t$ is a one-form on $\partial\Sigma$.
Whether we can add this term depends on what supersymmetry we want to preserve.  The two-dimensional sigma-model
with target $Z$ has two topologically twisted versions -- the $A$-model and the $B$-model.

In the $B$-model, there is no problem in adding the boundary interaction (\ref{wondo}).  However, the $B$-model
localizes on constant maps to $Z$, rather than on the nontrivial integration cycles associated to Morse theory that are of interest
in the present paper.  Hence the $B$-model with the boundary coupling (\ref{wondo}) is not a good generalization to two dimensions
of what we have said in section \ref{sigmaone}.

The $A$-model does lead to interesting integration cycles related to Morse theory in loop space.  However, the boundary coupling
(\ref{wondo}) does not work in the $A$-model.  The reason is that because of the $A$-model transformation law $\{Q,x^I\}=\psi^I$,
invariance of the boundary coupling requires us to set $\psi^{(1,0)}$
to zero on the boundary of $\Sigma$, just as we did in section \ref{sigmaone}.  The problem is that in two dimensions, unlike
one dimension, this is not a satisfactory boundary condition.   A quick explanation of why is that, in one dimension, setting the
boundary values of $\psi^{(1,0)}$ to zero amounts to a boundary insertion $\Omega_{i_1i_2\cdots i_n}\psi^{i_1}\psi^{i_2}\cdots\psi^{i_n}$
that has finite fermion number $\cF$.  In two dimensions, the analog would have to be formally
\begin{equation}\label{oxoc}\prod_{t\in\partial\Sigma} \Omega_{i_1i_2\cdots i_n}\psi^{i_1}(t)\psi^{i_2}(t)\cdots\psi^{i_n}(t),\end{equation}
now carrying an infinite $\cF$, as a result of which all correlation functions would vanish.

Here is a better explanation.  The fermion kinetic energy  of the $A$-model is
\begin{equation}\label{plond}i\int_\Sigma\left(\chi_{(1,0)}\frac{D}{D\bar z}\psi^{(1,0)}+\chi_{(0,1)}\frac{D}{Dz}\psi^{(0,1)}\right).\end{equation}
When we vary this kinetic energy, the boundary contributions are
\begin{equation}\label{lond}i\int_{\partial\Sigma}\left(\chi_{(1,0)}\delta\psi^{(1,0)}+\chi_{(0,1)}\delta\psi^{(0,1)}\right).\end{equation}
In general, any boundary condition sets to zero a middle-dimensional subspace of the fermion boundary values in such a way
that the boundary variations vanish.  To obey these conditions,  a boundary condition setting $\psi^{(1,0)}$ to zero must leave the
boundary values of $\chi_{(1,0)}$ unconstrained.  But this is not an elliptic boundary condition, and concretely, as the equation of motion
of $\chi_{(1,0)}$ is $\bar D\chi_{(1,0)}=0$, if we leave its boundary values unconstrained, $\chi_{(1,0)}$ will have infinitely many
zero modes, and all correlation functions will vanish.

To avoid this, all $A$-branes, both Lagrangian ones \cite{chernstring} and coisotropic ones \cite{KO}, involve a boundary condition
that sets to zero a linear combination of $\psi^{(1,0)}$ and $\psi^{(0,1)}$ (and similarly a linear combination of $\chi_{(1,0)}$ and
$\chi_{(0,1)}$).

{}From the standpoint of the present paper, the cure for the problem is that $S$ in eqn. (\ref{wondo}) should be holomorphic not in complex
structure $I$, but in some other complex structure $J$ which obeys $IJ=-JI$.  Thus, assuming that $I$ and $J$ are both integrable, the Kahler
structure of $Z$ should be extended to a hyper-Kahler structure and the sigma-model has eight supercharges rather than four.
(We can get by with an almost hyper-Kahler structure, as described at the end of section \ref{whati}.)    If $S$ is holomorphic in complex
structure $J$, then to ensure invariance of the boundary coupling (\ref{wondo}), we require vanishing of the boundary
values of $\psi^{(1,0;J)}$, that is the $(1,0)$ part of $\psi$ with respect to $J$. As $J$ anticommutes with $I$, there is no
component of $\psi$ that is of type $(1,0)$ with respect to both $I$ and $J$, and the problem encountered above disappears.  The rest
of the analysis of the boundary conditions given in the one-dimensional case in section \ref{sigmaone} has a straightforward extension
to the present two-dimensional case.  The boundary condition on $\chi$ must set to zero the boundary values of $\chi_{(0,1;J)}$.
On $x^I$, we take free boundary conditions (no restriction on the boundary values of $\delta x^I$).  The vanishing of the normal component
of the topological supercurrent is demonstrated in the same way as in section \ref{sigmaone}.

By taking $S$ to be holomorphic in a complex structure that anticommutes with $I$, and constructing a boundary condition as
just described, we ensure that the boundary coupling (\ref{wondo}) is compatible
with the $Q$-invariance of the $A$-model.  But we are still not out of the woods: we need to ask whether the $A$-model path integral
converges.  The bulk equations for supersymmetry of the $A$-model are flow equations of the general form
$\partial x^I/\partial t =-g^{IJ}\delta h/\delta x^J$, for a suitable functional $h$ which is determined by the bulk action of the
$A$-model.  To ensure convergence of the path integral, we must pick $S$ so that its real part has the same
asymptotic behavior as $h$ (they may differ by a correction  that grows too slowly to be problematical).

The last condition is hard to obey, since $h$ is determined entirely by the $A$-model defined with complex structure $I$
and is supposed to be related to the real part of a holomorphic function in some other complex structure $J$.
The constructions in sections \ref{quantint} and \ref{superham} are based on cases in which this can actually happen.   These cases,
assuming that $I$ and $J$ are both integrable, involve a hyper-Kahler target space and a doubling of supersymmetry relative
to the one-dimensional analysis of section \ref{sigmaone}.  The special nature of the
target space was explicit in sections   \ref{quantint} and
\ref{zono}.   And in section \ref{superham},
the Hamiltonians that work are related to the twisted masses \cite{twisted} which are precisely the potentials that preserve all the supersymmetry of a sigma-model with a hyper-Kahler target.
A similar story holds in section \ref{constraints} where we include gauge fields.

\section{Analogs With Gauge Fields}\label{cs}

In the present section, we repeat the analysis of section \ref{reverse}, this time in the presence of gauge interactions.

In section \ref{supergauge}, we add gauge fields to the one-dimensional supersymmetric sigma-models studied in section \ref{sigmaone}, taking
the target space of the sigma-model to be of finite dimension.

In section \ref{applics}, we consider the case that the target space of the
sigma-model is the infinite-dimensional space of gauge-connections on a three-manifold.  This leads to one of the main insights of the
present paper:  under certain conditions, the path integral of $\N=4$ super Yang-Mills theory on a half-space reduces to the path integral
of three-dimensional Chern-Simons gauge theory on the boundary of the half-space.

Finally, in section \ref{constraints}, we consider two-dimensional supersymmetric gauge theories with gauge fields.  This generalization of the
analysis of section \ref{quantint} leads to the construction of a new integration cycle for quantum mechanics with constraints.

\subsection{Supersymmetric Quantum Mechanics With Gauge Fields}\label{supergauge}

\subsubsection{Construction Of The Model}\label{construction}

Though the preliminary setup did not require this, our main application in section \ref{reverse} required the target space $Z$ of the sigma-model to be a complex manifold.   In the present discussion, we will want $Z$ to be a complex manifold for much the same reason, and we want $Z$ to be endowed
with a compatible symplectic structure so that we can define the moment map for a group action on $Z$.  So we will take $Z$ to be a Kahler
manifold to begin with.

The sigma-model with a Kahler target space has four supersymmetries. Here we will
consider a gauged version of the sigma-model, still with four supercharges.  We assume that $Z$ has a compact connected group $H$ of symmetries, and we will consider the combined theory of a map $\Phi:L\to Z$ together with vector multiplets that gauge the $H$ symmetry.

This theory can arise by dimensional reduction from four dimensions, and to understand some of its properties, it is convenient to start there.
We formulate the theory on $\RR^4$ with Euclidean signature and  spacetime coordinates $y^\mu$, $\mu=0,1,2,3$.
The only propagating  bosonic fields  are the gauge field $A=\sum_{\mu=0}^3 A_\mu\,\d
y^\mu$ and the map $\Phi:\RR^4\to Z$,
which we describe in terms of complex-valued fields $x^i$, $i=1,\dots,\mathrm{dim}_\C\,Z$ that correspond to local complex coordinates on $Z$.
The four supercharges are a spinor field $Q_\alpha,$ $\alpha=1,2$,
 of positive four-dimensional chirality and another spinor field $\tilde Q_{\dot\alpha}$, $\dot\alpha=1,2$ of negative chirality.
The algebra they generate is
\begin{equation}\label{trufo}\{Q_\alpha,\tilde Q_{\dot\alpha}\}=\sum_{\mu=0}^3\sigma^\mu_{\alpha\dot\alpha}P_\mu ,\end{equation}
where $\sigma^\mu_{\alpha\dot\alpha}$ are the Dirac matrices written in a chiral basis, and $P_\mu$ are the momentum generators.

Now we dimensionally reduce to two dimensions, taking the fields to be independent of $y^2$ and $y^3$.  The components $A_0,A_1$
of the four-dimensional gauge field survive as a two-dimensional gauge field, but the components $A_2$, $A_3$ become scalar
fields with values in the adjoint representation.  The supersymmetry algebra still takes the form (\ref{trufo}), but the momentum components
$P_2$ and $P_3$ are now simply the commutators with $A_2 $ and $A_3$, respectively.  It is convenient to combine $A_2$ and $A_3$ to
a complex scalar field $\sigma=A_2-iA_3$ with values in the adjoint representation.  We write $[\sigma,\cdot]$ for the infinitesimal
gauge symmetry generated by $\sigma$.

Upon dimensional reduction to two dimensions, the four-dimensional rotation group $SO(4)$ reduces to $SO(2)\times SO(2)'$, or equivalently
$U(1)\times U(1)'$, where the
first factor rotates $y^0,y^1$, and the second rotates $y^2,y^3$.
A spinor of positive chirality, such as $Q_\alpha$, has components of $U(1)\times U(1)'$ charges $\pm 1/2,\pm 1/2$, and a spinor
of negative chirality, such as $\tilde Q_{\dot \alpha}$, has components of charges $\pm 1/2,\mp 1/2$.  We denote the
components of $Q$ and $\tilde Q$ with these quantum numbers as $Q_{\pm\pm}$ and $\tilde Q_{\pm\mp}$.
If the superpotential vanishes (or more generally if it is quasihomogeneous), there is a third $U(1)$ symmetry, present already in four dimensions before dimensional reduction.  This is the $U(1)_R$ symmetry
under which $Q$ and $\tilde Q$ have respective charges $-1$ and $1$.
($U(1)_R$ may be anomalous in four dimensions, but not in the dimensional reduction to two dimensions.)
The usual supercharge of the two-dimensional $A$-model is
\begin{equation}\label{supch}Q=Q_{++}+\tilde Q_{-+}.\end{equation}
If $U(1)_R$ is a symmetry, then upon suitably twisting the theory and restricting to $Q$-invariant observables and correlation functions,
one gets a two-dimensional topological field theory that can be formulated on any oriented two-manifold  $\Sigma$. (One has to pick a metric
on $\Sigma$ to define the theory, but the results are independent of this metric.)  The starting point in twisting is that $Q$ is invariant under a certain modified rotation symmetry;
if $J$ is the generator of $U(1)$ and $R$ is the generator of $U(1)_R$, then $Q$ commutes with a linear combination $J'=J+R/2$.
In the twisted theory, the spins of all fields are  their $J'$ eigenvalues rather than their $J$ eigenvalues.
Moreover, since $Q$ is $J'$-invariant, its square does not generate a translation along $\RR^2$ (those translation generators have $J'=\pm 1$).
Rather, the supersymmetry algebra (\ref{trufo}) implies that
\begin{equation}\label{zonox} Q^2=[\sigma,\cdot].\end{equation}
In particular, this means that on gauge-invariant fields and states, $Q^2=0$ and one can define the cohomology of $Q$.   Upon restricting
states and operators to $Q$-invariant ones and projecting to the cohomology of $Q$, one obtains (in the $R$-symmetric case) a two-dimensional
topological field theory known as the $A$-model; it was analyzed in \cite{Phases}, where the
following formulas are  obtained and described in more detail  (with some minor differences in notation).  The $A$-model is $\Z$-graded by
$U(1)'$.
It is convenient to  introduce a generator $\cF$ of $U(1)'$ that is normalized so that $Q$ has $\cF=1$ and $\sigma$ has $\cF=2$, while
the fermions have $\cF=\pm 1$.   An important property of the model is that
\begin{equation}\label{trumy}[Q,\sigma]=0.\end{equation}
This reflects the fact that there is no elementary fermion of $\cF=3$, and is consistent with $Q^2=[\sigma,\cdot]$ since $[\sigma,\sigma]=0$.

After twisting, the fermionic fields in the vector multiplet
are an adjoint-valued one-form $\lambda$ of $\cF=1$ and adjoint-valued zero-forms $\eta,\rho$ of
$\cF=-1$.  Under the action of $Q$, the gauge field is contained in a multiplet that takes the form
\begin{equation}\label{clumsy}[Q,A_\mu]=\lambda_\mu,~~\{Q,\lambda_\mu\}=-D_\mu\sigma. \end{equation}
The rest of the vector multiplet becomes
\begin{equation}\label{umsy} [Q,\bar\sigma]=\eta, ~~\{Q,\eta\}=[\sigma,\bar\sigma],\end{equation}
and
\begin{equation}\label{tumsy}\{Q,\rho\}=D,~~[Q,D]=[\sigma,\rho].\end{equation}
All these formulas are consistent with $Q^2=[\sigma,\cdot].$
In (\ref{tumsy}), $D$ is an auxiliary field with values in the adjoint representation;
if we take the minimal form of the sigma-model action, the equation of motion of motion of $D$ is $D=\star F+\mu$, where $F$  is the curvature $\d A+A\wedge A$, $\star $ is the Hodge star, and $\mu$ is the moment map for the action of $H$ on $Z$.
The moment map is defined in the usual way by
\begin{equation}\label{morso}\frac{\partial\mu_a}{\partial x^i}=V_a^{\bar j}\omega_{\bar j i},~~\frac{\partial\mu_a}{\partial\bar{x^j}}=V_a^{i}\omega_{i\bar j }.\end{equation}
Here  $\omega$ is the Kahler form of $Z$ and $V_a$, $a=1,\dots,{\mathrm {dim}}\,H$ are the vector fields that generate the action of $H$ on $Z$.
We will write $V(\sigma)=\sum_a\sigma^a V_a$ for the vector field corresponding to $\sigma$.
After eliminating $D$ by its equation of motion, the first equation in (\ref{tumsy}) becomes
\begin{equation}\label{lumsy} \{Q,\rho\}=\star F+\mu.\end{equation}

As for the chiral multiplets, in the twisted theory the $\cF=1$ fermions are zero-forms $\psi^i$, $\psi^{\bar i}$ with values
in $\Phi^*(T^{(1,0)}Z)$ and $\Phi^*(T^{(0,1)}Z)$ (that is, the pullbacks to $\RR^2$ of the $(1,0)$ and $(0,1)$ parts of the tangent bundle of $Z$).
These fields obey
\begin{align}\label{flunky} [Q,x^i]&=\psi^i,~~ \{Q,\psi^i\}=V^i(\sigma) \\ \notag
                                           [Q,\bar{x^i}]&=\psi^{\bar i},~~\{Q,\psi^{\bar i}\}=V^{\bar i}(\sigma).\end{align}
The $\cF=-1$ fermi fields are a $(0,1)$-form $\chi^i$ valued in $\Phi^*(T^{(1,0)}Z)$, and a $(1,0)$-form $\chi^{\bar i}$ valued in
$\Phi^*(T^{(0,1)}Z)$. They transform in the familiar sort of multiplet $\{Q,\chi^i\}=\mathpzc{F}^i$, $ [Q,\mathpzc{F}^i]=[\sigma,\chi^i]$  where $\mathpzc{F}^i$ is an auxiliary field; there is a complex conjugate multiplet  $\bar{\phi^{j}},\chi^{\bar j},\bar{\mathpzc{F^j}}$. After eliminating $\mathpzc{F},\bar{\mathpzc{F}}$ by their equations of motion, we get
\begin{align}\label{zunky}\{Q,\chi^i\} & = \bar\partial_A x^i+g^{i\bar j}\frac{\partial\bar W}{\partial \bar{x^j}} \\ \notag
                                         \{Q,\chi^{\bar i} \}& =\partial_A\bar{x^i}+g^{\bar i j}\frac{\partial W}{\partial x^j}, \end{align}
where $g$ is the Kahler metric of $Z$ and we make the usual decomposition $\d_A=\partial_A+\bar\partial_A$ for the gauge-covariant exterior derivative $\d_A$ on the Riemann surface $\Sigma$.  (Explicitly if $w^q,~q=1,2$ are local coordinates on $\Sigma$, then
$\d_A$ is defined by $\d_A x^i=\d w^q(\partial_qx^i+A^a_q V_a^i)$.)

The equations for a supersymmetric configuration of the bosonic fields  are $\{Q,\Lambda\}=0$ for every fermionic field $\Lambda$.
In view of the above formulas, the supersymmetry conditions that involve $\sigma$
are
\begin{equation}\label{zsimga} D_\mu\sigma = V(\sigma)=[\sigma,\bar\sigma]=0.   \end{equation}
These equations say  that the gauge transformation generated by $\sigma$ leaves fixed  $A$, $x$, and $\bar\sigma$,
 so it is a symmetry of the whole
configuration.  These conditions are very restrictive, and in many applications they force $\sigma=0$.

The supersymmetry conditions for the other fields are
\begin{align}\label{simga} \star F+\mu & = 0 \\
                                          \bar \partial_Ax^i+g^{i\bar j}\frac{\partial \bar W}{\partial\bar{x^j}}&=0\notag ,\end{align}
along with the complex conjugate of the second equation.\footnote{In \cite{Phases}, instead of the second equation, two separate equations $\bar\partial_Ax^i=0=\partial W/\partial x^i$ were given.  This is because separate invariance was imposed
under $Q_{++}$ and $\tilde Q_{-+}$.  In the present paper, we will consider boundary conditions that conserve not $Q_{++}$ or
$\tilde Q_{-+}$ but only their sum $Q=Q_{++}+\tilde Q_{-+}$, so we  add the two equations.}

As we have already explained in section \ref{superham},  the second equation in (\ref{simga}) has two-dimensional symmetry only if $W$
is quasihomogeneous.  (A generic $W$ violates the $U(1)_R$ symmetry that was assumed in the construction of the twisted $A$-model.)
In the present section, we wish to consider a generic $W$, so we will have no two-dimensional symmetry.  For this reason
among others,\footnote{There is another difficulty in continuing the analysis in two dimensions: unless we double
the supersymmetry (as we will do in section \ref{constraints}),
we will run into difficulties analogous to those that were described in section \ref{sigmatwo}.} we make a further dimensional reduction to one dimension, where
the above equations are natural for any $W$.  In this reduction, the gauge field splits up into a one-dimensional gauge field $A_0$
and an adjoint-valued real scalar field $A_1$.  (We already generated two such fields $A_2$ and $A_3$ in the first stage of reduction; one
might think that there should be an  $SO(3)$ symmetry rotating these three fields, but this symmetry is spoiled by the choice of $Q$.)
We will return to the two-dimensional case in section \ref{constraints}.

The equations (\ref{simga}) are flow equations in the $y^0$ direction, up to a gauge transformation.  To agree with our previous
notation, we write $s$ for $y^0$.
In the gauge $A_0=0$, the equations (\ref{simga}) can be written
\begin{align}\label{nexo}\frac{\d A_1}{\d s}&=-\mu=-\frac{\partial h}{\partial A_1}  \\
                    \notag \frac{\d x^i}{\d s}&= -iA_1^aV_a^i-g^{i\bar j}\frac{\partial \bar W}{\partial\bar{x^j}}=-g^{i\bar j}\frac{\partial h}{\partial \bar{x^j}},\end{align}
with
\begin{equation}\label{gamag}h=A_1^a\mu_a+2\, \mathrm{Re}\,W.\end{equation}
(To verify that the gradient of $h$ is as claimed, one needs the usual relation $g_{i\bar j}=-i\omega_{i\bar j}$ between
the metric and the Kahler form.)   So as usual the equations for supersymmetry can be written as flow equations for a Morse function $h$.

\subsubsection{Geometrical Interpretation}\label{zgo}

\def\O{{\mathcal O}}
\def\W{{\mathcal W}}
\def\Sh{{{\mathrm {Sym}}^*(\frak h)}}
\def\eD{{\eusm D}}
\def\LV{\mathpzc{L}_{V(\sigma)}}
\def\OS{\Omega^{0,*}(\frak h_\C)}
Before trying to interpret the above formulas in differential geometry, we need a little background.  First consider a general manifold $Y$
with action of a connected Lie group $H$.   Let $\Omega^*(Y)$ be the space of differential forms on $Y$, graded in the usual way by the degree
of a form.  Introduce a variable $\sigma$, taking values in the Lie algebra $\frak h$ of $H$, and considered to be of degree 2.  Let ${\mathrm {Sym}}^*(\frak h)$ be the algebra of polynomial functions of $\sigma$.  The following natural operator acts on $\W=\Omega^*(Y)\otimes \Sh$:
\begin{equation}\label{morzo}\eD_0=\d +\iota_{V(\sigma)},\end{equation}
where $\d$ is the usual exterior derivative acting on $\Omega^*(Y)$,  $V(\sigma)$ is the vector field on $M$ corresponding to $\sigma\in\frak h$,
and $\iota_{V(\sigma)}$ acts by contraction of a differential form with $V(\sigma)$. Note that each term in $\eD_0$ has degree 1 (the contraction
operation has degree $-1$ but $\sigma$ has degree 2).  Evidently,
\begin{equation}\label{vorzo}\eD_0^2= \mathpzc{L}_{V(\sigma)},\end{equation}
where
\begin{equation}\label{torzo}\mathpzc{L}_{V(\sigma)}=\d\iota_{V(\sigma)}+\iota_{V(\sigma)}\d \end{equation}
is the Lie derivative with respect to the vector field $V(\sigma)$, or in other words, the generator of the symmetry of $\W$
that corresponds to $\sigma$.  The formula (\ref{vorzo}) corresponds to eqn. (\ref{zonox}) in the field theory construction.
It means that if we restrict to  the $H$-invariant subspace of $\W$, which we denote as $\W^H$, then $\eD_0^2=0$.
So we can define the cohomology of $\eD_0$ acting in that subspace.

To explain just how $\eD_0$ is related to the construction in section \ref{construction}, let $u^I$ be local (real)  coordinates on $Y$ and set
$\psi^I=\d u^I$.  Then we compute
\begin{equation}\label{pexer}[\eD_0, u^I]=\psi^I,~~ \{\eD_0,\psi^I\}=V^I(\sigma),~~[\eD_0,\sigma]=0.\end{equation}
This is in perfect parallel with eqns. (\ref{flunky}) and (\ref{trumy}), if we understand the $u^I$ to be the coordinates $x^i,\bar {x^i}$ of $Z$.  And we can find another multiplet of the same kind
if we consider the $u^I$ to be $A_1$ and interpret $\lambda_1$ as $\d A_1$; then (\ref{pexer})
matches (\ref{clumsy}).  So if we set $Y=Z\times \frak h$ (where $\frak h$ is parametrized by $A_1$),
then $\eD_0$ is very similar to the topological supercharge $Q$ that  was considered in section
\ref{construction}.

There are two essential differences between the construction in section \ref{construction} and what we have said so far here.
First of all, rather than being $\frak h$-valued, $\sigma$ in section \ref{construction} took values in the complexification $\frak h_\C$ of this Lie algebra.   The
formulas of section \ref{construction} also involve the complex conjugate $\bar\sigma$ of
$\sigma$, and a fermionic field $\eta$ such that $[Q,\bar\sigma]=\eta$.  To include these
fields in the construction, we view $\frak h_\C$ as a complex manifold (which is isomorphic
to $\C^n$ where $n=\mathrm{dim}\,H$).  We view the components of $\sigma$ as linear holomorphic functions on $\frak h_\C$,  we interpret $\eta$ as $\d \bar\sigma$, and we replace $\Sh$
with the space $\OS$ of $(0,q)$ forms on $\frak h_\C$, for $0\leq q \leq n$.  (We grade $\OS$ by considering $\sigma$, $\bar\sigma$,
and $\eta$ to have degrees $2,$ $-2$, and $-1$.)  Finally we define
\begin{equation}\label{toron}\eD_1=\d\bar\sigma^a\frac{\partial}{\partial\bar\sigma^a}+[\sigma,\bar\sigma]^a\iota_{\d\bar \sigma^a}
\end{equation}
and replace $\eD_0$ by
\begin{equation}\label{oron}\eD=\eD_0+\eD_1,\end{equation}
acting on  $\Omega^*(Y)\otimes \OS$.  This incorporates the commutation relations for $\bar\sigma$ and $\eta$ and
obeys
$\eD^2=\mathpzc{L}_{V(\sigma)}$ (where $\mathpzc{L}_{V(\sigma)}$ is now the symmetry generator in the bigger space).
Furthermore, the  cohomology of $\eD$
acting on the $H$-invariant part of $\Omega^*(Y)\otimes \OS$ coincides with the cohomology of $\eD_0$ acting on the smaller
space $\W^H$.   To prove this, a first observation is that if $\eD\Psi=0$, then the term in $\Psi$ of highest degree in $\d\bar\sigma$
is annihilated by $\d_\sigma=d\bar\sigma^a\partial/\partial\bar\sigma^a$.  This is so because $\d_\sigma$ is the part of $\eD$ of
highest degree in $\d\bar\sigma$ (namely degree 1).  Note that $\d_\sigma^2=0$, and that the cohomology of $\d_\sigma$ vanishes for states
of strictly positive degree in $\d\bar\sigma$.  Now suppose that $\eD\Psi=0$ and $\Psi$ has degree $k>0$ in $\d\bar\sigma$.  Then
by a transformation $\Psi\to\Psi'=\Psi+\eD\Lambda$,  for some $\Lambda$, we can replace $\Psi$ by a state $\Psi'$ that is of degree at most $k-1$ in $\d\bar\sigma$;
this is possible because the cohomology of $\d_\sigma$ vanishes for states of positive
degree in $\d\bar\sigma$.  Repeating this process,
we reduce to the case $k=0$, in other words the case that $\Psi$ is independent of $\d\bar\sigma$.  Then the condition $\eD\Psi=0$
tells us that $\Psi$ is holomorphic in $\sigma$.  To have finite degree, $\Psi$ must be polynomial in $\sigma$.
But then $\Psi$ can be regarded as an element of $\W^H$, and the action of $\eD$ on $\Psi$ coincides with the action of $\eD_0$.

In constructing the operator $\eD$, we have incorporated all of the bosons $\sigma,\bar\sigma,
A_1,x,$ and $\bar x$ that should be present in describing physical states.  (We have also properly taken into account
the time component $A_0$ of the gauge fields: in a Hamiltonian framework, this field is set to zero but is associated to
 a Gauss
law constraint that physical states should be $H$-invariant, and we have imposed this constraint.)  The fermions that we have incorporated
explicitly so far are $\eta=\d\bar\sigma$, $\psi^i=\d x^i$, $\psi^{\bar j}=\d\bar{x^j}$, and
$\lambda_1=\d A_1$.  We have not yet discussed explicitly the other fermions
$\lambda_0$, $\chi^i$, $\chi^{\bar j}$, and $\rho$, but they are present implicitly as they
 are canonically conjugate to the fermion fields that have been discussed.  However, we will need to make an important
adjustment to incorporate those fields correctly.  Fermions in this second group
are contraction operators, so they have definite commutation relations with $\eD$, but those commutation relations are not the ones we want.
For example, $\chi^i$ is canonically conjugate to $\psi^{\bar j}$, which appears in $\eD$
only in the term $\psi^{\bar j}\partial/\partial \bar{x^j}$.  So $\{\eD,\chi^i\}=g^{i\bar j}\partial
/\partial\bar{x^j}=-\d x^i/\d s$, while we want $\{Q,\chi^i\}=-(\d x^i/\d s+g^{i\bar j}\partial h/\partial
\bar {x^j})$.  To achieve this result, and its analogs for the other fields,
we conjugate $\eD$ by $\exp(-h)$ and define
\begin{equation}\label{polyno}Q=\exp(h)\eD\exp(-h).  \end{equation}
This is our final result for the description of $Q$ in terms of differential geometry.

Because $Q$ and $\eD$ are conjugate, computing the cohomology of $Q$ is equivalent to computing that of $\eD$.
However, we have to be careful to describe the class of wavefunctions in which we wish to take the cohomology.   What we ultimately want to do with a wavefunction $\Psi$ that represents
a cohomology class of $\eD$ is to pair it as in eqn. (\ref{tonso}) below with another wavefunction
$\tilde \Psi$ that we will allow to have exponential growth for $h\to\infty$.  To ensure that
such integrals converge, we want $\Psi$ to be supported on a region on which $h$
is bounded above. (Without changing anything essential, we could instead require $\Psi$ to
decay faster than $\exp(-h)$ for $h\to+\infty$.)  We refer  to a region in which $h$ is bounded above as a region with $h<<\infty$,
and in the above definitions,  with $Y=Z\times\frak h$, we replace $\Omega^*(Y)$ by what we will call
$\Omega_{h<<\infty}^*(Y)$, the space of differential forms on $Y$ on whose
support $h$ is bounded above.

  First we will take a mathematical approach to the cohomology, and
then a physics-based approach.
For the mathematical approach, we begin with the fact that, by arguments already explained
(these arguments are unaffected by the support condition), the
desired cohomology of $\eD$ is  the same as the cohomology of $\eD_0$ acting
on $\W^H_{h<<\infty}$, the $H$-invariant part of $\Omega_{h<<\infty}^*(Y)\times \Sh$.
Ignoring the support condition for a moment, the cohomology of $\eD_0$ acting on $\Omega^*(Y)\otimes \Sh$ is known as the $H$-equivariant cohomology of $Y$.
In general, if $Y$ is any space with action of $H$, the complex $\Omega^*(Y)\otimes \Sh$ with action of the differential $\eD_0$
is known as the Cartan model of the $H$-equivariant cohomology of $Y$.  For example, if $H$ acts freely on $Y$, the equivariant cohomology of $Y$
is the same as the ordinary cohomology of the quotient $Y/H$.  At the opposite extreme, if $Y$ is a point, then $\eD_0=0$ and its cohomology
is just $\Sh^H$, the algebra of $H$-invariant polynomials on $\frak h$.

In the present case, the support condition means that we  actually want not the usual equivariant cohomology of $Y$, but the equivariant cohomology with values in differential forms supported at $h<<\infty$.   This  version of the equivariant cohomology is naturally computed by
using the $H$-invariant Morse function $h$ as an equivariant Morse function in the sense
of \cite{abott}.  The reason for this is that $h$ is bounded above on any cycle generated
by flows with respect to the Morse function $h$.

  First let us compute $\d h$:
\begin{align}\label{crito}\frac{\partial h}{\partial x^i}& =\frac{\partial W}{\partial x^i}+\omega_{\bar j i}V(A_1)^{\bar j}\\ \notag
                                      \frac{\partial h}{\partial A_1} & = \mu. \end{align}
Here $V(A_1)=A_1^aV_a$ is the vector field on $Z$ corresponding to $A_1\in\frak h$.  Contracting the first equation with $V(A_1)^i$ and using the fact that $V(A_1)^i\partial_i W=0$ (since $W$ is
$H$-invariant and holomorphic), we deduce that if $\partial h/\partial x^i=0$, then
$V(A_1)^iV(A_1)^{\bar j}\omega_{i\bar j}=0$, which implies (because of the relation of
$\omega$ to the Kahler metric $g$ and the positivity of $g$) that $V(A_1)=0$.  So the conditions for a critical point are
\begin{equation}\label{zonkeros}\frac{\partial W}{\partial x^i}=V(A_1)=\mu=0.\end{equation}

Since $W$ is $H$-invariant and holomorphic, it is invariant under the complexification $H_\C$ of the compact Lie group $H$.
The critical points of $W$ thus form $H_\C$ orbits.
We will assume that $W$ is sufficiently generic that it has only finitely many critical $H_\C$ orbits, which moreover are nondegenerate.
(We call a critical orbit nondegenerate if $W$ is nondegenerate in the directions transverse to the orbit.)   As explained in \cite{witten},
if $W$ is sufficiently generic, there are no flows between distinct critical orbits.   In this case,  $h$ is an equivariantly perfect Morse function in the sense of \cite{abott}, and
the desired equivariant cohomology is a direct sum of contributions from distinct critical orbits.

Among the critical orbits of $W$, only those that admit points with $\mu=0$ contribute to the equivariant cohomology of $Y$; this is clear
from the fact that $\mu=0$ is one of the conditions for a critical point of $h$.   We call a critical
orbit unstable if it contains no point with $\mu=0$. (In \cite{witten},  it was shown that for understanding Stokes phenomena, such unstable orbits can  consistently be excluded.)  If an orbit $\O_\C$ does contain points with $\mu=0$, then it is contractible to its subspace with $\mu=0$,
and this subspace is an orbit $\O$ of the compact group $H$.    In this case, $\O_\C$ is equivalent topologically to $T^*\O$.  The $H$ orbit
$\O$ is isomorphic to $H/P$ for some subgroup $P\subset \O$.  We call the orbit stable if $P$ is a finite group, and (strictly) semi-stable
otherwise.

If $\O$ is stable, then the equivariant cohomology of $\O$ is the ordinary cohomology of the quotient
$\O/H$.  That quotient is a point, so a stable orbit contributes a one-dimensional space to the equivariant cohomology of $Y=Z\times\frak h$. This contribution
appears in a degree that is equal to the Morse index of the Morse function $h$ at the orbit $\O$; for $h$ of the form we have
considered, this index is always one-half of the real dimension of $Z$.  This will be explained shortly.
In general, if $\O\cong H/P$, where $P$ is a Lie group of Lie algebra $\frak p$, then the equivariant
cohomology of $\O$  is the same as the equivariant cohomology of $P$ acting on a point.  That equivariant cohomology is the cohomology of $\eD_0=0$ acting on the $P$-invariant
part of $\mathrm{Sym}^*(\frak p)$; in other words, it is the ring of $P$-invariant polynomials on $\frak p$.   The contribution of such an orbit to the equivariant cohomology of $Y$ is this polynomial
ring, shifted in degree by the Morse index of $h$, which again is $\frac{1}{2}\mathrm{dim}\,Z$.
This means that such an orbit contributes to the equivariant cohomology of $Y$  a single  class of degree $\frac{1}{2}\mathrm{dim}\,Z$, and infinitely many classes of higher degree.

To show that the Morse index of $h$ at a nondegenerate critical orbit is always $\frac{1}{2}\mathrm{dim}\,Z$ (where the real dimension is understood),
independent of the dimension of the stabilizer $P$,
we first observe that the function $h_0=2\,\mathrm{Re}\,W$, understood as a Morse function on $Z$, has Morse index equal to one-half the real codimension of the critical orbit $\O_\C$, or $\frac{1}{2}\mathrm{dim}\,Z-\mathrm{dim}\,\O$.  Now the critical orbit $\O_\C\subset Z$ of the Morse
function $h_0$ pulls back, under the projection $Z\times\frak g\to Z$, to $\O_\C\times \frak g\subset Z\times\frak g$.
Let us regard $ h_1=A_1^a\mu_a$ as a function on $\O_\C\times\frak g$; it has a critical set, defined by the equations $\mu=V(A_1)=0$,
that is a $\frak p$ bundle over $\O$ (where $\O\subset\O_\C$ is defined by $\mu=0$).    The Morse index of $h_1$ is $\mathrm{\dim}\,\O$,
and the Morse index of $h=h_0+h_1$ is the sum of the Morse indices of $h_0$ and of $h_1$, or $\frac{1}{2}\mathrm{dim}\,Z$.

\def\mV{{\mathpzc{V}}}

Now let us explain these facts from a physical point of view.  The potential energy of the model is
\begin{align}\label{tusco}\mV&= |\d h|^2  +|V(\sigma)|^2 + |[A_1,\sigma]|^2+ |[\sigma,\bar\sigma]|^2\\
  &=2|\d W|^2+|\mu|^2+ |V(A_1)|^2+|V(\sigma)|^2+|[A_1,\sigma]|^2+ |[\sigma,\bar\sigma]|^2.\notag\end{align}
(In any expression such as $|\d h|^2$, the symbol $|~|$ refers to the norm with respect to the appropriate metric
on the space in question.)
In the first line, we write $\mV$ as the sum of $|\d h|^2$ (a contribution that is familiar in supersymmetric quantum mechanics without
gauge fields) and additional terms whose origins in dimensional reduction of gauge fields from four dimensions are hopefully
recognizable.  (These terms are the squares of the $\sigma$-dependent quantities that appear in (\ref{zsimga}) and
must vanish for unbroken supersymmetry.)  To go to the second line, we used the explicit form of $\d h$, and we also used again the identity
$V(A_1)^i\partial_iW=0$ to eliminate a cross term.   Not coincidentally, $\mV$ is invariant under $SO(3)$ rotations acting on
$A_1,A_2,A_3$ (where $\sigma=A_2-iA_3$).  The underlying $SO(3)$ symmetry of the reduction of the gauge theory from four dimensions to one
has been broken by the choice of $Q$, but this does not affect the formula for $\mV$.

In the classical approximation, a supersymmetric vacuum corresponds to a zero of $\mV$.  Such zeroes evidently correspond to
semistable critical orbits of $W$, that is to $H_\C$ orbits $\O_\C$  on which $\d W=0$ and we can also solve the equation $\mu=0$.
  Consider first the case of a free or semi-free critical orbit.
 In this case, since the vector fields $V_a$ are linearly independent along such an orbit, the terms $|V(A_1)|^2+|V(\sigma)|^2$ in the potential
energy give masses to all components of $A_1 $ and $\sigma$.  The $|\mu|^2$ term in the energy, when restricted to $\O_\C$,
 vanishes only on an $H$-orbit
$\O\subset\O_\C$ and gives a mass to those fluctuations away from $\O$ that lie in $\O_\C$; nondegeneracy of $W$ means that the
$|\d W|^2$ term gives mass to fluctuations normal to $\O_\C$.  So up to a gauge transformation (that is, up to the action of $H$),
a stable critical orbit contributes one classical vacuum, and in expanding around this vacuum,
all fluctuations are massive.   Because of the mass gap, there is no subtlety in quantization:  a stable critical
orbit contributes one supersymmetric state to the cohomology.  By much the same
computation as in supersymmetric quantum mechanics without gauge fields, the eigenvalue of $\cF$ for this state is the Morse
index of the function $h$ at the given critical orbit $\O_\C$.   For a Morse function of the type described in eqn. (\ref{gamag}), this
index, as explained above, is always one-half of the real dimension of $Z$.

More generally, we consider a semistable critical  orbit $\O_\C$ such that the locus $\O\subset \O_\C$ with $\mu=0$ breaks the gauge symmetry from $H$ to a subgroup
$P$ of positive dimension.  Fluctuations normal to $\O$ are still massive. All classical vacua corresponding to points in $\O$ are gauge-equivalent,
and by partially fixing the gauge symmetry, we can select a particular point $p\in \O$, leaving a residual unbroken gauge group $P$.   But $A_1$ and $\sigma$ now have flat directions. The conditions
$V(A_1)=V(\sigma)=0$ now tell us that $A_1$ and $\sigma$ must lie in the Lie algebra $\frak p$ of  $P$.
 The operator $Q$ whose cohomology we wish to compute reduces at low energies to the operator $\eD$ defined in eqn. (\ref{oron}),
except that the symmetry group $H$ is replaced by $P$ and the space $Y$ with action of $P$ is just the Lie algebra $\frak p$, parametrized
by $A_1$.  As has already been discussed, the cohomology of this operator is the ring of $P$-invariant polynomials on the complex
Lie algebra $\frak p_\C$, graded in such a way that the linear functions $\sigma$ on this Lie algebra are of degree 2.  ($A_1$ does not contribute to this cohomology, since it takes values
in an equivariantly contractible space.)  The contribution
of a semistable orbit $\O_\C$ to the space of supersymmetric states is hence the indicated ring of $P$-invariant polynomials, shifted in degree by the Morse
index of $h$, which again is one-half the real dimension of $Z$.

In the last paragraph, we did not worry about whether the cohomology classes of $Q$ that we described are square-integrable.
In fact, polynomial functions of $\frak p_\C$ are not square-integrable in the natural metric on that Lie algebra.  It seems quite likely,
though not completely clear, that the $Q$-cohomology classes associated to strictly semistable critical orbits (with $P$ of positive dimension) have no
square-integrable representatives.  In a somewhat similar compact situation, the cohomology of $Q$ would coincide with the space of normalizable states
annihilated by the Hamiltonian $H=QQ^\dagger+Q^\dagger Q$, but the case in which the classical potential admits noncompact flat directions at zero energy is quite different and it seems natural to conjecture that the
kernel of $H$, in the space of square-integrable wavefunctions, has a basis corresponding to the stable critical orbits only.

There is, however, a natural dual of the version of equivariant cohomology that we have considered so far;
the dual consists of the equivariant cohomology with compact support along the $A_1$ and $\sigma$ directions, but with no growth condition for $h\to+\infty$, and decaying exponentially
for $h\to-\infty$.  Cohomology classes
of this type can be described conveniently if one allows distributional wavefunctions.  The
most basic $Q$-invariant delta function is
\begin{equation}\label{zongo}\delta(A_1)\delta(\lambda_1)\delta(\sigma)\delta(\bar\sigma)\delta(\eta),\end{equation} which certainly has compact support in the $A_1$ and $\sigma$ directions.
(Here $\delta(\sigma)\delta(\bar\sigma)$ is an alternative notation for  $\delta(\mathrm{Re}\,\sigma)\delta(\mathrm{Im}\,\sigma)$.) This
delta function is of degree zero, since $\delta(\lambda_1)$ and $\delta(\eta)$ have opposite degrees.  A
more general $Q$-invariant distributional wavefunction is as follows.  Take any $H$-invariant polynomial
$\mathcal P$ on $\frak h^*$ (the dual of $\frak h$)  and consider the wavefunction
\begin{equation}\label{ozongo}{\mathcal {P}}(\partial/\partial\sigma)\,\delta(A_1)\delta(\lambda_1)\delta(\sigma)\delta(\bar\sigma)\delta(\eta)\end{equation}
which again is $Q$-invariant. It has degree $-2k$ if $\mathcal P$ is homogeneous of degree $k$.
To satisfy the decay condition along $Y$, we should multiply one of these delta functions by a differential
form on $Y=Z\times \frak h$ that vanishes rapidly for $h\to-\infty$.  For instance, if $Z$ is a Calabi-Yau manifold with $H$-invariant holomorphic
volume form $\Omega$, and $h=\mathrm{Re}\,W$, then an example of a class $\tilde\Psi$
in the dual version of the equivariant cohomology
is obtained by multiplying  $\exp(W)\Omega$ by the basic delta function given in (\ref{zongo}).  This product is annihilated by $\eD$ and has degree $\frac{1}{2}\mathrm{dim}\,Z$.

Now we can define a duality pairing.
If $\Psi$ is a class in the ``ordinary'' equivariant cohomology of $Z\times\frak h$  (with polynomial growth allowed in the $\sigma$ directions,
but supported at $h<<\infty$) , and $\tilde\Psi$ is a class in the
equivariant cohomology  of $Z\times \mathfrak h$ with compact support in the $A_1$ and $\sigma$ directions  (with no growth condition for $h\to\infty$, and decaying at $h\to-\infty$), then there is a natural pairing
\begin{equation}\label{tonso}(\tilde\Psi,\Psi)=\int_{Z\times \frak h\times \frak h_\C} \d A_1\,\d\lambda_1\,
\d\sigma\,\d\bar\sigma\,\d\eta\,\d  x\, \d\bar x\,\d\psi\,\d\bar\psi   ~\Psi\tilde\Psi.\end{equation} If $\Psi$ and $\tilde\Psi$
have definite degree, the sum of these degrees must equal the real dimension of $Z$ in order for this integral to be nonzero.   A typical such case is that $\Psi$ is a class associated
to a stable critical orbit, and $\tilde\Psi$ is as described in the last paragraph; both are of degree
$\frac{1}{2}\mathrm{dim}\,Z$.

Going back to the ``ordinary'' equivariant cohomology,
one  lesson from the above analysis is that in general, any cohomology class of $\eD$ can be represented by a wavefunction
$\Psi(x,\bar x, \psi,\sigma)$, independent of the other fields.  In addition, cohomology classes associated to semi-free orbits have
representatives that are independent of $\sigma$.

\subsubsection{Gauge-Invariant Integration Cycles}\label{gauginv}

Next we will analyze    the gauge-invariant version of something familiar from section \ref{sigmaone}: how to represent
ordinary integrals over cycles obtained by solving flow equations by one-dimensional path integrals.

 We consider
the path integral of the supersymmetric gauge theory that we have described above on a one-manifold $L$. We take $L$ to be the half-line
$s\leq 0$.

We define the boundary condition at $s=-\infty$ by saying that, for $s\to-\infty$, $x(s)$ approaches a specified $H$-orbit $\O\subset Z$ consisting of critical points of $W$
on which $\mu=0$.    The path integral on $L$ with this behavior imposed at $s\to-\infty$, when viewed
as a function of the boundary values of the fields at $s=0$, gives a state $\Psi$
in the equivariant cohomology of $Y=Z\times\frak h$.
This class can be localized on the $H$-invariant space  $\CC_\O
\subset Z\times \frak h$ that parametrizes solutions of the flow equations that start on $\O$.  (In particular, therefore, it is supported for $h<<\infty$.) The
codimension of $\CC_\O$ is $\frac{1}{2}\mathrm{dim}\,Z$, the index of the Morse function $h$.

The description we have given of the behavior at $s=-\infty$ is sufficiently precise if $\O$ is a stable orbit, for then the
fields $A_1$ and $\sigma$ are massive in expanding around $\O$ and naturally vanish at $s=-\infty$.  More fundamentally,
there is no essential  ambiguity about the initial conditions because  a critical orbit of this
type is associated to an equivariant cohomology class that is unique, up to a constant multiple.  The path integral actually
fixes the multiple in a natural way, generating, just as in section \ref{sigmaone}, the Poincar\'e dual of $\CC_\O$ rather
than a multiple of this.

If, however, $\O$ is strictly semi-stable, then the potential energy for $A_1$ and $\sigma$
has flat directions in expanding around $\O$.  The input conditions at $s=-\infty$
require choosing a $Q$-invariant wavefunction on the space of flat directions.  This wavefunction
is not supposed to have distributional support at $\sigma=0$ (this would land us in the wrong
sort of equivariant cohomology), but rather will have polynomial dependence on $\sigma$.
The initial conditions at $s=-\infty$ depend on the choice of such a wavefunction.  The state
$\Psi$ obtained from the path integral on $L$ is much more complicated in the strictly semi-stable
case, because its support has more than one branch.  This is because of eqn. (\ref{zsimga}),
according to which a $Q$-invariant field configuration on $L$ must lie on the locus in $Z\times\frak h$ that is invariant under the symmetry generated by $\sigma$.  This condition is trivial
if $\sigma=0$, in which case all flows are allowed, but becomes non-trivial as soon as $\sigma$ is
non-zero.    The branches of the space of $Q$-invariant field configurations are classified
by the conjugacy class of the  subgroup of $P$ that is left unbroken by $\sigma$.

Having picked a critical orbit $\O$ (and some additional data if $\O$ is only semi-stable) to fix the initial conditions at $s=-\infty$,  there are, as in section \ref{sigmaone}, several slightly different path
integrals that we can do on $L$.  If we take the action of the sigma-model to be the physical action of the supersymmetric gauge theory --
a gauge-extended version of (\ref{tomxo}) --
then the state defined by the path integral at $s=0$ is annihilated by $Q=\exp(h/\epsilon)\eD\exp(-h/\epsilon)$.  Just as in eqn. (\ref{omxo}),
it  is slightly more convenient
to add to the action $I$ an exact form
\begin{equation}\label{polno}I\to I+\frac{1}{\epsilon}\int_L\d s\,\frac{\d h}{\d s}.\end{equation}
 The modified action can be written more naturally in terms of the supersymmetric flow equations,
as in eqn. (\ref{omxo}).  The modification has the effect of multiplying the quantum state at $s=0$ by $\exp(-h/\epsilon)$, so that it is annihilated
by $\eD$ rather than $Q$.  We will use this approach.

The path integral on $L$, as a function of the values of the fields at $s=0$, will determine
a state $\Psi$ in the equivariant cohomology.  This state will be localized on the cycle $\CC_\O$
in $Z\times \frak h$ obtained by solving the flow equations.
However, using the path integral to compute a state associated to a critical orbit $\O$ is not really what we want to do. Our real aim is  to
represent an ordinary integral over a middle-dimensional cycle in $Z$ as a one-dimensional path integral,
just as we did in section \ref{sigmaone} in the absence of gauge-invariance.

As in eqn. (\ref{zonkers}) and the example given after eqn. (\ref{ozongo}), we assume that $Z$ is a Calabi-Yau manifold with holomorphic volume-form $\Omega$.  We also pick on $Z$
an $H$-invariant holomorphic function $S$.  The path integral that we will describe is always convergent if $S$ coincides with the superpotential
$W$ of the sigma-model.  In general, we want to pick $S$ to be close enough to $W$ so that the path integral will be convergent.  (For example, if $Z=\C^n$
and $W$ is a polynomial whose terms of highest degree are sufficiently generic, then as in eqn. (\ref{polz}),  it will suffice if $S$ differs
from $W$ by subleading terms.)  The ordinary integral that we want to represent by a one-dimensional path integral is
\begin{equation}\label{horsefly}\int_{\CC_\O\cap\{A_1=0\}}\Omega\,\exp(S).  \end{equation}
The meaning of this is as follows.   The cycle $\CC_\O\subset Z\times\frak h $  is of codimension
$\frac{1}{2}\mathrm{dim}\,Z$, and hence of dimension $\frac{1}{2}\mathrm{dim}\,Z+\mathrm{dim}\,
H$.  By intersecting this cycle with the locus $A_1=0$, whose codimension is $\mathrm{dim}\,H$, we get the cycle $\CC_\O\cap\{A_1=0\}
\subset Z$.  The dimension of this cycle is $\frac{1}{2}\,\mathrm{dim}\,Z$, the correct dimension
for the integral (\ref{horsefly}) to make sense.   The restriction to $A_1=0$ will come from the
boundary condition at $s=0$, as discussed presently.

In fact, $\Gamma_\O=\CC_\O\cap\{A_1=0\}$ is an $H$-invariant, middle-dimensional cycle in $Z$ which
was introduced in \cite{witten} in a more naive fashion.  There, it was characterized as the
cycle that parametrizes points in $Z$ that can be reached at $s=0$ by solving the flow equations
for the Morse function $h_0=2\,\mathrm{Re}\,W$ on $Z$, starting at $s=-\infty$ on the  $H$-orbit
$\O$.  In the present paper, the gauge theory machinery has led us to consider not the Morse
function $h_0$ on $Z$, but the more sophisticated Morse function $h=h_0+A_1^a\mu_a$ on
$Z\times\frak h$. However, upon setting $A_1=0$, we reduce to the cycle in $Z$ that was
analyzed in \cite{witten}.

We now want to show that it is possible to pick a supersymmetric boundary condition at $s=0$ so that the path integral on the half-line
will indeed compute the integral (\ref{horsefly}).  Just as in section \ref{sigmaone}, part of the construction will be an insertion at $s=0$ of the
operator
\begin{equation}\label{thop} \exp(S(x(0)))\,\Omega_{i_1i_2\dots i_n}\psi^{i_1}(0)\psi^{i_2}(0)
\dots\psi^{i_n}(0),\end{equation}
and this implies a boundary condition
\begin{equation}\label{port}\left.\psi^{(1,0)}\right|_{s=0}=0.\end{equation}
The boundary conditions on other fermions in the chiral multiplets are the same as before,
and for similar reasons:
we set $\chi_{(0,1)}=0$ at $s=0$ and leave other components of $\psi,\chi$ unrestricted.

What is new relative to the previous analysis is that we have to find the right boundary conditions for the vector multiplet.
One essential point here is that since $\{Q,\psi^i\}=V^i(\sigma)$, where $V^i(\sigma)$ is generically non-zero,
the boundary insertion (\ref{thop}) is $Q$-invariant only
if we set $\sigma=0$ at $s=0$.  So also $\bar\sigma=0$ at $s=0$, and since $\{Q,\bar\sigma\}=\eta$, this is only consistent
if $\eta$ also vanishes at $s=0$.

There is another way to motivate the boundary condition $\sigma(0)=0$.  We will describe a formalism that works for semistable
as well as stable critical orbits $\O$.  In the semistable case, the conditions at $s=-\infty$ allow $\sigma$ to be unbounded,
so the boundary condition $\sigma(0)=0$ is needed for convergence of the path integral.

The same reasoning motivates us to pick a boundary condition $A_1(0)=0$, which in any case
is needed if we hope to arrive at the integral (\ref{horsefly}).  For consistency, as $[Q,A_1]=\lambda_1$,
we must also set $\lambda_1(0)=0$.

At this point, we have set to zero the boundary values of half the fermions in the gauge multiplet, namely $\eta$ and $\lambda_1$,
so we must leave unconstrained the other fermions in that multiplet, namely $\rho$ and $\lambda_0$.   This boundary condition
is $Q$-invariant, for reasons similar to what was explained at the end of section \ref{sigmaone}.  As there, some contributions to $Q$ vanish because of the fermionic boundary conditions.
For example, one contribution to $Q$ from the vector multiplet is $\sum_a\,\eta_a\,\d \sigma^a/\d s$, which upon quantization becomes $-\sum_a\,\eta_a\,\partial/\partial\bar\sigma^a$.
This vanishes at $s=0$, because $\eta $ does.

{}From the point of view of classical differential geometry, what the path integral with these
boundary conditions computes is the pairing (\ref{tonso}) between a state $\Psi$  in
the  equivariant cohomology of $Z$ supported at $h<<\infty$, defined by the boundary condition at $s=-\infty$, and a state $\tilde\Psi$ in the equivariant
cohomology with compact support in the $A_1$ and $\sigma$ directions, defined by the boundary condition at $s=0$.  (These two dual forms of equivariant cohomology were defined more
precisely in section \ref{zgo}.)
The state $\tilde\Psi$ is given by (\ref{thop}) multiplied by $\delta(\chi_{(0,1)})$ and  the delta function in (\ref{zongo}):
\begin{equation}\label{gonf}\tilde\Psi=  \exp(S(x(0))\,\delta(\psi^{(1,0)})\delta(\chi_{(0,1)})\delta(A_1)\delta(\lambda_1)\delta(\sigma)\delta(\bar\sigma)\delta(\eta).\end{equation}
(Here we have written $\Omega_{i_1i_2\dots i_n}\psi^{i_1}(0)\psi^{i_2}(0)
\dots\psi^{i_n}(0) $ as $\delta(\psi^{(1,0)})$.) The bosonic boundary condition (\ref{peroxide}) is needed in verifying $Q$-invariance
of $\delta(\chi_{(0,1)})$.

The pairing $(\tilde\Psi,\Psi)$ defined by the path integral (or in differential geometry by
eqn. (\ref{tonso})) reduces to the desired integral (\ref{horsefly}).  The localization on
$\CC_\O$ comes from $\Psi$, and the localization at $A_1=0$ comes from $\tilde\Psi$.  The naturalness of this pairing gives another motivation for the boundary
condition that we have placed on the vector multiplet.

The reason that we have given such a detailed explanation of this gauge-invariant generalization of the result of section \ref{sigmaone}
is that it has an interesting application to three-dimensional Chern-Simons gauge theory, to which we turn next. We will be able to
express the path integral of Chern-Simons gauge theory in three dimensions in terms of a path integral of $\N=4$ super Yang-Mills theory
in four dimensions.  This in turn has an interesting application: it gives a new
perspective on the relation \cite{GSV} between Khovanov homology \cite{K} and spaces of BPS states.  We leave the details
for \cite{wittenfive} and remark here only that once one re-expresses the Chern-Simons path integral in terms of $\N=4$ super Yang-Mills
theory, one can study it using standard techniques of electric-magnetic duality and related string theory dualities.

\subsection{Application To Chern-Simons Gauge Theory}\label{applics}

\subsubsection{The Chern-Simons Form As A Superpotential}\label{superform}
\def\B{{\mathcal B}}
\def\BB{{\mathfrak B }}
We will apply what we have learned in section \ref{supergauge} to the following special situation.  We compactify $\N=4$ super Yang-Mills
theory from four dimensions to one on a three-manifold $W$.  Thus, we formulate the theory on the four-manifold $M=\RR\times W$,
making an $R$-symmetry twist so that some
supersymmetry is preserved.  The twist is accomplished by embedding the holonomy
group $SO(3)$ of $W$ in the $SO(6)$ $R$-symmetry group in the obvious way (the vector representation of  $SO(6)$ transforms as the vector representation of
$SO(3)$ plus three singlets) .  Of the six adjoint-valued scalar fields $\Phi$ of the $\N=4$
theory, the twist converts three, which we will call
$\phi$, into an adjoint-valued one-form on $M$, while the other three, which we will call $\tilde\phi$, remain as scalar fields.
The twist ensures that the compactified theory is independent of the choice of a metric $g$ on $W$, but we do need to make a choice
to define $\N=4$ super Yang-Mills theory and to make the following constructions.

For gauge group $G=U(n)$, the twisted theory can be realized
in string theory.  One considers Type IIB superstring theory on $\RR^4\times T^*W$ (here $T^*W$ may
be replaced by any Calabi-Yau three-fold $X$ that admits $W$ as a special Lagrangian submanifold), with $n$ D3-branes wrapped on
$M=\RR\times W\subset \RR^4\times T^*W$.   Then, by arguments similar to those in \cite{BSV}, upon scaling up the metric of $T^*W$
relative to the string scale, the low energy theory along $M$
is the twisted theory described in the last paragraph. To get the twisted $\N=4$ theory on a half-space (as we will want presently),
one can let the $n$ D3-branes end on an NS5-brane that is supported on $\RR^3\times W$.
  For $G=SO(n)$ or $Sp(n)$, a similar construction is possible using an orientifold
three-plane supported on $\RR\times W$.
These brane constructions will be used in \cite{wittenfive}, but are not needed for our present purposes.

In the reduction from four dimensions to one, the four-dimensional gauge field $A$ splits up as a one-dimensional gauge field
$A_0\,\d s$ (we parametrize the first factor of $M=\RR\times W$ by $s$) and a gauge field tangent to $W$ that we will call $B$.
The twisted compactification on $W$ leaves four unbroken supersymmetries and the resulting theory can be interpreted as  a one-dimensional supersymmetric
gauge theory of precisely the sort studied in section \ref{construction}, but with an infinite-dimensional gauge group.
If we assume that the underlying four-dimensional theory is formulated on a trivial $G$-bundle, then the gauge group in the reduced
one-dimensional theory is
$H=\mathrm{Maps}(W,G)$, the space of maps from $W$ to the finite-dimensional gauge group $G$.  Somewhat more intrinsically, if the
compactified theory on $W$ is defined in terms of connections on a $G$-bundle $E\to W$, then $H=\mathrm{Aut}(E)$ is the group
of bundle automorphisms of $E$.

The bosonic fields of the vector multiplet of the $H$ gauge symmetry are the untwisted fields $A_0$ and $\tilde\phi$.   (Once we make
a particular choice of a topological supercharge $Q$, $\tilde\phi$ splits up into the fields called $A_1$, $\sigma$, and $\bar\sigma$ in
section \ref{construction}.)  The other bosonic fields are usefully combined to a complex-valued field $\B=B+i\phi$ on $W$,
which we can think of as a  connection on a bundle $E_\C\to W$ whose structure group is the complexification $G_\C$ of $G$. (If $E$ was
understood as a principal $G$-bundle, then $E_\C$ is its complexification,which is a principal $G_\C$-bundle.) The fields $\B$ are the bosonic components of chiral supermultiplets that are acted on by the group $H$ and in fact by its complexification $H_\C$, consisting of $G_\C$-valued
gauge transformations.
Let $\BB$ be the space of all possible $\B$ fields.
Once we endow $W$ with a Riemannian metric $g$, which we write in local coordinates as $g=\sum_{a,b=1}^3g_{ab}\d y^a\d y^b$, and pick a gauge
coupling constant $e$,
$\BB$  acquires a Kahler metric
\begin{equation}\label{zonk}\d s^2=-\frac{1}{2e^2}\int_W\d^3y\sqrt g g^{ab}\Tr\,\delta \B_a\otimes \delta \bar\B_b \end{equation}
and an associated symplectic form that is readily written in terms of the real fields $B,\phi$:
\begin{equation}\label{ronk}\omega=-\frac{1}{e^2}\int_W\d^3y\sqrt{g}g^{ab}\Tr\,\delta B_a\wedge\delta \phi_b.\end{equation}
Relative to this symplectic form, the moment map for the action of $H$ on $\BB$ is
\begin{equation}\label{dorf}\mu=-\frac{1}{e^2}\star\d_B\star\phi=-\frac{1}{e^2}D_a\phi^a,\end{equation}
where $\star$ is the Hodge star operator on $W$, $\d_B=\d+[B,\cdot]$ is the gauge-covariant exterior derivative defined with the real
connection $B$, and similarly $D_i$ is the covariant derivative with respect to $B$.
A gauge theory with four supercharges in general has also a superpotential, which is a gauge-invariant holomorphic function $\eusm W$
on the space that parametrizes the chiral superfields -- in our case, this space is $\BB$.  To be more precise, $\eusm W$ is not
quite uniquely determined as a complex-valued function.  It is defined only up to an additive constant, because only derivatives of
$\eusm W$ appear in the action, and only up to an arbitrary phase factor, which is possible because the mapping from the underling
$\N=4$ theory in four dimensions to a low energy description with vector and chiral multiplets (and four supercharges) is only
uniquely determined up to an $R$-symmetry rotation, which can change the phase of $\eusm W$.  In our case, the superpotential is the integral of the complex Chern-Simons form
\begin{equation}\label{superz}\eusm W=-\frac{\exp(i\alpha)}{e^2}\int_W\d^3y\,\epsilon^{abc}\Tr\,\left(\B_a\partial_b\B_c+\frac{2}{3}\B_a\B_b\B_c\right).
\end{equation}  where $\exp(i\alpha)$ is the arbitrary phase.

To verify that this is the right superpotential, one may for example start with the potential energy of the $\N=4$ theory in four
dimensions and express it in the language of the reduced one-dimensional theory.  In  ten-dimensional notation,
the bosonic part of the action of $\N=4$ super Yang-Mills theory is $-(1/2e^2)\int \d^4y\sum_{I,J=0}^9\Tr\,[D_I,D_J]^2$, where
$D_I$ is a covariant derivative (if $I=0,\dots,3$) or the commutator with a scalar field (if $I=4,\dots,9$).  The part of this
action that involves only the chiral superfields is the time integral of  a potential energy function $\mathcal V$ on $\BB$; this function  can be written (after some integration by parts) as $\mathcal V=|\d\eusm W|^2+|\mu|^2$, using the above
formulas for $\eusm W$, $\mu$, and the metric on $\BB$.    This is the expected form, independent of the phase $\alpha$, and gives one way to verify the claimed formula for $\eusm W$.

The fact that  the Chern-Simons function arises in this way as a superpotential has been important in recent work
\cite{DGH} (this paper actually involves reduction from five to two dimensions rather than from four to one), and related observations have been made, for example, in \cite{issues}.

In section 4.1 of \cite{witten}, the gradient flow equations were studied for the flow on $\BB$ with the Morse function
\begin{equation}\label{uperz} h = 2\,\mathrm{Re}\,\mathrm{\eusm W}.\end{equation}
Of course, these equations depend on the phase $\alpha$.    It was shown that these flow equations have a natural interpretation in
$\N=4$ super Yang-Mills theory.  They express invariance under the topological supercharge $Q$ of twisted $\N=4$ super Yang-Mills
theory with the twist related to geometric Langlands that was studied in \cite{KW}.   The supercharge $Q$ depends on a twisting
parameter, which was called $t$ in \cite{KW} and $u$ in \cite{witten}.   According to eqn. (4.9) of \cite{witten}, the twisting parameter
 is related to the phase $\alpha$ of $\eusm W$ by
\begin{equation}\label{umiko} t=\frac{1-\cos\alpha}{\sin\alpha}.\end{equation}

\subsubsection{The Chern-Simons Path Integral From Four Dimensions}\label{omoxo}

By simply adapting what we said in section \ref{gauginv}, we can now represent a Chern-Simons path integral in three dimensions -- over
a suitable integration cycle -- in terms of a four-dimensional path integral.

\def\top{{\mathrm{top}}}
We  formulate the $\N=4$ theory on $M=\RR_+\times W$, where $\RR_+$ is the half-line $s\geq 0$. As long as the gauge
theory theta-angle vanishes (we defer the generalization to \cite{wittenfive}),  the theory can be regarded as an infinite-dimensional version of the one-dimensional gauge-invariant theory
studied in section \ref{gauginv}. In what follows, we simply imitate that analysis.  So, as $s\to\infty$, we require
$\B$ to approach a stable critical orbit of the superpotential.  (As we discuss in detail presently, these orbits correspond to complex
flat connections obeying a mild restriction.)  At $s=0$, we impose the boundary conditions that were used in section \ref{gauginv} to interpret an ordinary integral
with gauge symmetry in terms of a path integral.  In particular, to  maintain $Q$-invariance at $s=0$, we  make the modification in the action that
was described in eqn. (\ref{polno}), replacing the usual $\N=4$ action $I_{\N=4}$ by a $Q$-invariant action $I_{\N=4}^\top=I_{\N=4}
+h(0)$.  Here $h(0)$ is simply $h$ evaluated at $s=0$.  (It does not matter whether we subtract from the action the value of $h$ at $s=-\infty$,
because the boundary condition there fixes $h(s=-\infty)$ to be a constant, and in particular $Q$-invariant. We omit here a factor $1/\epsilon$
multiplying $h$ in (\ref{polno}); it has already been incorporated as the factor of $1/e^2$ in the definition of $\eusm W$.)

Just as in section \ref{gauginv},  we can now use the path integral of the $\N=4$ theory on $M$ to
 compute the integral of $\exp(\eusm W)$  over a real cycle $\Gamma_\O\subset\BB$:
\begin{equation}\label{zonko}\int DA\,D\Phi\,D\lambda\exp(-I^{\top}_{\N=4})\left.\exp(\eusm W)\right|_{s=0}=\int_{\Gamma_\O} D\B\,\exp( \eusm W).\end{equation}
On the left of eqn. (\ref{zonko}), $A$, $\Phi$, and $\lambda$ are the bose and fermi fields of $\N=4$ super Yang-Mills theory, and $\eusm W$ is to be evaluated
at $s=0$.  The boundary conditions on fermions at $s=0$ are those of section \ref{gauginv} and are discussed in detail, in the present
infinite-dimensional context, in an appendix. On the right hand side, the integral is over a middle-dimensional cycle $\Gamma_\O$ in the space of complex-valued connections $\B$.
This cycle, which corresponds to $\CC_\O\cap\{A_1=0\}$ in (\ref{horsefly}),
 is found by solving the gradient flow equations with respect to the Morse function $h=2\,\mathrm{Re}\,\eusm W$, starting  at $s=\infty$ from the critical orbit $\O$ that is used
 to define the boundary conditions on the left hand side of (\ref{zonko}).  Because $\eusm W$ is a multiple of the Chern-Simons action of the complex-valued
connection $\B$,  the integral on the right hand side of (\ref{zonko}) is simply the Chern-Simons path integral, carried out over an
integration cycle that differs from the usual one. (The usual integration cycle is defined by setting $\phi=0$ or in other words
by taking $\B$ to be real.)  As explained in \cite{witten},
the usual quantization of the Chern-Simons coupling constant, which is required to make sense of the integral over the usual integration
cycle, is not necessary for defining the integral over a cycle obtained by gradient flow,
and indeed this quantization is not satisfied in the present context, as the superpotential contains an arbitrary factor $\exp(i\alpha)/e^2$.

An important generalization of (\ref{zonko}) -- discussed at several points in this paper -- involves
including on both sides an additional factor $\eusm T$, where $\eusm T$ is a holomorphic function on $\BB$
 that grows too slowly at infinity to affect the convergence of the path integral.  In the present context,
there is a natural class of such functions.  Let $K$ be a knot (an embedded oriented circle) in $W$ and let $R$ be an irreducible  representation of $G$, which we analytically
continue  to a holomorphic representation of $G_\C$.  Then we define the holonomy function or Wilson loop operator
\begin{equation}\label{zoom}W_{K,R}(\B)=\Tr_R\,P\exp\oint_K\B.\end{equation}
$W_{K,R}(\B)$ is, roughly speaking, the exponential of a linear function on $\BB$, while the superpotential $\eusm W$ is cubic.  So inclusion of a factor
$W_{K,R}(\B)$ does not affect the convergence of the path integral.  In the $\N=4$ theory, to preserve the topological symmetry, this factor must be inserted at $s=0$ (where the boundary condition $\psi^{(1,0)}=0$ ensures $Q$-invariance of any holomorphic function).  So we get
a generalization of (\ref{zonko}):
\begin{equation}\label{izonko}\int DA\,D\Phi\,D\lambda\exp(-I^{\top}_{\N=4})\,\left.\bigl(\exp(\eusm W)W_{K,R}(\B)\bigr)\right|_{s=0}=
\int_{\Gamma_\O} D\B\,\exp( \eusm W)\,W_{K,R}(\B).\end{equation}
This has an obvious generalization involving a link -- a union of disjoint  embedded oriented circles $K_i\subset W$ -- rather than a knot.
Labeling the $K_i$ by irreducible representations $R_i$, we have
\begin{equation}\label{izonkox}\int DA\,D\Phi\,\D\lambda\exp(-I^\top_{\N=4})\,\left.\bigl(\exp(\eusm W)\prod_iW_{K_i,R_i}(\B)\bigr)\right|_{s=0}=
\int_{\Gamma_\O} D\B\,\exp( \eusm W)\,\prod_iW_{K_i,R_i}(\B).\end{equation}   There is a further extension in which links
are replaced by a certain class of labeled graphs,
as discussed for example in \cite{graphs}.
Another extension involving the theta-angle of the four-dimensional gauge theory is important in relation to Khovanov homology and
will be described in \cite{wittenfive}.

Now let us discuss a few more details concerning these formulas.
Since the superpotential is the integral of the complex Chern-Simons form,
a critical point is a complex-valued flat connection, characterized by the vanishing of the curvature
\begin{equation}\mathcal F=\d\B+\B\wedge \B.\end{equation}
The monodromies of such a flat connection give a representation of the fundamental group of $W$
into the finite-dimensional gauge group $G_\C$, or in other words a homomorphism $\rho:\pi_1(W)\to G_\C$.  The flat connections associated
with a given monodromy form
an orbit for the group $H_\C$ of $G_\C$-valued gauge transformations. A critical orbit is nondegenerate if the corresponding representation
of the fundamental group has no first-order deformations.   In terminology introduced in section \ref{zgo},
a  critical orbit is called semistable if it admits points at which the
moment map $\mu$ (defined in eqn. (\ref{dorf})) vanishes.  It is called stable if in addition the subgroup of $G_\C$ that commutes with the
monodromies of the flat connection is a finite group.   A critical orbit containing no zero of $\mu$ is unstable.  As shown in \cite{corlette},
the unstable critical orbits are exactly those that correspond to strictly triangular representations of the fundamental group.

If $W$ is compact, the space $M=\RR_+\times W$ is macroscopically one-dimensional.  In this case, if there are flat directions in the bosonic potential
of the theory, their quantum fluctuations are important and need to be treated carefully in defining the path integral. This tends
to cause complications in applications of (\ref{izonkox}).   Such infrared questions
are entirely absent if the boundary condition in the path integral at $s=-\infty$ is set by a stable and nondegenerate flat connection.  For a stable
flat connection corresponds -- just as in the  context of section \ref{gauginv} with finitely many degrees of freedom -- to a massive vacuum of the one-dimensional theory that arises by compactification on $W$.

There is another way to avoid infrared subtleties: one can take $W$ to be noncompact.  The simplest choice is $W=\RR^3$.
In field theory on $\RR^3$ (or $\RR^3\times\RR_+$) it is natural to consider only fields and gauge transformations that are trivial  at infinity.  So
we define $H$ to be the group of $G$-valued gauge transformations on $\RR^3$ that are 1 at infinity,\footnote{Since the fields $\tilde\phi$
-- or $A_1$ and $\sigma$ -- take values in the adjoint representation of $H$, this definition of $H$ implies that those fields vanish at infinity
on $\RR^3$.} and we require the complex-valued connection $\B$
to vanish at infinity on $\RR^3$. $H$ acts on the space $\BB$ of such connections, and we  run the above theory with this choice of $H$ and $\BB$.   Since the fundamental group of $\RR^3$ is trivial, the only
critical orbit is the one that contains the trivial connection.  This critical orbit is nondegenerate and stable.  Nondegeneracy means that
after gauge fixing,
in expanding around the trivial flat connection, there are no zero modes that vanish at infinity.    Stability expresses the fact
that  $H$ -- and  its complexification $H_\C$ -- act freely on the orbit in $\BB$ that contains the trivial flat connection.
Indeed, $H$ was defined as  a group of gauge transformations that are 1 at infinity; no non-trivial  gauge transformation obeying this
condition  leaves fixed the trivial flat connection.

For $W=\RR^3$, the topological field theory under discussion here has not much to say in the absence of knots, and the formula (\ref{zonko}) is not terribly interesting -- both sides equal 1.  However,  knots in $\RR^3$ are interesting and (after being extended to include the gauge theory theta-angle) the formulas
(\ref{izonko}), (\ref{izonkox}) will be the starting points for the study
of Khovanov homology in \cite{wittenfive}.

A generalization, with  infrared divergences avoided in a similar way, is to let $W_0$ be a rational homology three-sphere with $p$ a point
in $W_0$ and to take
$W=W_0\backslash p$ (that is, $W$ is $W_0$ with $p$ omitted).  We take on $W$ a metric in which $p$ is projected to infinity, so that the
region near infinity in $W$ looks like the region near infinity in $\RR^3$.  We again define $H$ to consist of gauge transformations that
are 1 at infinity, and $\BB$ to parametrize complex-valued connections that vanish at infinity.  The orbit containing the trivial flat connection
is nondegenerate and stable.  For $G=SU(2)$, all critical orbits are stable (but not always nondegenerate); this is not necessarily so for $G$
of higher rank.

\subsubsection{Comparison To A Sigma-Model}\label{compsig}

A very illuminating example of the gauge theory construction that we have just considered arises if we take $W=S^1\times C$, where $C$ is
a Riemann surface.

We take the metric on $C$ to be much  larger than that on $S^1$, and then instead of simply compactifying on $W$ from four dimensions
to one, we can think in terms of compactification on $C$ from four to two dimensions, the two-manifold being in this
case $\RR_+\times S^1$.  Compactification of $\N=4$ super Yang-Mills
theory on a Riemann surface -- with precisely the same topological twist as in our present discussion -- was analyzed in \cite{KW},
following \cite{BV,MH}.  The low energy theory is a sigma-model with target space $\M_H(G,C)$, the moduli space of Higgs bundles
on $C$ with structure group $G$.  So the left hand side of eqn. (\ref{zonko}) reduces to a sigma-model path integral on $\RR_+\times S^1$,
with target $\M_H(G,C)$.

What about the right hand side of eqn. (\ref{zonko})?  Chern-Simons theory in three dimensions with gauge group $G$, when compactified on a Riemann
surface $C$, reduces to quantum mechanics with phase space the moduli space $\M(G,C)$ of flat (or holomorphic) $G$-bundles over $C$.
To define an exotic integration cycle in this quantum mechanics, we must first complexify the target space and then pick the integration
cycle.  The complexification of $\M(G,C)$ is precisely $\M_H(G,C)$,
and so the right hand side of (\ref{zonko}) is the quantum path integral of $\M(G,C)$ with
an exotic integration cycle.

Thus, the compactification of our present discussion on $C$ gives a special case of the construction in section \ref{quantint}.

\subsubsection{Elliptic Boundary Conditions}\label{elbo}

To arrive at (\ref{zonko}) and its generalizations, we took a finite-dimensional construction, described in section \ref{gauginv},
and adapted it to a case that the gauge group $H$ and the space $\BB$ that parametrizes the chiral superfields are infinite-dimensional.
In eqn. (\ref{wondo}), we have seen how such a generalization to infinite dimensions that may have sounded plausible can fail: it
implied a boundary condition that is not elliptic.   The formula (\ref{zonko}) is not afflicted with this sort of problem, basically
because the function $\eusm W$ that appears in the exponent has derivatives in the $W$ directions, which moreover are sufficiently
generic.  In this respect, (\ref{zonko}) is a generalization not of (\ref{wondo}), but of the sigma-model analysis of section
\ref{quantint} (to which it can be dimensionally reduced in some circumstances, as we noted in section \ref{compsig}).

In the appendix, we show in detail that the boundary condition that is implicit on the left hand side of eqn. (\ref{zonko}) is elliptic.
This  essentially means that the  expressions that one would encounter in expanding the left hand side of (\ref{zonko})
in perturbation theory have properties similar to what one would find in the finite dimensional construction of section
\ref{gauginv}, strongly suggesting that it would be possible to explicitly demonstrate the validity
of (\ref{zonko}) in perturbation theory.  (As always, a rigorous nonperturbative proof would be much harder.)

\subsection{Quantization With Constraints}\label{constraints}

Here we will very briefly consider the question of how one would describe integration
cycles in the Feynman integral of a quantum mechanics problem with first class constraints.

We recall the setting with which we began in section \ref{prelims}. $\M$ is a classical phase
space of dimension $2n$ with a real symplectic form that locally can be written
\begin{equation}\label{zomo}f=\sum_{r=1}^n\d p_r\wedge \d q^r.\end{equation}
The first class constraints are functions $\mu_a$ that, via their Poisson brackets,
\begin{equation}\label{ommo}\{\mu_a,\mu_b\}=f_{ab}^c\mu_c\end{equation}
generate the action on $\M$ of a Lie group $G$.   Classically, imposing the constraints means
restricting to the locus with $\mu_a=0$, $a=1,\dots,\mathrm{dim}\,G$, and dividing by $G$.  Quantum mechanically,
it means that one quantizes and restricts to the $G$-invariant subspace of the physical Hilbert space.
We want to approach this process from the point of view of an analytically continued path integral.

The constrained system can be described classically by the action
\begin{equation}\label{ompobo}S=\oint\left(p_r\d q^r-\phi^a\mu_a\d t\right),\end{equation}
where the $\phi^a$ are Lagrange multiplier fields that impose the constraints $\mu_a=0$.

The first step in describing new integration cycles is analytic continuation. We complexify
$\M$ to a complex symplectic manifold $\hat\M$, with complex structure $J$ and holomorphic
symplectic form $\Omega$.  (For more
details, see section \ref{anacon}.)  $p_r$ and $q^r$ analytically continue to holomorphic
functions $P_r$ and $Q^r$ on $\hat\M$, and similarly the $\mu_a$ analytically
continue to holomorphic functions $M_a$.  The action of $G$ on $\M$ analytically
continues to an action of the complex Lie group $G_\C$; the $M_a$ generate the action of
$G_\C$ via Poisson brackets (computed using the symplectic form $\Omega$).  The Lagrange
multiplier fields $\phi^a$ become complex fields that we denote $\varphi^a$, and the
complexified action is
\begin{equation}\label{zorox}\mathcal S=\oint\left(P_s\d Q^s-\varphi^a M_a\,\d t\right).\end{equation}
The integrand of the path integral is $\exp(i\mathcal S)$.  So to find a novel integration
cycle, we introduce a metric on $\hat\M$ and consider gradient flow with respect to the Morse function
\begin{equation}\label{flunkyx}h=\mathrm{Re}(i\mathcal S).\end{equation}

Without repeating all the steps from sections \ref{quantint} and \ref{superham}, we can
anticipate what will happen.  The gradient flow equation will involve a new almost
complex structure $I$ on $\hat\M$ that cannot coincide with $J$ but that we can usefully
take to anticommute with $J$.  The nicest case
will be that $I$ and $J$ and the holomorphic symplectic form $\Omega$ are all part of
a $G$-invariant hyper-Kahler structure on $\hat\M$.  In this case, the sigma-model with target $\hat \M$  has $\N=4$ supersymmetry in the
two-dimensional sense, with eight supercharges.  This is twice as much supersymmetry as was assumed in
our study in section \ref{gauginv} of an ordinary integral (as opposed to a path integral) with gauge symmetry.   All constructions
in this paper based on two-dimensional  path integrals, rather than one-dimensional ones, have involved this doubling of supersymmetry.
We gauge the $G$ symmetry preserving $\N=4$ supersymmetry.
 The flow equations of the resulting gauge-invariant sigma model
have full two-dimensional symmetry, and the sigma-model
gives a natural way to describe novel integration cycles in the path integral of the original
constrained quantum system.

Concretely,  to gauge the $G$ symmetry while preserving $\N=4$ supersymmetry, we introduce a vector multiplet, whose bosonic
components  are a two-dimensional gauge field $A$ and four scalars
in the adjoint representation of $G$.  Upon topological twisting, $A$ combines with two
of the scalars to a complex-valued gauge field $\A$, leaving a complex scalar $\sigma$.
On $S^1\times\RR_+$, with the two factors parametrized respectively by $t$ and $s$,
we write $\A=\A_t\d t+\A_s\d s$.  The complex field $\varphi$ of eqn. (\ref{zorox})
corresponds to $\A_t$.  As for the other bosonic fields $\A_s$ and $\sigma$, they were present in the case of an ordinary integral before
doubling of supersymmetry, and correspond
to the fields $A_0+iA_1$ and $\sigma$ in section \ref{supergauge}.

\appendix{}
\section{Details On The Four-Dimensional Boundary Condition}\label{details}

\def\D{{\mathcal D}}
Here we will describe in detail the boundary conditions on the
fermions of $\N=4$ super Yang-Mills theory that are implicit in
our main results such as eqn. (\ref{izonkox}) relating this theory
to Chern-Simons theory in three dimensions.  We will also verify
that these boundary conditions are elliptic.

In the notation of section 3.1 of \cite{KW}, the fermions of the
twisted theory are as follows.  The fermions of ${\cmmib F}=1$ are
one-forms $\psi$ and $\tilde\psi$, valued in the adjoint
representation.  And the fermions of ${\cmmib F}=-1$ are zero
forms $\eta$ and $\tilde\eta$ and a two-form $\chi$, all valued in
the adjoint representation.  We write $\chi=\chi^++\chi^-$, where
$\chi^+$ and $\chi^-$ are respectively selfdual and anti-selfdual.
The part of the fermion action that contains derivatives is (from
eqns. (3.39) and (3.46) of \cite{KW})
\begin{equation}\label{zmonkey}\frac{2i}{e^2}\int_M \Tr\,\left(\eta\,
\d_A\star\psi+\tilde\eta\, \d_A\star\tilde\psi
+\chi^+\d_A\psi+\chi^-\d_A\tilde\psi\right),\end{equation}
where $\star$ is the Hodge star, $\d_A$ is the gauge-covariant
exterior derivative, and wedge products of differential forms are
understood.  The reason that we write only the derivative part of
the fermion action is that nonderivative terms will not affect the
boundary  conditions or the condition of ellipticity.

It is convenient to re-express $\psi$ and $\tilde\psi$ in terms of
an adjoint-valued one-form $\psi_1$ and three-form $\psi_3$ by
$\psi=\psi_1+\star \psi_3$, $\tilde\psi=-\psi_1+\star\psi_3$.
Similarly, we re-express $\eta$ and $\tilde\eta$ in terms of an
adjoint-valued zero-form $\eta_0$ and four-form $\eta_4$ by
$\eta=(\eta_0+\star\eta_4)/2$, $\tilde\eta =(\eta_0-\star
\eta_4)/2$.  Then we combine the even degree forms to a field
$\Omega=\eta_0  +\chi+\eta_4$ that is a sum of all differential
forms of even degree.  And similarly we combine $\psi_1$ and
$\psi_3$ to a field $\Theta=\psi_1+\psi_3$ that is a sum of
differential forms of all odd degree.  Both $\Omega$ and $\Theta$
are adjoint-valued, of course.  The derivative part of the fermion
action becomes
\begin{equation}\label{zdonkey}\frac{2i}{e^2}\int_M \Tr\,\left(\Omega(\d_A+\star
\d_A\star)\Theta\right)=\frac{2i}{e^2}\int_M\Tr\,
\Omega\star(\star \d_A+\d_A\star)\Theta.\end{equation}
The operator whose boundary conditions we need to consider is
hence (modulo terms of order zero) the operator
\begin{equation}\label{ploo}\D =\star \d_A+\d_A\star \end{equation}
mapping adjoint-valued forms of odd degree to those of even
degree.

Now we want to view $\N=4 $ super Yang-Mills theory as a
topological field theory with topological supercharge $Q$. As
described in \cite{KW}, the choice of $Q$ depends on a parameter
$t=v/u$.
 From eqns. (3.23), (3.24) of \cite{KW}, and setting $u=1$, the transformation law
of the gauge field is $\delta
A=i(\psi+t\tilde\psi)=i((1-t)\psi_1+(1+t)\star\psi_3)$ and that of
the adjoint-valued one-form $\phi$ is
$\delta\phi=i(t\psi-\tilde\psi)=i((1+t)\psi_1-(1-t)\star\psi_3)$.
Introducing the complex-valued connection $\A=A+i\phi$, we find
that $\delta\A=(-1+i)(1+it)(\psi_1-i\star\psi_3)$.  To ensure
$Q$-invariance of the left hand side of (\ref{izonkox}), the
boundary condition must be such that $\delta\A|=0$.  Here, for any
differential form $\vartheta$, we write $\vartheta|$ for the restriction
of $\vartheta$ to the boundary.  (This is the restriction in the
sense of differential forms, so for example if $\vartheta$ is a
one-form, then $\vartheta|$ is the part of $\vartheta$ that is tangent
to the boundary.)  Since $\delta\A$ is a multiple of
$\psi_1-i\star\psi_3$, to ensure that $\delta\A|=0$, the boundary
condition must be, in part, that
\begin{equation}\label{turkox}\psi_1|-i\star\psi_3|=0.\end{equation}
Another condition comes from the fact that in the constructions of
section \ref{gauginv}, we want $\sigma|=0$ and hence also
$\bar\sigma|=0$.  To be compatible with this, the fermion boundary
condition must set $\delta\bar\sigma|=0$.  From eqn. (3.25) of
\cite{KW}, we have $\delta\bar\sigma=i(\eta+t\tilde
\eta)=(i/2)((1+t)\eta_0+(1-t)\star\eta_4)$, so the boundary
condition must tell us that
\begin{equation}\label{urkox}(1+t)\eta_0|+(1-t)\star\eta_4|=0.\end{equation}

When we vary the action (\ref{zmonkey}) with respect to $\Theta$, we
get a surface term
\begin{equation}\label{plondo}\frac{2i}{e^2}\int_{\partial
M}\Tr\,\left(\Omega\wedge \delta\Theta +\star\Omega\wedge\star\delta\Theta\right).
\end{equation}
The fermion boundary condition must ensure vanishing of
(\ref{plondo}), and it must ensure the vanishing of half of the
components of $\Omega$ and half of the components of $\Theta$.
Together with (\ref{turkox}) and (\ref{urkox}), these conditions
uniquely determine the boundary conditions for fermions.   For
example, we need $\chi|\wedge \psi_1|+\star\chi|\wedge
\star\psi_3|=0$.  (Here, for example, $\star\chi|$ is the
tangential part of $\star \chi$ at the boundary; in other words,
one acts with $\star$ before restricting to the boundary.)  In
view of (\ref{turkox}), this implies that the boundary condition
on $\chi$ is
\begin{equation}\label{jzombo}\chi|-i\star\chi|=0.\end{equation}
And finally, in addition to (\ref{turkox}), $\psi_1$ and $\psi_3$
must obey
\begin{equation}\label{ozero}(1+t)\star\psi_1|-(1-t)\psi_3|=0.\end{equation}

\subsection{Ellipticity}\label{elliptic}

\def\UU{{\eusm U}}
It remains to show that the boundary condition just described is
elliptic for generic $t$, in fact for $t\not=- i$. (As explained
in \cite{KW}, the model has properties similar to those of a
two-dimensional $A$-model for all values of $t$ except $t=\pm i$. A boundary condition
related to the one we consider by $\phi\to-\phi$ is elliptic for $t\not=i$.)
This is actually a straightforward exercise, the only difficulty
being that the criterion for a boundary condition to be elliptic
may not be familiar.

In general, consider fields $\Phi$ on a manifold $M$ with boundary
that obey an elliptic differential equation $\D\Phi=0$. (In our
application, the operator $\D$ is $\star \d_A+\d_A\star$, acting
from differential forms of odd degree to those of even degree;
this is a standard example of an elliptic operator.)   A general boundary
condition is given by an equation $\UU\Phi=0$, where $\UU$ is a linear map
from the space of boundary data of $\Phi$ to some linear space $V$.  (By boundary
data, we mean the boundary values if $\D$ is a Dirac-like equation, the boundary
values and normal derivatives at the boundary if $\D$ is a Laplace-like, etc.)

We can assume that $M$  looks near its boundary like $\partial
M\times \RR_+$, where $\RR_+$ is the half-line $y\geq 0$.
The notion of ellipticity is local along $\partial M$, so we can
take $\partial M=\RR^3$.  We can then also work in momentum space,
that is consider wavefunctions of definite momentum $\vec k$ along $\RR^3$.
A boundary condition is called elliptic if for every  $f\in V$ with nonzero momentum $\vec k$,
there is a unique $\Phi$ of momentum $\vec k$ that obeys  $\D\Phi=0$, decays
exponentially with increasing $y$, and also obeys a shifted version of the boundary
condition
\begin{equation}\label{zelk}\UU\Phi=f.\end{equation}
 The
importance of this criterion is that an elliptic differential
equation on a manifold with boundary, endowed with an elliptic
boundary condition, has properties similar to those of an elliptic
differential equation on a manifold without boundary.  As a
result, quantum perturbation theory has similar properties to
those that are familiar in the absence of a boundary.  (One
expects the same to be true for the nonperturbative quantum
theory, though this is certainly much harder to prove.)

Before considering our problem, let us practice with the case of Dirichlet
or Neumann boundary conditions for the scalar Laplace equation $\d\star\d\phi=0$, where $\phi$ is a scalar
field.  Let $\vec u$ be
Euclidean coordinates along $\partial M=\RR^3$, and $y$ a normal coordinate.  A general solution of $\d\star\d\phi=0$ whose dependence on $\vec u$ is $\exp(i\vec k\cdot \vec u)$ (for some non-zero momentum vector
$\vec k$) and that vanishes exponentially with increasing $y$ is a multiple of $\phi_{\vec k}=\exp(i\vec k\cdot\vec u)\exp(-|k|y)$.  Dirichlet boundary conditions mean that the map $\UU$ takes a solution of the scalar
Laplace equation on the half-space $\RR^3\times \RR^+$ to its restriction to $y=0$.  So the criterion for
ellipticity is as follows: for any constant $c$, and any nonzero $\vec k$, there must exist a unique
solution $\phi$ of the scalar wave equation, decaying exponentially with $y$, and with $\phi|_{y=0}=c
\exp(i\vec k\cdot \vec u)$.  Clearly, these conditions have the unique solution  $\phi=c\phi_{\vec k}$, so Dirichlet boundary conditions are elliptic.  The case of Neumann boundary
conditions is similar.

Let us consider in the context of this definition the boundary
conditions for $\Theta=\psi_1+\psi_3$ in our problem. These boundary
conditions were given in eqns. (\ref{turkox}) and (\ref{ozero}):
\begin{align}\notag \label{zomozo} \psi_1|-i\star\psi_3| & = 0 \\
                                               (1+t)\star\psi_1|-(1-t)\psi_3|& =
0.\end{align}
We generalize these equations to include adjoint-valued forms
$\gamma_1,$ $\gamma_3$ of odd degree:
\begin{align}\notag \label{zomon} \psi_1|-i\star\psi_3| & = \gamma_1 \\
                                               (1+t)\star\psi_1|-(1-t)\psi_3|& =
\gamma_3.\end{align}
Concretely, for $\partial M=\RR^3$, we take $\gamma_1=\theta_1\exp(i\vec k\cdot  u)$,
$\gamma_3=\theta_3\exp(i\vec k\cdot \vec u)$, with constants $\theta_1$, $\theta_3$.  We have
to show that for any nonzero $\vec k$, and any $\theta_1,$ $\theta_3$, there is a unique choice of $\psi_1,\psi_3$, with plane wave dependence on $\vec u$ and exponential decay in $y$, obeying $\D\Theta=0$ along with (\ref{zomon}).  Since all fields have the same $\exp(i\vec k\cdot \vec u)$ dependence on $\vec u$, everything reduces to algebra.

The algebra can be carried out as follows.  First of all, we can
replace the first equation in (\ref{zomon}) by a pair of equations
saying that the difference between the left and right hand sides
is both closed and coclosed (in other words, annihilated by both
$\d$ and $\d\star_3$, where $\star_3$ is the Hodge star operator
of the boundary).  The complete set of conditions then becomes:
\begin{align}\label{omo}\notag\d\star_3(\psi_1|-i\star\psi_3|)& = \d\star_3(\theta_1\exp(i\vec k\cdot \vec u))\\
                                                  \d(\psi_1|-i\star\psi_3|)&=\d(\theta_1\exp(i\vec k\cdot \vec u))\\
  (1+t)\star\psi_1|-(1-t)\psi_3|& = \theta_3\exp(i\vec k\cdot \vec u).\notag
\end{align}

Now, for wavefunctions of momentum $\vec k$, we can solve the
first and third equations in (\ref{omo}), without affecting the
second one, as follows.  One class of solutions of $\D\Theta=0$ is
\begin{equation} \label{pongo}\psi_1=\d(a\phi_{\vec k}),~~\psi_3=\star\d(b\phi_{\vec
k}), \end{equation}
with $\phi_{\vec k}$ as above and constants $a,b$.  These constants can be adjusted in a unique
way to satisfy the first and third equations in (\ref{omo}), as
long as $(1-t)/(1+t)\not=i$ or $t\not=-i$.  (The condition
$t\not=-i$ is needed, since otherwise the first and third
equations in (\ref{omo}) depend on the same linear combination of
$a$ and $b$ and we cannot solve both equations.)

If we set $\psi_1=\pm\star\psi_3$, the equation $\D\Theta=0$ becomes
the selfdual or anti-selfdual Maxwell equations, together with a
Lorentz gauge condition (for a connection defined by a one-form
$\alpha$, the Lorentz gauge condition is $\d\star \alpha=0$).  The
selfdual and anti-selfdual equations are first order equations, so
initial data are given by prescribing a gauge connection on a
three-manifold.  In fact, any one-form $\gamma_1$ on $\RR^3$ can
be written uniquely  as $\gamma_1=a_++a_-+\d w$ where $a_+$ and
$a_-$ are coclosed and are the boundary values of solutions of the
selfdual and anti-selfdual Maxwell equations on $\RR^3\times
\RR_+$ that decay with increasing $y$, and $w$ is a zero-form.  Shifting $\psi_1$ and
$\star\psi_3$ by suitable linear combinations of $a_+$ and $a_-$,
we can solve the second equation in (\ref{omo}) with fields that
decay for increasing $y$. After doing this, the procedure of the
last paragraph can be used to solve the other two equations.

This establishes what we wanted for $\Theta$.
The boundary conditions for $\Omega$ can be treated similarly.

\subsection{A Generalization}\label{generalization}

One last remark is as follows.  In this appendix, we have
constructed a boundary condition that is compatible with the
topological supersymmetry of twisted $\N=4$ super Yang-Mills
theory and ensures that $\A|$ is invariant, where $\A=A+i\phi$.
Instead, we may pick any complex number $\kappa$ with a nonzero
imaginary part and ask for invariance of $\A_\kappa|$, where
$\A_\kappa=A+\kappa \phi$. $\A_\kappa$ can be regarded as a
complex-valued connection, just like $\A$.  In this appendix, we
have set $\kappa=i$ because this case seems natural in the context
of the derivation in section \ref{gauginv}. However, more general
values of $\kappa$ are useful in the application to Khovanov homology
\cite{wittenfive}.  The above derivation works for general
$\kappa$ with minor modifications.

\noindent {\it Acknowledgments}~ I thank S. Cherkis, R. Dijkgraaf, E. Frenkel, D. Gaiotto, S. Gukov, L. Hollands, R. Mazzeo, G. Moore, N. Nekrasov, and L. Rozansky for discussions  and the physics department of Stanford University for hospitality.

Research supported in part by NSF Grant PHY-0969448.
\bibliographystyle{unsrt}

\end{document}